\documentclass[a4paper,11pt,aps,prd,preprint,showpacs,amsmath,amssymb,superscriptaddress,nofootinbib,showkeys,tightenlines]{revtex4-1}
\usepackage[english]{babel}
\usepackage{graphicx}
\usepackage{bm}
\usepackage{braket}
\usepackage{slashed}
\usepackage{pbox}
\usepackage{xspace}
\usepackage{placeins}
\usepackage{hyperref}
\usepackage{graphicx}
\usepackage{subcaption}
\allowdisplaybreaks

\newcommand*{\cf}{cf.\ }
\newcommand*{\ie}{i.\,e.\ }
\newcommand*{\eg}{e.\,g.\ }
\newcommand*{\Eq}{Eq.\,}
\newcommand*{\Eqs}{Eqs.\,}
\newcommand*{\Sec}{Sec.\,}
\newcommand*{\Fig}{Fig.\,}
\newcommand*{\Figs}{Figs.\,}

\newcommand{\chic}{\ensuremath{\chi_{cJ}}\xspace}
\newcommand{\qqbar}{\ensuremath{\ket{Q \overline{Q}}}\xspace}
\newcommand{\qqbarg}{\ensuremath{\ket{Q \overline{Q} g}}\xspace}
\newcommand{\eq}[1]{\begin{align}#1\end{align}}
\newcommand{\bfeps}{\ensuremath{\boldsymbol{\varepsilon}}}
\newcommand{\bfa}{\ensuremath{\mathbf{a}}}
\newcommand{\bfb}{\ensuremath{\mathbf{b}}}
\newcommand{\bfA}{\ensuremath{\mathbf{A}}}
\newcommand{\bfB}{\ensuremath{\mathbf{B}}}
\newcommand{\bfD}{\ensuremath{\mathbf{D}}}

\newcommand{\bfP}{\ensuremath{\mathbf{P}}}
\newcommand{\bfk}{\ensuremath{\mathbf{k}}}

\newcommand{\bfp}{\ensuremath{\mathbf{p}}}
\newcommand{\bfq}{\ensuremath{\mathbf{q}}}
\newcommand{\bfE}{\ensuremath{\mathbf{E}}}

\newcommand{\bfSigma}{\ensuremath{\boldsymbol{\sigma}}}

\newcommand{\bfDel}{\ensuremath{\boldsymbol{\nabla}}}

\newcommand{\mbf}[1]{\mathbf{#1}}

\begin{document}

\preprint{TUM-EFT 68/15}

\title{Relativistic corrections to exclusive \texorpdfstring{$\chi_{cJ} + \gamma$}{chi cJ plus photon} production from  \texorpdfstring{$e^+ e^-$}{e+ e-} annihilation}

\author{Nora Brambilla}
\email{nora.brambilla@ph.tum.de}
\affiliation{Physik-Department, Technische Universit\"at M\"unchen, \\ James-Franck-Strasse 1, 85748 Garching, Germany \vspace{0.2cm}}
\affiliation{Institute for Advanced Study, Technische Universit\"at M\"unchen, Lichtenbergstrasse 2a, 85748 Garching, Germany \vspace{0.2cm}}
\author{Wen Chen}
\email{chenwen@ihep.ac.cn}
\affiliation{Institute of High Energy Physics and Theoretical Physics Center for Science Facilities, \\ Chinese Academy of Sciences,   Beijing 100049, China \vspace{0.2cm}}
\affiliation{School of Physics, University of Chinese Academy of Science, \\    Beijing 100049, China \vspace{0.2cm}} \vspace{0.2cm}
\author{Yu Jia}
\email{jiay@ihep.ac.cn}
\affiliation{Institute of High Energy Physics and Theoretical Physics Center for Science Facilities, \\ Chinese Academy of Sciences,   Beijing 100049, China \vspace{0.2cm}}
\affiliation{School of Physics, University of Chinese Academy of Science, \\    Beijing 100049, China \vspace{0.2cm}}
\affiliation{Center for High Energy Physics, Peking University, Beijing 100871, China \vspace{0.2cm}}
\author{Vladyslav Shtabovenko}
\email{vshtabov@zju.edu.cn}
\affiliation{Physik-Department, Technische Universit\"at M\"unchen, \\ James-Franck-Str. 1, 85748 Garching, Germany \vspace{0.2cm}}
\affiliation{Zhejiang  Institute of Modern Physics, Department of Physics, Zhejiang University, Hangzhou 310027, China\vspace{0.2cm}}
\author{Antonio Vairo}
\email{antonio.vairo@ph.tum.de}
\affiliation{Physik-Department, Technische Universit\"at M\"unchen, \\ James-Franck-Str. 1, 85748 Garching, Germany \vspace{0.2cm}}

\date{\today}
\begin{abstract}
  We calculate in the nonrelativistic QCD (NRQCD) factorization framework all leading relativistic corrections
  to the exclusive production of $\chi_{cJ}+\gamma$ in $e^+ e^-$ annihilation.
  In particular, we compute for the first time contributions induced by octet operators with a chromoelectric field.
  The matching coefficients multiplying production long distance matrix elements (LDMEs) are determined through perturbative matching between QCD and NRQCD at the amplitude level.
  Technical challenges encountered in the nonrelativistic expansion of the QCD amplitudes are discussed in detail.
  The main source of uncertainty comes from the not so well known LDMEs. 
  Accounting for it, we provide the following estimates for the production cross sections at $\sqrt{s} = 10.6\textrm{ GeV}$:
  $\sigma (e^+ e^- \to \chi_{ c0} + \gamma) = (1.6 \pm 0.6) \textrm{ fb}$,  $\sigma (e^+ e^- \to \chi_{ c1} + \gamma) = (12.1 \pm 5.5) \textrm{ fb}$,
  and $\sigma (e^+ e^- \to \chi_{ c2} + \gamma) = (3.4 \pm 2.1) \textrm{ fb}$.
\end{abstract}


\maketitle

\section{\label{sec:intro} Introduction}
More than four decades have passed since the experimental groups of Samuel Ting and Burton Richter \cite{Aubert:1974js,Augustin:1974xw} discovered the $J/\psi$, 
the first observed bound state formed  by a heavy (charm) quark and a heavy antiquark.
This event was of crucial importance for the establishment of Quantum Chromodynamics (QCD) as the theory of the strong interactions.
Even though heavy quarkonia are sometimes regarded as the ``hydrogen atoms'' of QCD, 
our present understanding of these hadronic systems is not as successful as that of the hydrogen atom in Quantum Electrodynamics (QED). 
The nonperturbative nature of QCD at low energy makes the theoretical description of heavy quarkonium  on the one hand more challenging -- and on the other hand also more interesting \cite{Brambilla:2004wf,Brambilla:2010cs,Brambilla:2014jmp}.

In particular, the theoretical description of the production mechanism for charmonia and bottomonia is complicated. 
While the creation of a heavy quark pair in a high energy collision can be described in perturbative QCD, 
this is not the case for the evolution of the pair into a heavy quarkonium, which is governed by nonperturbative long-distance effects.

Effective field theories (EFTs) provide an elegant way to treat this problem by exploiting the nonrelativistic nature of the system
and the separation of the relevant scales \cite{Brambilla:2004jw}. 
The EFT resulting from integrating out modes associated with the energy scale of the heavy-quark mass or larger is known as nonrelativistic QCD (NRQCD) \cite{Bodwin:1994jh}. 
It conjectures a factorization theorem allowing one to write the quarkonium production cross section as an expansion where each term 
is the product of a short-distance coefficient and a long-distance matrix element (LDME). 
The former is computed from matching to perturbative QCD, it is a series in the strong coupling $\alpha_{\rm s}$, and it incorporates effects from the hard scale $m$ (heavy quark mass) and above. 
The latter is of nonperturbative nature and can be obtained from fits to experimental data. 
The LDMEs are assumed to be universal and to obey power counting rules in the relative heavy quark velocity $v$, which in a nonrelativistic system is much smaller than one.
Hence, the importance of each term in the NRQCD-factorized cross section can be estimated by powers of $\alpha_{\rm s}$ and $v$. 
This counting gives NRQCD predictive power and allows for a systematic inclusion of radiative and relativistic corrections.

NRQCD also predicts that quarkonium formation is not limited to heavy quark pairs in a color singlet configuration, 
as it was assumed by the color singlet model \cite{Einhorn:1975ua,Ellis:1976fj,Carlson:1976cd}. 
Instead, the Fock state of a heavy quarkonium can be schematically understood as a sum over different contributions. 
For example, for the $P$-wave charmonium $\chi_{cJ}$ we have
\eq{
\ket{\chi_{cJ} (^3 P_J)} \sim \mathcal{O}(1) \ket{c \bar{c} (^3 P_J)} + \mathcal{O}(v) \ket{c \bar{c} (^3 S_1) g } + \mathcal{O}(v^2),
}
where higher Fock states that involve gluons are suppressed by powers of $v$. 
Nevertheless, in higher order calculations such contributions become relevant and must be included to obtain a consistent result.

Electromagnetic quarkonium production can be regarded as a relatively simple and clean process 
that makes it a perfect testing ground for verifying predictions made by NRQCD. 
Electromagnetic quarkonium production is an important subject of study at the BES-III experiment at the tau-charm factory BEPC-II in China,  
and at the Belle experiment at the KEKB asymmetric B-factory in Japan. 
For the latter, one particularly interesting electromagnetic quarkonium production process is the exclusive production of a heavy quarkonium 
and a hard ($|\bfk| \gtrsim m $) photon in electron-positron annihilations.

The scope of this work is to improve on the NRQCD calculation of the process $e^+ e^- \to \gamma^\ast \to \chi_{cJ} + \gamma$ by including higher Fock state contributions of $\mathcal{O} ( \alpha_{\rm s}^0 v^2)$.
The leading cross section of $\mathcal{O} ( \alpha_{\rm s}^0 v^0)$ was obtained in \cite{Chung:2008km}. 
Subsequently, corrections of $\mathcal{O} ( \alpha_{\rm s} v^0)$ \cite{Sang:2009jc,Li:2009ki}, partial corrections of $\mathcal{O} ( \alpha_{\rm s}^0 v^2)$ \cite{Li:2013nna,Chao:2013cca} 
and finally partial corrections of $\mathcal{O} ( \alpha_{\rm s} v^2)$ \cite{Xu:2014zra} were computed as well. 
One of the reasons why this process has attracted so much interest from the theory side in the past years is the anticipated connection between $C$-even quarkonia  and some exotic $XYZ$ particles
\cite{Brambilla:2010cs,Li:2009zu,Molina:2009ct,Li:2012vc}, although such identifications are still controversial \cite{Guo:2012tv,Olsen:2014maa}. 
In any case, there is clear experimental evidence \cite{Ablikim:2013dyn} that the $C$-even state $X(3872)$ can be measured in the same channel [$e^+ e^- \to X(3872) + \gamma$] as our process of interest. 
Therefore, irrespective of the true nature of $X(3872)$, in order to study this state in more detail, 
it is very useful to have precise and unambiguous predictions for $e^+ e^- \to \gamma^\ast \to \chi_{cJ} + \gamma$.
This is all the more important in view of the fact that no experimental measurements for the electromagnetic production of \chic and a hard photon are yet available, 
while good perspectives for this measurement exist at Belle II.

We emphasize that the above-mentioned NRQCD studies considered only operators that contribute through the dominant Fock state \qqbar. 
However, operators that project on the subleading Fock state \qqbarg already show up at $\mathcal{O}(\alpha_{\rm s}^0 v^2)$. 
The importance of this kind of operators was already noticed in the past.
In particular, they were explicitly incorporated in studies of the decay $\chi_{cJ} \to \gamma \gamma$ \cite{Ma:2002eva,Brambilla:2006ph}, 
which is a process very similar to $\gamma^\ast \to \chi_{cJ} + \gamma$.
In the present work, we will investigate these missing contributions to the production cross section 
by determining at $\mathcal{O}(\alpha_{\rm s}^0 v^2)$ in NRQCD the matching coefficients of octet operators containing chromoelectric fields.

The paper is organized in the following way. In \Sec\ref{sec:NRQCDcross} we discuss the relevant operators and LDMEs, 
and write the NRQCD-factorized cross sections for $e^+e^- \to \chi_{cJ} + \gamma$ at the desired precision. 
Different strategies for the determination of the NRQCD matching coefficients are discussed in \Sec\ref{sec:matching}. 
In \Sec\ref{sec:nrqcdside} we outline the calculation of the NRQCD amplitudes, while in \Sec\ref{sec:qcdside} we describe the QCD side of the matching 
and present the short-distance coefficients obtained with two independent methods. 
A long-standing discrepancy between the results of \cite{Ma:2002eva} and \cite{Brambilla:2006ph} 
regarding some matching coefficients that enter the NRQCD-factorized decay rates for $\chi_{cJ} \to \gamma \gamma$ at $\mathcal{O}(\alpha_{\rm s}^0 v^2)$ is resolved in \Sec\ref{sec:decay}. Sections \ref{sec:xsection} and \ref{sec:numerics} contain the final cross sections and  numerical predictions. 
Finally, a summary of the obtained results and their impact is presented in \Sec\ref{sec:summary}.
In Appendix \ref{sec:appendix0} we present additional  short-distance coefficients that arise from the matching. 
They are not relevant for our calculation but they are new and appear at $\mathcal{O}(v^4)$ contributing to the dominant Fock state. 
Appendix \ref{sec:appendix1} contains the details of the calculation made with the threshold expansion method of Braaten and Chen \cite{Braaten:1996jt}. 
Appendix \ref{sec:appendix3} contains useful formulas about polarization sums of hadrons.
Appendix \ref{a3} provides a derivation of the generalized Gremm-Kapustin relations for \texorpdfstring{$^3 P_J$}{3PJ} quarkonia.

\section{\label{sec:NRQCDcross} NRQCD factorization for the \texorpdfstring{$e^+e^- \to \chi_{cJ} + \gamma$}{chi cJ plus photon} cross section}

\subsection{Definition of NRQCD operators for exclusive quarkonium production \label{Sec:NRQCDdef}}
A generic NRQCD operator of naive scaling dimension $d_n$ (which is given by the dimension of the gluonic and fermionic fields only)
that describes the production of a heavy quarkonium $H$ can be written as \cite{Bodwin:1994jh}
\begin{equation}
\tilde{\mathcal{O}}_n = \chi^\dagger  \mathcal{K}_n \psi  \biggl ( \sum_X \sum_{m_J}  \ket{H +X } \bra{H + X} \biggr ) \psi^\dagger \mathcal{K}_n' \chi,
\end{equation}
where $\mathcal{K}^{(')}_n$ consists of a spin matrix (1 or $\bfSigma$), a color matrix (1 or $T^a$) 
and a polynomial in $\bfB^i \equiv \epsilon^{ijk} G^{kj}/2$, $\bfE^i \equiv G^{i0}$, and $\bfD = \bfDel - i g \bfA$ with $A^\mu$ being the gluon field, 
$G^{\mu \nu}$ the field-strength tensor of QCD  and $g$ the gauge coupling\footnote{
In the case of polarized production $\mathcal{K}^{(')}_n$ may also depend on polarization vectors \cite{Bodwin:1994jh}. 
Explicit NRQCD calculations of such processes can be found, e.\,g., in \cite{Braaten:1996jt}. 
In this work, we will restrict ourselves only to unpolarized production.}. 
Here and throughout the paper, bold fonts specify Cartesian 3-vectors. 
Furthermore, $\psi$ ($\chi$) denotes a Pauli field that annihilates (creates) a heavy quark (antiquark), 
while $\ket{ H +X }$ is a Fock state that contains a quarkonium $H$ and all other particles $X$ that appear in the final state,  
and have energy and momenta that are typically much smaller than the hard scale $m$. 
The summations run over all allowed $X$ states and the magnetic quantum number $m_J = -J, \ldots, J$, where $J$ denotes the spin of the quarkonium state. 
The operators $\tilde{\mathcal{O}}_n$ enter the NRQCD factorization formula as vacuum expectation values
\begin{equation}
\braket{0|\tilde{\mathcal{O}}_n|0} = \bra{0} \chi^\dagger  \mathcal{K}_n \psi  \biggl ( \sum_X \sum_{m_J}  \ket{H +X } \bra{H + X} \biggr ) \psi^\dagger \mathcal{K}_n' \chi \ket{0} \label{eq:ldme-prod},
\end{equation}
where $\ket{0}$ denotes the QCD vacuum and the quarkonium states have nonrelativistic normalization, i.\,e., 
\begin{equation}
\braket{H(\bfP)|H(\bfP')} = (2\pi)^3 \delta^{3} (\bfP-\bfP'),
\end{equation}
with $\bfP^{(')}$ being the total 3-momentum of the heavy quarkonium.

The sum over $X$ in \Eq\eqref{eq:ldme-prod} is one of the reasons why production matrix elements $\braket{0|\tilde{\mathcal{O}}_n|0}$ cannot easily be calculated in lattice simulations \cite{Bodwin:2003kc}. 
However, in exclusive electromagnetic production this sum is absent, such that the corresponding LDME reads
\begin{equation}
\braket{0|\mathcal{O}_n|0} = \bra{0} \chi^\dagger  \mathcal{K}_n \psi  \biggl ( \sum_{m_J}  \ket{H} \bra{H} \biggr ) \psi^\dagger \mathcal{K}_n' \chi \ket{0} \label{eq:ldme-prod-em}.
\end{equation}
Equation \eqref{eq:ldme-prod-em} can be further simplified using the rotational invariance of the matrix elements \cite{Bodwin:1994jh,Braaten:1996jt}. 
Since all matrix elements that differ only in the quantum number $m_J$ are identical, 
the sum over $m_J$ trivially reduces to $(2J+1)$ and the LDME factorizes into a product of two quarkonium-to-vacuum matrix elements
\begin{equation}
\braket{0|\mathcal{O}_n|0} = (2J+1) \bra{0} \chi^\dagger  \mathcal{K}_n \psi \ket{H} \bra{H}  \psi^\dagger \mathcal{K}_n' \chi \ket{0}.
\end{equation}

The matrix elements $\bra{0} \chi^\dagger  \mathcal{K}_n \psi \ket{H}$ and $\bra{H}  \psi^\dagger \mathcal{K}_n' \chi \ket{0}$ are the same 
that enter NRQCD factorization formulas for exclusive electromagnetic decays \cite{Bodwin:2002kk}. 
Hence, up to the prefactor $(2J+1)$, the LDME for exclusive electromagnetic production $\braket{0|\mathcal{O}_n|0}$ 
is identical to the corresponding LDME for electromagnetic decay $\braket{H|\mathcal{O}_{n\, \textrm{em}}|H}$
with $\mathcal{O}_{n\, \textrm{em}} = \psi^\dagger \mathcal{K}_n' \chi \ket{0} \bra{0} \chi^\dagger  \mathcal{K}_n \psi$:
\begin{equation}
\braket{0|\mathcal{O}_n|0}  = (2J+1) \braket{H|\mathcal{O}_{n\, \textrm{em}}|H}.
\end{equation}
For this reason from now on we will absorb the factor $(2J+1)$ into the matching coefficients~$F_n$:
\begin{equation}
F_n \to (2J+1) F_n, \qquad 
\braket{0|\mathcal{O}_n|0} \to \frac{1}{2J+1}\braket{0|\mathcal{O}_n|0} \,. 
\label{eq:redefineMatchingCoeff}
\end{equation}

\subsection{Color singlet and color octet production operators}
In NRQCD it is customary to classify production operators according to the color configuration of the $Q \overline{Q}$ pair in the leading Fock state 
through which these operators contribute to the process. 
In the dominant Fock state $\qqbar$ the heavy quark pair is in a color singlet configuration; hence operators that contribute dominantly through this state are denoted as singlet operators. 
An octet operator is an operator that gives a nonvanishing contribution only when the  $Q \overline{Q}$ pair is in a color octet configuration, such as in the subleading Fock state \qqbarg. 
An example for a color octet operator is
\begin{equation}
\chi^\dagger \bfSigma  T^a \psi  \biggl ( \sum_X \sum_{m_J}  \ket{H +X } \bra{H + X} \biggr ) \psi^\dagger \bfSigma T^a \chi \label{eq:prodincl-co},
\end{equation}
which contributes to the inclusive production of a spin-triplet $P$-wave quarkonium at leading order in $v$. 
In exclusive electromagnetic production there are no color octet operators similar to \Eq\eqref{eq:prodincl-co}. 
This is because a matrix element like  $\bra{0} \chi^\dagger \bfSigma  T^a \psi \ket{H}$ vanishes due to color conservation. 
However, if we replace $T^a$ with the chromoelectric field multiplied by the gauge coupling, we obtain
\begin{equation}
\bra{0} \chi^\dagger (g \bfE \cdot \bfSigma) \psi \ket{H},
\end{equation}
which gives a nonvanishing contribution to the exclusive electromagnetic production of a $P$-wave quarkonium at subleading order in $v$. 
The corresponding operator reads
\begin{equation}
\frac{1}{3} \left (  \chi^\dagger  ( - \frac{i}{2} \overset{\leftrightarrow}{\bfD} \cdot \bfSigma )   \psi
\ket{ H} \bra{H} \psi^\dagger  ( g \bfE \cdot  \bfSigma )  \chi
 + \textrm{H.c.} \right ), \label{eq:co-elprod}
\end{equation}
with $\psi^\dagger \overleftrightarrow{\bfD} \chi  \equiv \psi^\dagger (\bfD \chi) - (\bfD \psi)^\dagger\chi$, where H.c.\ stands for Hermitian conjugate. 
According to the usual NRQCD classification, the operator in \Eq\eqref{eq:co-elprod} is a color octet operator 
in the sense that this operator necessarily projects on the subleading Fock state \qqbarg.\footnote{ 
Many NRQCD practitioners prefer to use the words ``color octet'' to denote only operators similar to the one in \Eq\eqref{eq:prodincl-co}.
To avoid misinterpretations we will speak in the following of operators that contribute through the leading or subleading quarkonium Fock state \qqbar or \qqbarg respectively.} 

To our knowledge, the number of studies that considered contributions from operators with chromoelectric or chromomagnetic fields to NRQCD-factorized cross sections and decay rates is very small. 
One of such studies was concerned with the electromagnetic and hadronic decays of the $S$-wave heavy quarkonia \cite{Bodwin:2002hg}, 
where LDMEs containing one chromoelectric field were of relative order $v^4$ as compared to the leading order LDMEs. 
The authors, however, managed to eliminate such LDMEs from their basis by using the Gremm-Kapustin relations \cite{Gremm:1997dq}. 
An explicit determination of the matching coefficients multiplying \bfE-field dependent LDMEs was carried out in \cite{Ma:2002eva}, 
where the authors considered electromagnetic decays of the $P$-wave spin triplet quarkonia. 
There, the relevant LDMEs contributed already at relative order $v^2$. 
Matching coefficients of various decay LDMEs with chromoelectric and chromomagnetic fields were determined in \cite{Brambilla:2006ph,Brambilla:2008zg}, 
where the authors systematically studied relativistic corrections to electromagnetic and hadronic decays of heavy quarkonia up to relative order $v^7$.

As far as the production of heavy quarkonia is concerned, one should mention the study \cite{Bodwin:2012xc} of higher order relativistic corrections 
to the gluon fragmentation into a spin-triplet $S$-wave quarkonium. 
One of the LDMEs contributing at relative order $v^4$ involves a combination of a covariant derivative and a chromoelectric field. 
Also in this case the authors could eliminate it from their basis via a field redefinition. 
We believe that the present work is the first study, in which matching coefficients multiplying LDMEs with chromoelectric fields are explicitly computed for a heavy quarkonium production process
and cannot be traded with other operators.\footnote{
As shown in \cite{Brambilla:2006ph}, using field redefinitions the relevant LDMEs with chromoelectric fields could, in principle, 
be traded for LDMEs that involve a time derivative acting on fermion and gluon fields. We have chosen not to include such operators in our basis.
}

\subsection{Production cross section and power counting rules}
\label{sec:powercounting}
After having clarified the NRQCD framework and the terminology of this work, let us proceed with applying NRQCD factorization to the exclusive process $e^+ e^- \to \gamma^\ast \to \chi_{cJ} + \gamma$. 
To assess the relative importance of the NRQCD matrix elements we adopt the power counting rules from \cite{Bodwin:1994jh}. 
To have a homogeneous counting, we count $\alpha_{\rm s} (m) \sim v^2$, which will be important when combining our results with the existing radiative corrections. 
Each of our formulas depends on three LDMEs, where two of them are $v^2$ suppressed as compared to the leading LDME. 
When we refer to relative $\mathcal{O}(v^2)$ corrections, we always do so with respect to the absolute size of the leading order matrix element. 
Hence, at relative $\mathcal{O}(v^2)$ the NRQCD-factorized cross sections for $e^+ e^- \to \gamma^\ast \to \chi_{cJ} + \gamma$ can be written as
\begin{subequations}
\eq{
&\sigma (e^+ e^- \to \chi_{ c_0} + \gamma) = \frac{F_1(^3 P_0)}{3 m^4} \braket{0| \chi^\dagger \left  ( - \frac{i}{2} \overset{\leftrightarrow}{\bfD} \cdot \bfSigma \right ) \psi
| \chi_{c0}} \braket{\chi_{c0}| \psi^\dagger  \left ( - \frac{i}{2} \overset{\leftrightarrow}{\bfD} \cdot \bfSigma \right ) \chi
| 0} \nonumber \\
&+ \frac{G_1(^3 P_0)}{6 m^6} \left [ \braket{0| \chi^\dagger  \left ( - \frac{i}{2} \overset{\leftrightarrow}{\bfD} \cdot \bfSigma \right )   \psi
| \chi_{c0}} \braket{\chi_{c0}| \psi^\dagger  \left ( - \frac{i}{2} \overset{\leftrightarrow}{\bfD} \cdot \bfSigma \right )  \left ( - \frac{i}{2} \overset{\leftrightarrow}{\bfD}  \right )^2 \chi
| 0} + \textrm{H.c.} \right ] \nonumber\\
& + \frac{T_8(^3 P_0)}{3 m^5}  \left [i  \braket{0| \chi^\dagger \left ( - \frac{i}{2} \overset{\leftrightarrow}{\bfD} \cdot \bfSigma \right )   \psi
| \chi_{c0}} \braket{\chi_{c0}| \psi^\dagger  ( g \bfE \cdot  \bfSigma )  \chi
| 0} + \textrm{H.c.} \right ] \nonumber \\
& \equiv \frac{F_1(^3 P_0)}{m^4} \braket{0| \mathcal{O}_1 (^3 P_0)|0} + \frac{G_1(^3 P_0)}{m^6} \braket{0| \mathcal{P}_1 (^3 P_0)|0} + \frac{T_8(^3 P_0)}{m^5} \braket{0| \mathcal{T}_8 (^3 P_0)|0},
}
\eq{
& \sigma (e^+ e^- \to \chi_{c1} + \gamma) = \frac{F_1(^3 P_1)}{2 m^4}  \braket{0| \chi^\dagger  \left ( - \frac{i}{2} \overset{\leftrightarrow}{\bfD} \times \bfSigma   \right ) \psi
| \chi_{c1}} \cdot\braket{\chi_{c1}| \psi^\dagger \left ( - \frac{i}{2} \overset{\leftrightarrow}{\bfD} \times \bfSigma  \right ) \chi
| 0} \nonumber \\
&+ \frac{G_1(^3 P_1)}{4 m^6} \left [ \braket{0| \chi^\dagger \left ( - \frac{i}{2} \overset{\leftrightarrow}{\bfD} \times \bfSigma  \right )   \psi
| \chi_{c1}}\cdot \braket{\chi_{c1}| \psi^\dagger  \left ( - \frac{i}{2} \overset{\leftrightarrow}{\bfD} \times \bfSigma \right  ) \left ( - \frac{i}{2} \overset{\leftrightarrow}{\bfD}  \right )^2 \chi
| 0} + \textrm{H.c.} \right ]\nonumber \\
& + \frac{T_8(^3 P_1)}{2 m^5}   \left [i  \braket{0| \chi^\dagger  \left ( - \frac{i}{2} \overset{\leftrightarrow}{\bfD} \times \bfSigma  \right )   \psi
| \chi_{c1}}\cdot \braket{\chi_{c1}| \psi^\dagger  ( g \bfE \times  \bfSigma )  \chi
| 0} + \textrm{H.c.} \right ] \nonumber  \\
& \equiv \frac{F_1(^3 P_1)}{m^4} \braket{0| \mathcal{O}_1 (^3 P_1)|0} + \frac{G_1(^3 P_1)}{m^6} \braket{0| \mathcal{P}_1 (^3 P_1)|0} + \frac{T_8(^3 P_1)}{m^5} \braket{0| \mathcal{T}_8 (^3 P_1)|0},
}
\eq{
& \sigma (e^+ e^- \to \chi_{c2} + \gamma) = \frac{F_1(^3 P_2)}{m^4} \braket{0| \chi^\dagger \left ( - \frac{i}{2} \overleftrightarrow{\bfD}^{(i} \bfSigma^{j)} \right) \psi | \chi_{c2}} \braket{\chi_{c2}| \psi^\dagger \left ( - \frac{i}{2} \overleftrightarrow{\bfD}^{(i} \bfSigma^{j)} \right  ) \chi
| 0} \nonumber \\
&+ \frac{G_1(^3 P_2)}{2 m^6} \left [ \braket{0| \chi^\dagger \left ( - \frac{i}{2} \overleftrightarrow{\bfD}^{(i} \bfSigma^{j)}  \right )   \psi
| \chi_{c2}} \braket{\chi_{c2}| \psi^\dagger \left ( - \frac{i}{2} \overleftrightarrow{\bfD}^{(i} \bfSigma^{j)}  \right )  \left ( - \frac{i}{2} \overset{\leftrightarrow}{\bfD} \right  )^2 \chi
| 0} + \textrm{H.c.} \right ] \nonumber\\
& + \frac{ T_8(^3 P_2)}{m^5}  \left [i  \braket{0| \chi^\dagger  \left ( - \frac{i}{2} \overleftrightarrow{\bfD}^{(i} \bfSigma^{j)} \right )   \psi
| \chi_{c2}} \braket{\chi_{c2}| \psi^\dagger   (g \bfE^{(i} \bfSigma^{j)})  \chi
| 0} + \textrm{H.c.} \right ] \nonumber \\
& \equiv \frac{F_1(^3 P_2)}{m^4} \braket{0| \mathcal{O}_1 (^3 P_2)|0} + \frac{G_1(^3 P_2)}{m^6} \braket{0| \mathcal{P}_1 (^3 P_2)|0} + \frac{T_8(^3 P_2)}{m^5} \braket{0| \mathcal{T}_8 (^3 P_2)|0},
}
\label{eq:cs}
\end{subequations}
with $\displaystyle \bfa^{(i} \bfb^{j)} \equiv \frac{\bfa^i \bfb^j + \bfa^j \bfb^i}{2} - \frac{\delta^{ij}}{3} \bfa \cdot \bfb$. 
The subscript $1$ labels operators contributing through the dominant Fock state \qqbar, 
while the subscript $8$ labels operators that project on the subleading Fock state \qqbarg only.  

According to the NRQCD power counting rules, the scaling of $\braket{0| \mathcal{O}_1 (^3 P_J)|0}$ is $v^5$, while  $\braket{0| \mathcal{P}_1 (^3 P_J)|0}$ scales as $v^7$ 
due to the presence of two additional covariant derivatives. 
In comparison to $\braket{0| \mathcal{O}_1 (^3 P_J)|0}$ the matrix element $\braket{0| \mathcal{T}_8 (^3 P_J)|0}$ contains one covariant derivative less, 
but involves a chromoelectric field, which scales as $v^3$ so that the absolute size of this LDME is also $v^7$\cite{Bodwin:1994jh}.

It is worth noting that, in general, expectation values of operators like \Eq\eqref{eq:prodincl-co}, which contribute through the subleading Fock state \qqbarg, 
contain an additional suppression in $v$, if the $Q \overline{Q}$ pair in the color octet configuration is produced with a different angular momentum than the $Q \overline{Q}$ pair in the dominant Fock state. 
The reason for this is that in the corresponding short-distance process we create a color octet $Q \overline{Q}$ pair that emits a soft gluon during its nonperturbative evolution into a heavy quarkonium. 
The emission of the gluon that changes the orbital angular momentum of the pair by one unit and converts it into a color singlet corresponds to an electric dipole (E1) transition, which accounts for the extra suppression.
However, according to the power counting rules from \cite{Bodwin:1994jh}, for LDMEs that contain chromoelectric fields, 
the additional $\mathcal{O}(v)$ suppression is already accounted for in the power counting rule of $\bfE$.\footnote{
We thank G.\,Bodwin for communications on this point; \cf also \cite{Bodwin:2002hg} where the authors explicitly apply this prescription when estimating the importance of LDMEs with chromoelectric fields.} 
This is why $\braket{0| \mathcal{T}_8 (^3 P_J)|0}$ is only $\mathcal{O}(v^2)$ suppressed as compared to $\braket{0| \mathcal{O}_1 (^3 P_J)|0}$ and hence parametrically as important as $\braket{0| \mathcal{P}_1 (^3 P_J)|0}$.

It is well known that the power counting of \cite{Bodwin:1994jh} is not the most conservative choice in the framework of NRQCD. 
In particular, the underlying assumption that the typical hadronic scale $\Lambda_{\textrm{QCD}}$ is of order $m v^2$ may not be satisfied in all quarkonium production and decay processes. 
A more cautious choice would be to assume that $\Lambda_{\textrm{QCD}} \sim m v$ as was done  \eg in \cite{Brambilla:2006ph} 
(\cf also \cite{Brambilla:2004jw} and references therein for a detailed discussion of different countings in NRQCD).
 In the resulting conservative power counting, each chromoelectric field scales only as $v^2$ (not as $v^3$) but there is also an additional power of $v$ coming from the subleading Fock state $\ket{Q \overline{Q} g}$. 
It is therefore interesting to observe that $\braket{0| \mathcal{T}_8 (^3 P_J)|0}$ scales as $v^7$ both in the perturbative counting of \cite{Bodwin:1994jh} and in the conservative counting of \cite{Brambilla:2006ph}.

Equations \eqref{eq:cs} depend on nine matching coefficients. 
The matching coefficients $F_1(^3 P_J)$ and $G_1(^3 P_J)$ at $\mathcal{O}(\alpha_{\rm s}^0)$ were computed in \cite{Chung:2008km} and \cite{Li:2013csa,Chao:2013cca} respectively. 
In \cite{Li:2013csa,Chao:2013cca} (as well as in the order $\alpha_{\rm s}$ analysis of \cite{Xu:2014zra}) the authors, although working at $\mathcal{O}(v^2)$, 
did not include the matrix element $\braket{0| \mathcal{T}_8 (^3 P_J)|0}$ in their cross section formula.
Therefore the corresponding matching coefficients $T_8(^3 P_J)$ remained unknown.
In this work, we compute for the first time the matching coefficients $T_8(^3 P_J)$ at $\mathcal{O}(\alpha_{\rm s}^0)$. 
This is the last missing piece to have the complete $\mathcal{O}( v^2)$ corrections for exclusive electromagnetic production of $\chi_{cJ}$ and a hard photon. 
The calculation is essentially a tree level calculation, but nontrivial, as we need to work with a quarkonium composed of two heavy quarks and a gluon. 
How to determine $T_8(^3 P_J)$ from matching QCD to NRQCD will be discussed in the next section.

\section{\label{sec:matching} Matching NRQCD production coefficients}
\subsection{Matching conditions}
There exist different approaches to calculate the NRQCD matching coefficients in a quarkonium production process. 
In all of them the matching is done between perturbative QCD and perturbative NRQCD, relying on the same behavior of both theories at low energies. 
The NRQCD matrix elements are evaluated in such a way that the quarkonium Fock states are replaced by the perturbative Fock states containing on-shell quarkonium constituents, \eg
\begin{equation}
\braket{0| \chi^\dagger  \mathcal{K}_n \psi | H  } \braket{H|  \psi^\dagger \mathcal{K}_n' \chi
| 0}  \to \braket{0| \chi^\dagger  \mathcal{K}_n \psi | Q \overline{Q}  } \braket{ Q \overline{Q}|  \psi^\dagger \mathcal{K}_n' \chi
| 0}. \label{eq:nrqcdRepl}
\end{equation}
The explicit calculation of matrix elements on the right-hand side of \Eq\eqref{eq:nrqcdRepl} in perturbative NRQCD will be explained in \Sec\ref{sec:nrqcdside}. 
To derive the values of the coefficients multiplying such matrix elements, we need to impose a relation between suitable quantities (e.\,g., Green's functions, cross sections, on-shell amplitudes) in QCD and NRQCD. 
In EFTs such relations are called matching conditions.

In \cite{Bodwin:1994jh} the matching condition for quarkonium production was given as
\eq{
\sigma ( Q \overline{Q} ) |_{\textrm{pert QCD}} = \sum_n \frac{F_n}{m^{d_n-4}} \braket{0|\mathcal{O}_n^{Q \overline{Q}}|0} |_{\textrm{pert NRQCD}} \label{eq:factFormCrossSection},
}
which imposes an equality between the partonic cross section to produce an on-shell heavy quark pair in perturbative QCD 
and the sum of suitable LDMEs multiplied by matching coefficients $F_n$ and inverse powers of the heavy quark mass in perturbative NRQCD.
Assuming that the relative momentum of the heavy quarks in the rest frame, $\bfq$, is small, one can  expand both sides of \Eq\eqref{eq:factFormCrossSection} in $|\bfq|/m$. 
Within this nonrelativistic expansion one can read off the values of the matching coefficients $F_n$ and then substitute them into the NRQCD-factorized production cross section. A more technical explanation of this approach can be found in \cite{Maksymyk:1997rq}. 

One of the difficulties related to practical applications of \Eq\eqref{eq:factFormCrossSection} is the necessity to perform a nonrelativistic expansion of the phase-space measure on the QCD side of the matching. 
In processes, where heavy quarkonium is produced together with other particles, such an expansion tends to become complicated and requires great care.

In case of exclusive reactions (such as our process of interest) it is also possible to employ the NRQCD factorization at the amplitude level \cite{Braaten:1998au,Braaten:2002fi},
\eq{
\mathcal{A}_{\textrm{pert QCD}} = \sum_n c^{i_1 \ldots i_k}_n \bra{Q \overline{Q}}  \psi^\dagger \mathcal{K}^{i_1 \ldots i_k}_n \chi \ket{0} \equiv \mathcal{A}_{\textrm{pert NRQCD}} \label{eq:factFormAmp},
}
where $c^{i_1 \ldots i_k}_n$ is a short-distance coefficient. Equation \eqref{eq:factFormAmp} is an equality between the on-shell amplitude to produce a heavy quark pair in perturbative QCD and a sum of quarkonium-to-vacuum matrix elements
multiplied or contracted with short-distance coefficients in perturbative NRQCD. 
Once the short-distance coefficients $c_n$ are known, they can be substituted into the NRQCD-factorized production amplitude
\eq{
\mathcal{A}_{\textrm{NRQCD}} = \sum_n c^{i_1 \ldots i_k}_n \bra{H}  \psi^\dagger \mathcal{K}^{i_1 \ldots i_k}_n \chi \ket{0} \label{eq:npNRQCDAmp}.
}
Squaring \Eq\eqref{eq:npNRQCDAmp} and integrating over the phase space of the physical quarkonium we obtain the NRQCD production cross sections as in \Eqs\eqref{eq:cs}. 
Notice that in general not all matrix elements in $\mathcal{A}_{\textrm{pert NRQCD}}$ give a nonvanishing contribution to $\mathcal{A}_{\textrm{NRQCD}}$. 
Since QCD amplitudes do not have definite angular momentum (\cf \Sec\ref{sec:qcdside}), in the matching we will determine more short-distance coefficients than required in \Eqs\eqref{eq:cs}.
The advantage of this method is that the matching can be done in a much simpler way, since it does not require one to compute the QCD matrix element squared and expand the phase space measure in $|\bfq|/m$. 
In this work we calculate our matching coefficients by employing the NRQCD factorization at the amplitude level.

\subsection{Kinematics}
\label{sec:kinematics}
We now make explicit the kinematics that we will use throughout the matching calculation. 
We distinguish between two frames of reference. 
The laboratory frame is the center of mass (CM) frame of the colliding leptons, where the heavy quarkonium and the photon fly apart from the interaction point in opposite directions. 
The rest frame is the frame in which the heavy quarkonium is at rest.

We denote the 4-momenta of the heavy quarks and the gluon in the laboratory frame with $p_1$, $p_2$ and $p_g$ respectively; 
$k$ stands for the 4-momentum of the photon, while the momenta of the colliding leptons are labeled by $l_1$ and $l_2$.
We will also make use of the polarization vectors for the photon $\varepsilon_\gamma$, for the gluon $\varepsilon_g$ and for the dilepton system $L^\mu \equiv e \bar{v}(l_2) \gamma^\mu u(l_1)$. 
The direction of the photon 3-momentum is referred to as $\hat{\bfk} \equiv \bfk / |\bfk| $. 
In the matching all the external momenta are put on-shell and the masses of the leptons are neglected as compared to the CM energy $\sqrt{s}$. Hence,
\eq{
p_1^2 &= p_2^2 = m^2, \\
l_1^2 &= l_2^2 = k^2 = p_g^2 = 0, \\
\varepsilon_\gamma \cdot k  & =  \varepsilon_g \cdot p_g = (l_1 + l_2) \cdot L =  0.
}
Finally, the 4-momentum of the heavy quarkonium is denoted by $P$. 
In the perturbative matching $P$ is expressed as a sum of the heavy quark and gluon momenta, \ie $P = p_1 + p_2$ for a $Q\overline{Q}$ system and $P = p_1 + p_2 + p_g$ for a $Q\overline{Q} g$ system. 
In the physical process, however, $P$ is the 4-momentum of the quarkonium with $P^2 = M_{\chi_{cJ}}^2$, where $M_{\chi_{cJ}}$ denotes the mass of \chic measured in experiments. 
We will explicitly state the meaning that we assign to $P$ at the different stages of this work.

To distinguish a laboratory frame vector from a rest frame vector, the latter will be assigned an additional subscript $R$. 
For example, when considering the $Q\overline{Q}$ system, it is convenient to introduce the relative momentum of the heavy quarks in the rest frame, defined as $\bfq \equiv ( \bfp_{1,R} - \bfp_{2,R})/2$. 
In the case of the $Q\overline{Q} g$ system, two relative momenta are needed, given by
\begin{equation}
\bfq_1 = \frac{1}{2} \left ( \bfp_{1,R} - \bfp_{2,R} \right ), \quad \quad
\bfq_2 = \frac{1}{6} \left ( 2 \bfp_{g,R} - \bfp_{1,R} - \bfp_{2,R}\right ).
\end{equation}
More details regarding the kinematics of the $Q\overline{Q}$ and $Q\overline{Q}g$ systems can be found in Appendix~\ref{sec:appendix1}.

\subsection{Automatized expansions}
As all matching prescriptions involve nonrelativistic expansions of the perturbative QCD amplitudes, let us also briefly discuss different strategies to realize this task.

In \cite{Bodwin:1994jh}, matching calculations for decays were done in such a way that the expanded QCD amplitudes were explicitly rewritten in terms of energies, Cartesian 3-vectors, Pauli matrices and Pauli spinors.
Since such objects naturally arise on the NRQCD side of the matching, having both sides of the matching equation expressed as products of 3-vectors and Pauli structures facilitates the extraction of the matching coefficients.
This approach to NRQCD matching calculations was developed further and generalized to be applicable both to production and decays in \cite{Braaten:1996jt} and \cite{Braaten:1996rp}.
Following \cite{Braaten:1996rp} we will denote it as the threshold expansion method. 
Applying the technique developed by Braaten and Chen is conceptually simple, yet the calculations tend to become rather cumbersome when one goes beyond tree level or leading order in $v$. 
In particular, the necessity to work with noncovariant objects on the QCD side of the matching makes such
calculations quite tedious not only when done by hand but also when automatized with \textsc{FORM} \cite{Vermaseren:2000nd} or similar symbolic manipulation systems.

The manifest Lorentz covariance of the QCD amplitudes can be preserved if one uses the covariant projector technique \cite{Bodwin:2002hg}. 
This property is especially useful when the evaluation is automatized using existing software for symbolic calculations in relativistic Quantum Field Theories (QFTs) (\eg \textsc{FeynCalc} \cite{Mertig:1990an,Shtabovenko:2016sxi},
\textsc{Redberry} \cite{Bolotin:2013qgr} or \textsc{FDC} \cite{Wang:2004du} to name those that are often used in NRQCD calculations). 
This is one of the main reasons why almost all modern NRQCD matching calculations are done using projectors.

For the present calculation, however, we will not use this approach. 
The reason is that there is no literature on how to apply the projector technique to extract matching coefficients
multiplying matrix elements of the type $\braket{0| \mathcal{T}_8 (^3 P_J)|0}$. 
On the other hand, the calculation of the matching coefficients $T_8(^3 P_J)$ using the threshold expansion method and matching at the amplitude level is straightforward, albeit tedious.
Similar calculations have already been done in the investigation of the electromagnetic decay $\chi_{cJ} \to \gamma \gamma$ \cite{Ma:2002eva,Brambilla:2006ph}, 
such that one can benefit from the existing knowledge on the subject.
To automatize our matching calculations we employ \textsc{FeynArts} \cite{Hahn:2000kx}, \textsc{FeynCalc}, \textsc{FORM} and \textsc{FeynCalcFormLink} \cite{Feng:2012tk} as well as self-written \textit{Mathematica} codes. The software framework for automatizing nonrelativistic EFT calculations  using \textsc{FeynCalc} will be presented elsewhere \cite{Brambilla:2018}.

\section{NRQCD amplitudes} \label{sec:nrqcdside}
\subsection{NRQCD-factorized production amplitudes and cross sections}
In order to apply the threshold expansion method at the amplitude level, we introduce NRQCD production amplitudes at $\mathcal{O}(v^2)$ relative to the scaling of the leading order quarkonium-to-vacuum matrix element
\begin{subequations}
\eq{
 \mathcal{A}_{\textrm{NRQCD}}^{J=0} &=  \frac{c_1^{J=0}}{m^2} \braket{\chi_{c0} |\psi^\dagger \left ( - \frac{i}{2} \overleftrightarrow{\bfD} \cdot \bfSigma \right ) \chi|0}  \nonumber \\
 & + \frac{c_3^{J=0}}{m^4} \braket{\chi_{c0} |\psi^\dagger \left ( - \frac{i}{2} \overleftrightarrow{\bfD} \cdot \bfSigma \right )
\left ( - \frac{i}{2} \overleftrightarrow{\bfD}  \right )^2  \chi|0}  +
\frac{d_1^{J=0}}{m^3} \braket{\chi_{c0} |\psi^\dagger( g \bfE  \cdot \bfSigma ) \chi|0} ,  \label{eq:amp1} \\
\nonumber \\
 \mathcal{A}_{\textrm{NRQCD}}^{J=1} &= \frac{(c_1^{J=1})^i}{m^2}  \braket{\chi_{c1} |\psi^\dagger \left ( - \frac{i}{2} \overleftrightarrow{\bfD} \times \bfSigma \right )^i \chi|0} \nonumber \\
& + \frac{(c_3^{J=1})^i}{m^4} \braket{\chi_{c1} |\psi^\dagger \left ( - \frac{i}{2} \overleftrightarrow{\bfD} \times \bfSigma \right )^i
\left ( - \frac{i}{2} \overleftrightarrow{\bfD}  \right )^2  \chi|0} +
\frac{(d_1^{J=1})^i}{m^3} \braket{\chi_{c1} |\psi^\dagger (g \bfE  \times \bfSigma)^i  \chi|0}, \\
\nonumber \\
 \mathcal{A}_{\textrm{NRQCD}}^{J=2} &= \frac{(c_1^{J=2})^{ij}}{m^2} \braket{ \chi_{c2} |\psi^\dagger \left ( - \frac{i}{2} \overleftrightarrow \bfD^{(i}  \bfSigma^{j)} \right ) \chi|0} \nonumber \\
&  + \frac{(c_3^{J=2})^{ij}}{m^4} \braket{ \chi_{c2} |\psi^\dagger \left ( - \frac{i}{2} \overleftrightarrow \bfD^{(i}  \bfSigma^{j)} \right )
\left ( - \frac{i}{2} \overleftrightarrow{\bfD}  \right )^2  \chi|0} +
\frac{(d_1^{J=2})^{ij}}{m^3} \braket{\chi_{c2} |\psi^\dagger g \bfE^{(i} \bfSigma^{j)}  \chi|0}.
}
 \label{eq:amp}
\end{subequations}
The scaling of these matrix elements in $v$ is estimated according to the power counting rules discussed in \Sec\ref{sec:powercounting}. 
For example, the matrix element $\braket{\chi_{c0} |\psi^\dagger \left ( - \frac{i}{2} \overleftrightarrow{\bfD} \cdot \bfSigma \right ) \chi|0}$ scales as $v^{5/2}$, 
while the scaling of $\braket{\chi_{c0} |\psi^\dagger ( g \bfE  \cdot \bfSigma ) \chi|0}$ is $v^{9/2}$. 

As explained in \Sec\ref{sec:matching}, to arrive at production cross sections in the laboratory frame, 
we need to square the amplitudes in \Eqs\eqref{eq:amp}, average over the polarizations of the leptons, 
sum over the polarizations of the photon and the heavy quarkonium and integrate over the phase space for producing a heavy quarkonium and a photon,
\begin{equation}
 \sigma (e^+ e^- \to \chi_{cJ} + \gamma) = 2 M_{\chi_{cJ}} (2J+1)\int d \Phi_J \frac{1}{4}\sum_{\textrm{pols}} |\mathcal{A}_{\textrm{NRQCD}}^{J}|^2, \label{eq:xsectionnrqcd}
\end{equation}
where \textrm{pols} refers to the photon polarizations and the summation over the  polarizations of the heavy quarkonium is implicit in the amplitude squared.
The prefactor $(2J+1)$ appears because of our redefinition of the matching coefficients in \Eq\eqref{eq:redefineMatchingCoeff} 
and will be precisely canceled  by $1/(2J+1)$ in the formula for the sum over the heavy quarkonium polarizations (\cf Appendix \ref{sec:appendix3}). 
The prefactor $2 M_{\chi_{cJ}}$ accounts for the fact that our matrix elements have nonrelativistic normalization.\footnote{
An easy way to understand where $2 M_{\chi_{cJ}}$ comes from is to compute the cross section from an NRQCD amplitude with relativistic normalization of the matrix elements 
and then convert to the nonrelativistic normalization via the well-known relation \cite{Braaten:1998au}
$\braket{H| \psi^\dagger  \mathcal{K} \chi | 0}_{\textrm{rel norm}} = \sqrt{2 M_{\chi_{cJ}}} \braket{H| \psi^\dagger  \mathcal{K} \chi | 0}$.} 
Finally, the phase space measure is given by
\begin{equation}
d \Phi_J  = \frac{s - M_{\chi_{cJ}}^2}{64 \pi^2 s^2} d \Omega_{k}, \label{eq:phasespace}
\end{equation}
with the integration done over the solid angle of the photon 3-momentum \bfk.  
This yields
\begin{subequations}
\eq{
& \sigma (e^+ e^- \to \chi_{c0} + \gamma) = 2 M_{\chi_{c0}}  \int d \Phi_0 \frac{1}{4} \sum_{\textrm{pols}} \biggl ( 3 \frac{|c_1^{J=0}|^2}{m^4} \braket{0|\mathcal{O}_1 (^3 P_0)|0} \nonumber \\
  &\qquad\qquad
  + 6 \frac{c_1^{J=0} (c_3^{\ast J=0})}{m^6} \braket{0|\mathcal{P}_1 (^3 P_0)|0}  + 3 i \frac{ c_1^{J=0} (d_1^{\ast J=0}) }{m^5} \braket{0|\mathcal{T}_8 (^3 P_0)|0} \biggr ),  \\
\nonumber \\
& \sigma (e^+ e^- \to \chi_{c1} + \gamma) = 2 M_{\chi_{c1}} \int d \Phi_1 \frac{1}{4} \sum_{\textrm{pols}} \biggl ( 2 \frac{(c_1^{J=1})^i (c_1^{\ast J=1})^i}{m^4} \braket{0|\mathcal{O}_1 (^3 P_1)|0} \nonumber \\
&\qquad\qquad
+ 4 \frac{(c_1^{J=1})^i (c_3^{\ast J=1})^{ i}}{m^6} \braket{0|\mathcal{P}_1 (^3 P_1)|0}  +  2 i \frac{(c_1^{J=1})^i (d_1^{\ast  J=1})^{i}  }{m^5} \braket{0|\mathcal{T}_8 (^3 P_1)|0} \biggr ), \\
\nonumber \\
&  \sigma (e^+ e^- \to \chi_{c2} + \gamma)  = 2 M_{\chi_{c2}} \int d \Phi_2 \frac{1}{4} \sum_{\textrm{pols}} \biggl (
\frac{1}{m^4}
 \left [(c_1^{J=2})^{ij} (c_1^{\ast J=2})^{ij}  - \frac{1}{3} (c_1^{J=2})^{ii} (c_1^{\ast J=2})^{jj} \right ]   \nonumber \\
 & \qquad\qquad
 \times \braket{0|\mathcal{O}_1 (^3 P_2)|0}  + \frac{2}{m^6}
 \left [(c_1^{J=2})^{ij} (c_3^{\ast J=2})^{ij} - \frac{1}{3} (c_1^{J=2})^{ii} (c_3^{\ast J=2})^{jj} \right ] \braket{0|\mathcal{P}_1 (^3 P_2)|0} \nonumber \\
 & \qquad\qquad
 + \frac{i}{m^5}
\left [(c_1^{J=2})^{ij} (d_1^{\ast J=2})^{ij}   - \frac{1}{3} (c_1^{J=2})^{ii} (d_1^{\ast J=2})^{jj}   \right ] \braket{0|\mathcal{T}_8 (^3 P_2)|0} \biggr ),
}
\label{eq:mcs}
\end{subequations}
where we used that $c_1 c_3^\ast  =  c_3 c_1^\ast$  and $c_1^\ast d_1 = - c_1 d_1^\ast$.\footnote{
These relations between products of the short distance coefficients 
are given in anticipation of our explicit results listed in \Sec\ref{sec:qcdside}.}
Comparing \Eq\eqref{eq:mcs} with \Eq\eqref{eq:cs} we can express the matching coefficients $F_1(^3 P_J)$, $G_1(^3 P_J)$ and $T_8(^3 P_J)$ through the short-distance coefficients $c_1$, $c_3$ and $d_1$
\begin{subequations}
\eq{
F_1(^3 P_0) &= 2 M_{\chi_{c0}} \int d \Phi_0 \frac{1}{4} \sum_{\textrm{pols}} 3 \, |c_1^{J=0}|^2, \\
G_1(^3 P_0) &= 2 M_{\chi_{c0}} \int d \Phi_0 \frac{1}{4} \sum_{\textrm{pols}} 6 \, c_1^{J=0} (c_3^{\ast J=0}), \\
T_8(^3 P_0) &= 2 i M_{\chi_{c0}} \int d \Phi_0 \frac{1}{4} \sum_{\textrm{pols}} 3 \, c_1^{J=0}  (d_1^{\ast J=0}), \\
\nonumber \\
F_1(^3 P_1) &= 2 M_{\chi_{c1}} \int d \Phi_1 \frac{1}{4} \sum_{\textrm{pols}} 2 \, (c_1^{J=1})^i (c_1^{\ast J=1})^i, \\
G_1(^3 P_1) &= 2 M_{\chi_{c1}} \int d \Phi_1 \frac{1}{4} \sum_{\textrm{pols}} 4 \, (c_1^{J=1})^i (c_3^{\ast J=1})^{ i},\\
T_8(^3 P_1) &= 2 i M_{\chi_{c1}} \int d \Phi_1 \frac{1}{4} \sum_{\textrm{pols}} 2 \, (c_1^{J=1})^i (d_1^{\ast  J=1})^{i}, \\
\nonumber \\
F_1(^3 P_2) &= 2 M_{\chi_{c2}} \int d \Phi_2 \frac{1}{4} \sum_{\textrm{pols}}  \left (
(c_1^{J=2})^{ij} (c_1^{\ast J=2})^{ij}  - \frac{1}{3} (c_1^{J=2})^{ii} (c_1^{\ast J=2})^{jj}\right) ,  \\
G_1(^3 P_2) &= 2 M_{\chi_{c2}} \int d \Phi_2 \frac{1}{4} \sum_{\textrm{pols}} 2 \left ( (c_1^{J=2})^{ij} (c_3^{\ast J=2})^{ij} - \frac{1}{3} (c_1^{J=2})^{ii} (c_3^{\ast J=2})^{jj} \right ),  \\
T_8(^3 P_2) &= 2 i M_{\chi_{c2}} \int d \Phi_2 \frac{1}{4} \sum_{\textrm{pols}} \left ( (c_1^{J=2})^{ij} (d_1^{\ast J=2})^{ij}   - \frac{1}{3} (c_1^{J=2})^{ii} (d_1^{\ast J=2})^{jj} \right ).
}
\label{eq:mcoeff}
\end{subequations}
Our task is to calculate all matching coefficients and in particular $T_8(^3 P_J)$ at $\mathcal{O}(\alpha_{\rm s}^0)$, which can be inferred from the knowledge of $c_1$ and $d_1$. 
To determine the short-distance coefficients $c_1^J$, $c_3^J$ and $d_1^J$ we need to carry out the matching between perturbative QCD and perturbative NRQCD at the amplitude level.

\subsection{Computation of the NRQCD amplitudes}
The perturbative NRQCD matrix elements are most conveniently evaluated in the operator formalism by rewriting heavy quark fields in terms of Pauli spinors 
and single fermion creation and annihilation operators, as explained in \cite{Cho:1995ce}. 
The evaluation of matrix elements that enter perturbative NRQCD amplitudes is straightforward. For example,
replacing $\ket{H}$ with $\ket{Q (\bfp_{1,R})\overline{Q} (\bfp_{2,R})}$ and $\ket{Q (\bfp_{1,R})\overline{Q} (\bfp_{2,R}) g (\bfp_{g,R})}$ in $\mathcal{A}_{\textrm{NRQCD}}^{J=0}$
from \Eq\eqref{eq:amp1} we find
\eq{
\braket{Q (\bfp_{1,R})\overline{Q} (\bfp_{2,R}) |\psi^\dagger \left ( - \frac{i}{2} \overleftrightarrow{\bfD} \cdot \bfSigma \right ) \chi|0} 
\!& = \xi^\dagger (\bfSigma \cdot \bfq) \eta, \\
\braket{Q (\bfp_{1,R})\overline{Q} (\bfp_{2,R}) |\psi^\dagger \left ( - \frac{i}{2} \overleftrightarrow{\bfD} \cdot \bfSigma \right ) \left ( - \frac{i}{2} \overleftrightarrow{\bfD}  \right )^2  \chi|0} 
\!&= \bfq^2 \xi^\dagger (\bfSigma \cdot \bfq) \eta, \\
\braket{Q (\bfp_{1,R})\overline{Q} (\bfp_{2,R}) |\psi^\dagger (g \bfE  \cdot \bfSigma ) \chi|0} \!&= 0, \\
\braket{Q (\bfp_{1,R})\overline{Q} (\bfp_{2,R}) g (\bfp_{g,R})  |\psi^\dagger \left ( - \frac{i}{2} \overleftrightarrow{\bfD} \cdot \bfSigma \right ) \chi|0} 
\!&= -g \xi^\dagger (\bfSigma \cdot \bfeps^{\ast}_{g,R}) \eta, \\
\braket{Q (\bfp_{1,R})\overline{Q} (\bfp_{2,R}) g (\bfp_{g,R}) |\psi^\dagger \left ( - \frac{i}{2} \overleftrightarrow{\bfD} \cdot \bfSigma \right ) \left ( - \frac{i}{2} \overleftrightarrow{\bfD}  \right )^2  \chi|0} 
\!&= -g \xi^\dagger \bfSigma^i \eta \left [ 2 \bfq^i (\bfq \cdot  \bfeps^{\ast}_{g,R}) +  \bfeps^{\ast i}_{g,R}\bfq^2 \right ], \\
\braket{Q (\bfp_{1,R})\overline{Q} (\bfp_{2,R}) g (\bfp_{g,R}) |\psi^\dagger (g \bfE  \cdot \bfSigma)  \chi|0} \!&= - i g |\bfp_g|  \xi^\dagger (\bfSigma \cdot \bfeps^{\ast}_{g,R}) \eta.
}
Notice that, although matrix elements made only of covariant derivatives contribute through the leading Fock state \qqbar,
because of the gluon field in the covariant derivative they also give a nonvanishing contribution to the subleading Fock state \qqbarg.

As explained in \Sec\ref{sec:matching}, to match the QCD amplitudes we need, however, more operators and matching coefficients on the NRQCD sides than needed to compute production cross sections alone. 
For example, apart from $P$-wave spin triplet quarkonium (\chic) we may also produce $S$-wave spin singlet quarkonium ($\eta_c$). 
Even though we are actually not interested in the latter, we still have to determine the short-distance coefficients of the corresponding operators. 
In fact, this property of the threshold expansion method provides an important consistency check of the whole matching calculation: 
if the short-distance coefficients were determined correctly, the difference $\mathcal{A}_{\textrm{pert QCD}} - \mathcal{A}_{\textrm{pert NRQCD}}$ would necessarily vanish order by order in the expansion parameters. 
Notice that the perturbative matching between QCD and NRQCD does not rely on any specific power counting. 
Instead, we compare $\mathcal{A}_{\textrm{pert QCD}}$ to $\mathcal{A}_{\textrm{pert NRQCD}}$ order by order in $\bfq_1$ and $\bfq_2$. 
This follows the approach adopted in \cite{Brambilla:2006ph}, where the matching for electromagnetic decays was done order by order in~$1/m$.

The NRQCD amplitudes relevant for the perturbative matching (including additional matrix elements and Lagrangian insertions allowed by the symmetries) read
\begin{subequations}
\eq{
& \mathcal{A}_{\textrm{pert NRQCD}}^{J=0} = 
\frac{c_0^{J=0}}{m} \braket{H |\psi^\dagger \chi|0} + \frac{c_1^{J=0}}{m^2} \braket{H |\psi^\dagger \left ( - \frac{i}{2} \overleftrightarrow{\bfD} \cdot \bfSigma \right ) \chi|0} \nonumber \\
&+  \frac{c_2^{J=0}}{m^3} \braket{H |\psi^\dagger \left ( - \frac{i}{2} \overleftrightarrow{\bfD}  \right )^2 \chi|0}  
+ \frac{c_3^{J=0}}{m^4} \braket{H |\psi^\dagger \left ( - \frac{i}{2} \overleftrightarrow{\bfD} \cdot \bfSigma \right )
\left ( - \frac{i}{2} \overleftrightarrow{\bfD}  \right )^2  \chi|0} \nonumber \\
& + \frac{d_0^{J=0}}{m^3} \braket{H |\psi^\dagger (g \bfB  \cdot \bfSigma)  \chi|0} 
+ \frac{d_1^{J=0}}{m^3} \braket{H |\psi^\dagger (g \bfE  \cdot \bfSigma)  \chi|0} + \frac{c_0^{J=0}}{m}\braket{H |\psi^\dagger \chi|0}{}_{\mathcal{L}_{2-f}}  \nonumber \\
& + \frac{c_1^{J=0}}{m^2} \braket{H |\psi^\dagger \left ( - \frac{i}{2} \overleftrightarrow{\bfD} \cdot \bfSigma \right ) \chi|0}{}_{\mathcal{L}_{2-f}} 
+  \frac{c_2^{J=0}}{m^3} \braket{H |\psi^\dagger \left ( - \frac{i}{2} \overleftrightarrow{\bfD}  \right )^2 \chi|0}{}_{\mathcal{L}_{2-f}},   \nonumber \\ \\
&\mathcal{A}_{\textrm{pert NRQCD}}^{J=1} = \frac{(c_1^{J=1})^i}{m^2}  \braket{H |\psi^\dagger \left ( - \frac{i}{2} \overleftrightarrow{\bfD} \times \bfSigma \right )^i \chi|0} \nonumber \\
&+ \frac{(c_3^{J=1})^i}{m^4} \braket{H |\psi^\dagger \left ( - \frac{i}{2} \overleftrightarrow{\bfD} \times \bfSigma \right )^i
\left ( - \frac{i}{2} \overleftrightarrow{\bfD}  \right )^2  \chi|0} + \frac{(d_0^{J=1})^i}{m^3} \braket{H |\psi^\dagger (g \bfB  \times \bfSigma)^i  \chi|0} \nonumber \\
& + \frac{(d_1^{J=1})^i}{m^3} \braket{H |\psi^\dagger (g \bfE  \times \bfSigma)^i  \chi|0} 
+ \frac{(c_1^{J=1})^i}{m^2}  \braket{H |\psi^\dagger \left ( - \frac{i}{2} \overleftrightarrow{\bfD} \times \bfSigma \right )^i \chi|0}{}_{\mathcal{L}_{2-f}}, \\
\nonumber \\
& \mathcal{A}_{\textrm{pert NRQCD}}^{J=2} = \frac{(c_1^{J=2})^{ij}}{m^2} \braket{H |\psi^\dagger \left ( - \frac{i}{2} \overleftrightarrow \bfD^{(i}  \bfSigma^{j)} \right ) \chi|0} 
+  \frac{(c_2^{J=2})^{ij}}{m^3} \braket{H |\psi^\dagger \left ( - \frac{i}{2} \right)^2 \overleftrightarrow{\bfD}^{(i}  \overleftrightarrow{\bfD}^{j)}   \chi|0} \nonumber \\
& + \frac{(c_3^{J=2})^{ij}}{m^4} \braket{H |\psi^\dagger \left ( - \frac{i}{2} \overleftrightarrow \bfD^{(i}  \bfSigma^{j)} \right )
\left ( - \frac{i}{2} \overleftrightarrow{\bfD}  \right )^2  \chi|0} \nonumber \\
& + \frac{(d_0^{J=2})^{ij}}{m^3} \braket{H |\psi^\dagger g \bfB^{(i} \bfSigma^{j)}  \chi|0} + \frac{(d_1^{J=2})^{ij}}{m^3} \braket{H |\psi^\dagger g \bfE^{(i} \bfSigma^{j)}  \chi|0} \nonumber \\
& + \frac{(c_1^{J=2})^{ij}}{m^2} \braket{H |\psi^\dagger \left ( - \frac{i}{2} \overleftrightarrow \bfD^{(i}  \bfSigma^{j)} \right ) \chi|0}{}_{\mathcal{L}_{2-f}}
+ \frac{(c_2^{J=2})^{ij}}{m^3} \braket{H |\psi^\dagger \left ( - \frac{i}{2} \right)^2 \overleftrightarrow{\bfD}^{(i}  \overleftrightarrow{\bfD}^{j)}   \chi|0}{}_{\mathcal{L}_{2-f}}.
}
\label{eq:nrqcdamps}
\end{subequations}
Some explanations are in order. 
The matrix elements with the subscript $\mathcal{L}_{2-f}$ contain Lagrangian insertions, where the relevant part of the NRQCD Lagrangian (2-fermion sector only) is
\eq{
\mathcal{L}_{2-f} &= \psi^\dagger \left ( \frac{\bfD^2}{2m} + \frac{g \bfB \cdot \bfSigma }{2 m} + \frac{[\bfD \cdot, g \bfE]}{8 m^2} + \frac{i \bfSigma \cdot [\bfD \times, g \bfE]}{ 8 m^2} 
+ \frac{\bfD^4}{8 m^3} + \frac{\{ \bfD^2, g \bfB \cdot \bfSigma  \}}{8 m^3}\right ) \psi \nonumber \\
& + \chi^\dagger \left ( -\frac{\bfD^2}{2m} - \frac{g \bfB \cdot \bfSigma }{2 m} + \frac{[\bfD \cdot, g \bfE]}{8 m^2} + \frac{i \bfSigma \cdot [\bfD \times, g \bfE]}{ 8 m^2} 
- \frac{\bfD^4}{8 m^3} + \frac{\{ \bfD^2, g \bfB \cdot \bfSigma  \}}{8 m^3}\right ) \chi
\label{eq:lnrqcd}.
}
In the diagrammatic representation of the NRQCD quarkonium-to-vacuum matrix elements, Lagrangian insertions correspond to the emission of a gluon from one of the external heavy quark lines. 
For example, for $\braket{H |\psi^\dagger \chi|0}$ and $\braket{H |\psi^\dagger \chi|0}_{\mathcal{L}_{2-f}}$ we have
\begin{subequations}
\eq{
\begin{minipage}[b]{0.1\linewidth}\parbox{15mm}{
 \includegraphics[scale=0.5]{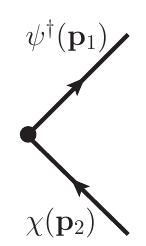}
}
\end{minipage}
 & \enspace = i \braket{Q (\bfp_{1,R})\overline{Q} (\bfp_{2,R}) |\psi^\dagger \chi|0} \\
\begin{minipage}[b]{0.1\linewidth}\parbox{15mm}{
 \includegraphics[scale=0.5]{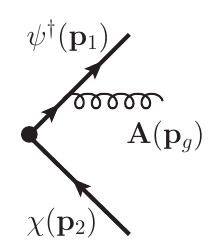}
}
\end{minipage} +
\begin{minipage}[b]{0.1\linewidth}\parbox{10mm}{
 \includegraphics[scale=0.5]{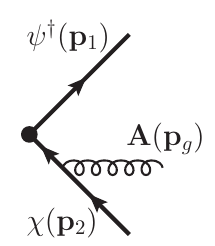}
}
\end{minipage}
 & \enspace = i \braket{Q (\bfp_{1,R})\overline{Q} (\bfp_{2,R}) g (\bfp_{g,R}) |\psi^\dagger \chi|0}_{\mathcal{L}_{2-f}},\label{eq:gluonEmissions}
 }
\end{subequations}
where the quark-gluon vertex on the heavy quark (antiquark) line is understood as a sum of all six vertices from quark-gluon (antiquark-gluon) interaction terms in \Eq\eqref{eq:lnrqcd}. 
Hence, to calculate the diagrams on the left-hand side of \Eq\eqref{eq:gluonEmissions} we need to apply Feynman rules for the $\psi^\dagger \chi$ operator and for the interaction terms in \Eq\eqref{eq:lnrqcd}.

The derivation of such Feynman rules in the operator approach is a simple QFT exercise: we sandwich the operator between the vacuum and a Fock state that contains our fields 
(so that all momenta are ingoing), calculate the matrix element, amputate external states (Pauli spinors and polarization vectors) and multiply the result by $i$. 
For example, for the operator $\psi^\dagger \{ \bfD^2, \bfSigma \cdot g \bfB \}/(8 m^3) \psi$ we obtain
\begin{equation}
\begin{minipage}[b]{0.1\linewidth}\parbox{20mm}{
 \includegraphics{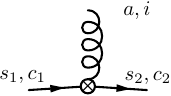}
}
\end{minipage} \quad \quad \quad =  \frac{1}{8 m^3} \left[ 2 (\bfp_{g,R} \cdot \bfp_{1,R} ) + 2 \bfp_{1,R}^2 + \bfp_{g,R}^2 \right] \epsilon^{ijk }\bfSigma^j_{s_1 s_2} \bfp_{g,R}^k T^a_{c_1 c_2}.
\end{equation}

Once we have the full expressions for the amplitudes in \Eq\eqref{eq:nrqcdamps} with $\ket{H}$ replaced by the perturbative \qqbar and \qqbarg Fock states, 
we can expand them in the relative momenta of the quarkonium constituents. 
For the $Q \overline{Q}$ case we need to expand up to the third order in $|\bfq|/m$, which allows us to determine the values of $c_0$, $c_1$, $c_2$ and $c_3$. 
For the $Q \overline{Q} g$ case it is sufficient to expand the amplitude up to the first order in $|\bfq_1|/m$ and $|\bfq_2|/m$, so that we can extract $d_0$ and $d_1$.

\section{QCD amplitudes and matching} \label{sec:qcdside}
\begin{figure}[ht]
\centering
\includegraphics[width=7cm,clip]{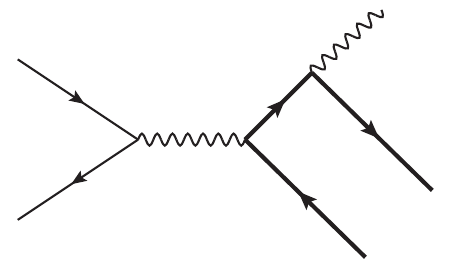}
\includegraphics[width=7cm,clip]{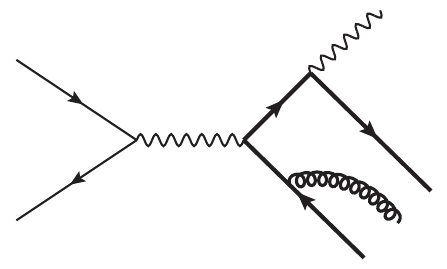}
\caption{Representative QCD diagrams for the processes $e^+ e^- \to Q \overline{Q}  + \gamma$ and $e^+ e^- \to$  $Q \overline{Q} g +$  $\gamma$ that are relevant for the matching to NRQCD. 
For the production of $Q \overline{Q}$ there are only two diagrams in total: the one displayed here and a second one where the photon is emitted from the heavy antiquark line. 
The production of $Q \overline{Q} g$ is described by six diagrams that differ in the quark (antiquark) line that emits the photon and/or the gluon.
}
\label{fig:qcd}
\end{figure}

In this section, we compute the nonrelativistic expansions for the relevant QCD amplitudes that describe exclusive electromagnetic production of $Q \overline{Q}$, \ie
\begin{equation}
e^- (l_1) \, e^+ (l_2) \to Q(p_1) \,  \overline{Q}(p_2) + \gamma (k), \label{eq:singlet}
\end{equation}
and $Q \overline{Q} g$
\begin{equation}
e^- (l_1)   e^+ (l_2) \to Q(p_1) \,  \overline{Q}(p_2) \, g(p_g)  + \gamma (k).\label{eq:octet}
\end{equation}
The process in \Eq\eqref{eq:singlet} is described by two QCD Feynman diagrams (\cf left panel of \Fig\ref{fig:qcd})  so that
\eq{
\mathcal{A}_{\textrm{pert QCD}}^{Q \overline{Q}} = - \frac{i e^2 e_Q^2 g}{s} L_\mu 
& \left [\frac{\bar{u}(p_1) \slashed{\varepsilon}^\ast_\gamma (  \slashed{p}_1 + \slashed{k} + m) \gamma^\mu  v(p_2)}{(p_1 + k)^2 - m^2} \right.
\nonumber \\
& \left. + \frac{\bar{u}(p_1)   \gamma^\mu  (  -\slashed{p}_2 - \slashed{k} + m) \slashed{\varepsilon}^\ast_\gamma  v(p_2)}{(p_2 + k)^2 - m^2} \right ],
}
while the reaction in \Eq\eqref{eq:octet} involves six diagrams (\cf right panel of \Fig\ref{fig:qcd}) and the corresponding amplitude reads
\eq{
\mathcal{A}_{\textrm{pert QCD}}^{Q \overline{Q} g} = - \frac{i e^2 e_Q^2 g}{s} T^a L_\mu  &\left[
\frac{\bar{u}(p_1) \slashed{\varepsilon}^\ast_g ( \slashed{p}_1 +  \slashed{p}_g  + m) \slashed{\varepsilon}^\ast_\gamma  ( \slashed{p}_1 +  \slashed{p}_g  + \slashed{k} + m) \gamma^\mu  v(p_2)
}{((p_1 + p_g)^2 - m^2) ((p_1 + p_g + k)^2 - m^2)} \right.\nonumber \\
& + \frac{\bar{u}(p_1) \slashed{\varepsilon}^\ast_\gamma ( \slashed{p}_1 +  \slashed{k}   + m) \slashed{\varepsilon}^\ast_g  ( \slashed{p}_1 +  \slashed{p}_g  + \slashed{k} + m) \gamma^\mu  v(p_2)
}{((p_1 + k)^2 - m^2) ((p_1 + p_g + k)^2 - m^2)} \nonumber \\
& + \frac{\bar{u}(p_1) \slashed{\varepsilon}^\ast_g ( \slashed{p}_1 +  \slashed{p}_g + m)  \gamma^\mu (  - \slashed{p}_2 -\slashed{k}  + m) \slashed{\varepsilon}^\ast_\gamma v(p_2)
}{((p_1 + p_g)^2 - m^2) ((p_2 + k )^2 - m^2)}   \nonumber \\
& + \frac{\bar{u}(p_1) \slashed{\varepsilon}^\ast_\gamma (  \slashed{p}_1 + \slashed{k}  + m)  \gamma^\mu (  - \slashed{p}_2 -\slashed{p}_g  + m) \slashed{\varepsilon}^\ast_g v(p_2)
}{((p_1 + k)^2 - m^2) ((p_2 + p_g)^2 - m^2) } \nonumber \\
& + \frac{\bar{u}(p_1) \gamma^\mu  (  - \slashed{p}_2  -\slashed{p}_g -\slashed{k}  + m) \slashed{\varepsilon}^\ast_\gamma ( -\slashed{p}_2 -  \slashed{p}_g + m) \slashed{\varepsilon}^\ast_g  v(p_2)
}{((p_2 + p_g + k )^2 - m^2) ((p_2 + p_g)^2 - m^2) } \nonumber \\
& \left. + \frac{\bar{u}(p_1) \gamma^\mu  (  - \slashed{p}_2  -\slashed{p}_g -\slashed{k}  + m) \slashed{\varepsilon}^\ast_g ( -\slashed{k} -  \slashed{p}_2 + m) \slashed{\varepsilon}^\ast_\gamma  v(p_2)
}{((p_2 + p_g + k )^2 - m^2) ((p_2 + k)^2 - m^2) } \right],
}
where $L_\mu$ has been defined in \Sec\ref{sec:kinematics} and $e_Q$ denotes the heavy quark electric charge in units of $e$. 
In the following sections where we discuss the phenomenology of electromagnetic $\chi_{cJ}$ production, $e_Q$ should be always understood as the charge of the charm quark \ie $e_Q \equiv e_c = +2/3$. The amplitudes $\mathcal{A}_{\textrm{pert QCD}}^{Q \overline{Q}}$ and $\mathcal{A}_{\textrm{pert QCD}}^{Q \overline{Q} g}$ can be evaluated in the laboratory frame or in the rest frame.
Both frames have their advantages and disadvantages. 
In the course of this study we carry out the calculations in both frames and verify that we arrive at the same results for the final matching coefficients.

It is convenient to decompose all Dirac structures involving heavy quarks into scalar, pseudoscalar, vector, axial-vector and tensor components. 
Since this is a tree level calculation in four dimensions, such decomposition can be done in an unambiguous way. Then the amplitudes can be written as
\eq{
\mathcal{A}_{\textrm{pert QCD}}^{Q \overline{Q}} &= - \frac{i e^2 e_Q^2 g}{s}  \left [ j_1^\mu \bar{u}(p_1) \gamma_\mu v(p_2) + j_2^\mu \bar{u}(p_1) \gamma_\mu \gamma_5 v(p_2) \right ], \\
\mathcal{A}_{\textrm{pert QCD}}^{Q \overline{Q} g} &= - \frac{i e^2 e_Q^2 g}{s} T^a \left [ j_3^\mu \bar{u}(p_1) \gamma_\mu v(p_2) + j_4^\mu \bar{u}(p_1) \gamma_\mu \gamma_5 v(p_2) \right ],
}
where the coefficients $j_i^\mu$ do not depend on the heavy quark spinors. 
The vector and axial-vector structures $\bar{u}(p_1) \gamma_\mu v(p_2)$ and  $\bar{u}(p_1) \gamma_\mu \gamma_5 v(p_2)$ can be rewritten 
in terms of Pauli matrices and spinors using formulas given in Appendix \ref{sec:appendix1}. 
Furthermore, all the propagators and scalar products in $j_i^\mu$ have to be expressed in terms of $\bfq$ (for $Q \overline{Q}$) or of $\bfq_1$ and $\bfq_2$ (for $Q \overline{Q} g$). 
As in \Sec\ref{sec:nrqcdside}, we expand up to the third order in $|\bfq|/m$ and up to the first order in $|\bfq_1|/m$ and $|\bfq_2|/m$.

The NRQCD amplitudes in \Eq\eqref{eq:nrqcdamps} are given for specific values of the total angular momentum $J$, 
but $\mathcal{A}_{\textrm{pert QCD}}^{Q \overline{Q}}$ and $\mathcal{A}_{\textrm{pert QCD}}^{Q \overline{Q} g}$ should be understood as sums of contributions from different values of $J$.  
Therefore, on the QCD side one has to identify Cartesian tensors that correspond to the matrix elements on the NRQCD side and decompose them into irreducible spherical tensors
along the lines of \cite{Coope1970}. 
For example, a term that contains the rank 2 tensor $\bfSigma^i \bfq^j$ and no further occurrences of $\bfq$ contributes to
\begin{equation}
\frac{c_1^{J=0}}{m^2} \braket{Q \overline{Q} |\psi^\dagger \left ( - \frac{i}{2} \overleftrightarrow{\bfD} \cdot \bfSigma \right ) \chi|0},
\end{equation}
from $\mathcal{A}_{\textrm{pert NRQCD}}^{J=0}$ but also to
\begin{equation}
\quad \frac{(c_1^{J=1})^i}{m^2}  \braket{Q \overline{Q} |\psi^\dagger \left ( - \frac{i}{2} \overleftrightarrow{\bfD} \times \bfSigma \right )^i \chi|0},
\end{equation}
from $\mathcal{A}_{\textrm{pert NRQCD}}^{J=1}$ and
\begin{equation}
\frac{(c_1^{J=0})^{ij}}{m^2} \braket{ Q \overline{Q} |\psi^\dagger \left ( - \frac{i}{2} \overleftrightarrow \bfD^{(i}  \bfSigma^{j)} \right ) \chi|0},
\end{equation}
from $\mathcal{A}_{\textrm{pert NRQCD}}^{J=2}$. 
We can disentangle these contributions by explicitly projecting out the $J=0,1,2$ components of the tensor, i.\,e.,
\begin{equation}
\bfSigma^i \bfq^j \to
\begin{cases}
\displaystyle \frac{1}{3} \delta^{ij} (\bfSigma \cdot \bfq) \quad \textrm{for } J=0\\
\\
\displaystyle \frac{\bfSigma^i \bfq^j - \bfSigma^j \bfq^i}{2} \quad \textrm{for } J=1\\
\\
\displaystyle \frac{\bfSigma^i \bfq^j + \bfSigma^j \bfq^i}{2} - \frac{1}{3 } \delta^{ij} (\bfSigma \cdot \bfq) \quad \textrm{for } J=2
\end{cases}. \label{eq:cartProj}
\end{equation}
Applying such projections to the whole QCD amplitudes allows us to do the matching between QCD and NRQCD for each $J$-value separately 
and successfully extract the short-distance coefficients order by order in the relative momenta.

Before presenting our results for the short-distance coefficients, 
we would like to mention that on the QCD side of the matching we encounter terms that are singular in the limit $|\bfp_g| \to 0$, i.\,e., proportional to $1/|\bfp_g|$. 
Such singularities arise when the gluon is emitted from one of the external heavy quark lines and should cancel in the matching. 
Indeed, matrix elements with the Lagrangian insertions given by \Eq\eqref{eq:lnrqcd} on the NRQCD side of the matching generate terms that precisely reproduce the singularities of the QCD amplitude. 
The identical infrared behavior of the full QCD and NRQCD amplitudes follows from general principles underlaying the construction of low-energy effective field theories. 
The cancellation of the $1/|\bfp_g|$ singularities on both sides of the matching has been explicitly checked in both the laboratory frame and the rest frame. 
A more detailed discussion on this subject can be found in \cite{Brambilla:2006ph} and \cite{Brambilla:2008zg}.

\subsection{Matching in the rest frame}
Rest frame short-distance coefficients carry an additional subscript $R$ to distinguish them from the coefficients obtained in the laboratory frame. They read
\begin{subequations}
\eq{
c_{0,R}^{J=0} &= -\lambda(\boldsymbol{L}_R \cdot( {\hat{\bfk}_{R}} \times \bfeps_{\gamma,R}^\ast)),\\
c_{1,R}^{J=0} &=  \frac{i }{3} (1 - 3 r) r_{-} \lambda(\boldsymbol{L}_R\cdot\bfeps_{\gamma,R}^\ast), \\
c_{2,R}^{J=0} &=\frac{2}{3} \lambda (\boldsymbol{L}_R\cdot( {\hat{\bfk}_{R}} \times \bfeps_{\gamma,R}^\ast)),\\
c_{3,R}^{J=0} &=- \frac{i}{30} (9 - 24 r + 35 r^2) \lambda(\boldsymbol{L}_R\cdot\bfeps_{\gamma,R}^\ast),\\
d_{0,R}^{J=0} &= - \lambda r_{-}(\boldsymbol{L}_R \cdot( {\hat{\bfk}_{R}} \times \bfeps_{\gamma,R}^\ast)), \\
d_{1,R}^{J=0} & =\frac{1}{6}  (1 + 3 r) r_{-} \lambda(\boldsymbol{L}_R\cdot\bfeps_{\gamma,R}^\ast),  \\
\nonumber \\
(c_{1,R}^{J=1})^i &= i\lambda\left (r_{-}(\boldsymbol{L}_R\times\bfeps_{\gamma,R}^\ast)^i-r_{+}(\boldsymbol{L}_R\cdot\hat{\bfk}_{R})(\hat{\bfk}_{R}\times\bfeps_{\gamma,R}^\ast)^i\right), \\
(c_{3,R}^{J=1})^i &= -\frac{i}{5}\lambda\left((4-9r)r_{-}^2(\boldsymbol{L}_R\times\bfeps_{\gamma,R}^\ast)^i-(4+9r)r_{+}^2(\boldsymbol{L}_R\cdot\hat{\bfk}_{R})(\hat{\bfk}_{R}\times\bfeps_{\gamma,R}^\ast)^i\right), \\
(d_{0,R}^{J=1})^i & =  \lambda r_{-}\left(-(\boldsymbol{L}_R\cdot\bfeps_{\gamma,R}^\ast)\hat{\bfk}_{R}^i+2rr_{+}(\boldsymbol{L}_R\cdot\hat{\bfk}_{R})\bfeps_{R}^{\ast i}\right),\\
(d_{1,R}^{J=1})^i &= \lambda \left((1+r^2)r_{-}^2(\boldsymbol{L}_R\times\bfeps_{\gamma,R}^\ast)^i-r_{-}r_{+}^2(1 - r + 3r - r^3)(\boldsymbol{L}_R\cdot\hat{\bfk}_{R})(\hat{\bfk}_{R}\times\bfeps_{\gamma,R}^\ast)^i\right),  \\
\nonumber \\
(c_{1,R}^{J=2})^{ij} &= i\lambda \left (rr_{-} (\boldsymbol{L}_R^i \bfeps_{R}^{\ast j} + i \leftrightarrow j  ) 
+ rr_{+}(\boldsymbol{L}_R\cdot\hat{\bfk}_{R}) (\hat{\bfk}_{R}^i \bfeps_{R}^{\ast j}  +i \leftrightarrow j) 
-(\boldsymbol{L}_R\cdot\bfeps_{\gamma,R}^\ast)\hat{\bfk}_{R}^i\hat{\bfk}_{R}^j \right),\\
(c_{2,R}^{J=2})^{ij} &= - \lambda (\boldsymbol{L}_R\cdot( {\hat{\bfk}_{R}} \times \bfeps_{\gamma,R}^\ast)) \, \hat{\bfk}_{R}^i\hat{\bfk}_{R}^j, \\
(c_{3,R}^{J=2})^{ij} &=\frac{i}{5} \lambda\left (5r^2r_{-}^2(\boldsymbol{L}_R^i  \bfeps_{R}^{\ast j} + i \leftrightarrow j) 
-5r^2r_{+}^2(\boldsymbol{L}_R\cdot\hat{\bfk}_{R}) (\hat{\bfk}_{R}^i \bfeps_{R}^{\ast j} +i \leftrightarrow j)  \right. \nonumber \\
& \left.  \hspace{8cm}
+3 (\boldsymbol{L}_R\cdot\bfeps_{\gamma,R}^\ast)\hat{\bfk}_{R}^i \hat{\bfk}_{R}^j \right) ,\\
(d_{0,R}^{J=2})^{ij} &=0, \\
(d_{1,R}^{J=2})^{ij} &= \lambda \left ( 2r^2r_{-}r_{+}^2(\boldsymbol{L}_R\cdot\hat{\bfk}_{R}) (\hat{\bfk}_{R}^i \bfeps_{R}^{\ast j} +i \leftrightarrow j) 
- r_{-}(\boldsymbol{L}_R\cdot\hat{\bfk}_{R}) \hat{\bfk}_{R}^i \hat{\bfk}_{R}^j \right),
}
\label{eq:restframesds}
\end{subequations}
with $\lambda \equiv e^2 e_Q^2/s$ and $r_{\pm}\equiv(1\pm r)^{-1}$, where $r \equiv 4 m^2/s$ is the kinematic suppression factor. 
$\boldsymbol{L}_R$ is the spatial part of the leptonic current in the rest frame defined in \Sec\ref{sec:kinematics}.

The connection to the laboratory frame can be established  by means of the following Lorentz boost transformations
\begin{equation}
\bfk_{R} = \frac{\sqrt{s}}{M_{\chi_{cJ}}} \bfk, \quad \bfeps^{\ast}_{R} = \bfeps^{\ast}, \quad \boldsymbol{L}_R 
= \boldsymbol{L} +  \frac{(M_{\chi_{cJ}} - \sqrt{s})^2}{2 M_{\chi_{cJ}} \sqrt{s}} (\boldsymbol{L} \cdot \hat{\bfk}) \, \hat{\bfk},
\end{equation}
where the appearance of the physical quarkonium mass stems from the fact that we have employed the kinematics of the physical quarkonium, \ie used $P^2 = M_{\chi_{cJ}}^2$. 
In \Sec\ref{sec:GKrels} we will show how the dependence on $M_{\chi_{cJ}}$ can be eliminated up to the desired order in $v$.

Matching in the rest frame of the heavy quarkonium is very similar to the calculation of the $\mathcal{O}(\alpha_{\rm s}^0 v^2)$ corrections to the decay process $\chi_{cJ} \to \gamma \gamma$. 
The production of $\chi_{cJ}$ at rest can be regarded as the decay of a heavy vector boson $\gamma^\ast$ into a photon and a quarkonium. 
As has already been observed in \cite{Chung:2008km}, in the limit $s \to 0$, the production short-distance coefficients from \Eqs\eqref{eq:restframesds} 
should reduce into the short distance coefficients for the decay of $\chi_{cJ}$ into two photons. In our calculation this is indeed the case.

\subsection{Matching in the laboratory frame}
In the laboratory frame it is straightforward to carry out the phase space integrations in \Eq\eqref{eq:mcoeff} and thus arrive at the final matching coefficients. 
The more complicated part of the calculation is the manipulation of the QCD amplitudes. 
Obviously, potentially large laboratory frame momenta of the heavy quarks and gluon are not suitable parameters for doing a nonrelativistic expansion. 
The correct expansion can be done only after these momenta have been rewritten in terms of small rest frame momenta, which greatly increases the number of intermediate terms.

Lorentz boost transformations that relate potentially large laboratory frame momenta of a $Q\overline{Q}$ system to the small rest frame momenta in a way that is useful for NRQCD matching calculations 
were introduced in \cite{Braaten:1996jt}.  
In this work we generalize those formulas for a $Q\overline{Q} g$ system and summarize them in Appendix \ref{sec:appendix1}.

The matching procedure yields the following values for the short-distance coefficients:
\begin{subequations}
\eq{
c_0^{J=0} & = - \lambda (\boldsymbol{L} \cdot (\hat{\bfk} \times \bfeps^{\ast}_{\gamma})), \\
c_1^{J=0} & =  \frac{i }{3} (1 - 3 r) r_{-}  \lambda   (\boldsymbol{L} \cdot \bfeps^{\ast}_{\gamma}), \\
c_2^{J=0} & = \frac{2}{3}  \lambda ( \boldsymbol{L} \cdot (\hat{\bfk} \times \bfeps^{\ast}_{\gamma})), \\
c_3^{J=0} & = - \frac{i}{30} (9 - 24 r + 35 r^2)   r_{-}^2 \lambda  (\boldsymbol{L} \cdot \bfeps^{\ast}_{\gamma}), \\
d_0^{J=0} & = -  r_{-}  \lambda  (\boldsymbol{L} \cdot (\hat{\bfk} \times \bfeps^{\ast}_{\gamma})), \\
d_1^{J=0} & =  \frac{1}{6}  (1 + 3 r) r_{-} \lambda  (\boldsymbol{L} \cdot \bfeps^\ast_\gamma), \\
\nonumber \\
(c_1^{J=1})^i & =  - i \lambda r_{-} \left (  (1 - \sqrt{r}) (\boldsymbol{L} \cdot \hat{\bfk}) (\hat{\bfk} \times \bfeps^{\ast}_{\gamma})^i  - (\boldsymbol{L} \times \bfeps^{\ast}_{\gamma})^i        \right ), \\
(c_3^{J=1})^i & =   \frac{i}{10} \lambda r_{-}^2 \left (  (1 - \sqrt{r})^2 (8 + 13 \sqrt{r} ) (\boldsymbol{L} \cdot \hat{\bfk}) (\hat{\bfk} \times \bfeps^{\ast}_{\gamma})^i 
- 2 (4 - 9 r) (\boldsymbol{L} \times \bfeps^{\ast}_{\gamma})^i        \right ), \\
(d_0^{J=1})^i &= -  \lambda r_{-} \left   (\sqrt{r} \, \bfeps^{\ast i}_\gamma  (\boldsymbol{L} \cdot \hat{\bfk}) -
\hat{\bfk}^i (\boldsymbol{L} \cdot \bfeps^\ast_\gamma)  \right),  \\
(d_1^{J=1})^i &=  \frac{1}{2} \lambda r_{-}   \left   (
(2 - \sqrt{r}) (\boldsymbol{L} \cdot \hat{\bfk}) (\hat{\bfk} \times \bfeps^\ast_\gamma)^i - 2(  \boldsymbol{L} \times \bfeps^\ast_\gamma)^i  \right ),   \\
\nonumber \\
(c_1^{J=2})^{ij} & = -i \lambda r_{-} \left (  (1- \sqrt{r})\sqrt{r} (\boldsymbol{L} \cdot \hat{\bfk})
(\hat{\bfk}^i \bfeps^{\ast j}_\gamma + i \leftrightarrow j)  + r (\boldsymbol{L}^i \bfeps^{\ast j}_\gamma + i \leftrightarrow j) \right. \nonumber  \\
& \left. \hspace{3cm}
- (1 - r) (\boldsymbol{L} \cdot \bfeps^{\ast}_{\gamma} ) \hat{\bfk}^i \hat{\bfk}^j \right ), \\
(c_2^{J=2})^{ij} & = - \lambda (\boldsymbol{L} \cdot (\hat{\bfk} \times \bfeps^{\ast}_{\gamma})) \hat{\bfk}^i \hat{\bfk}^j, \\
(c_3^{J=2})^{ij} & = \frac{i}{10} \lambda r_{-}^2 \left ( 5 \sqrt{r} (1+ 2\sqrt{r}) (1-\sqrt{r})^2  (\boldsymbol{L} \cdot \hat{\bfk}) (\hat{\bfk}^i \bfeps^{\ast j}_\gamma + i \leftrightarrow j ) \right. \nonumber \\
& \left. \hspace{3cm}
- 10 r^2 ( \boldsymbol{L}^i  \bfeps^{\ast j}_\gamma + i \leftrightarrow j ) - 6  (1-r)^2 (\boldsymbol{L} \cdot \bfeps^{\ast}_{\gamma}) \hat{\bfk}^i \hat{\bfk}^j
 \right ), \\
(d_0^{J=2})^{ij} & = 0, \\
(d_1^{J=2})^{ij} & =  \frac{1}{2} \lambda r_{-}   \left (
\sqrt{r} (\boldsymbol{L} \cdot \hat{\bfk} )   (\hat{\bfk}^i \bfeps^{\ast j}_\gamma + i \leftrightarrow j  )
- 2
\hat{\bfk}^i \hat{\bfk}^j ( \boldsymbol{L} \cdot \bfeps^\ast_\gamma)
\right ).
} \label{eq:sdcoeffs}
\end{subequations}

\subsection{Relations between matrix elements and heavy quarkonium masses} \label{sec:GKrels}
From the NRQCD point of view, the LDMEs should be regarded as independent nonperturbative parameters. 
Nevertheless, symmetries and approximations can be used to establish relations between different LDMEs that are valid up to a certain order in $v$.

Particularly useful relations can be derived from the equations of motion of NRQCD and are known as Gremm-Kapustin relations \cite{Gremm:1997dq}. 
Since our LDMEs directly follow from those that appear in the electromagnetic decay $\chi_{cJ} \to \gamma \gamma$, we can use the Gremm-Kapustin relations\footnote{The factor $1/2$ in front of $m \braket{0| \mathcal{T}_8 (^3 P_J)|0}$ accounts for the fact that $\braket{0| \mathcal{T}_8 (^3 P_J)|0} = 2 \braket{\chi_{cJ}| \mathcal{F}_{\textrm{EM}} (^3 P_J)|\chi_{cJ}}$, with $\braket{\chi_{cJ}| \mathcal{F}_{\textrm{EM}} (^3 P_J)|\chi_{cJ}}$ being the chromoelectric LDME from \cite{Ma:2002eva}.} that were first obtained in \cite{Ma:2002eva},
\begin{equation}
\braket{0| \mathcal{P}_1 (^3 P_J)|0}  = m E_{\chi_{cJ}} \braket{0| \mathcal{O}_1 (^3 P_J)|0} + \frac{m}{2} \braket{0| \mathcal{T}_8 (^3 P_J)|0}, \label{eq:gkrel}
\end{equation}
where $E_{\chi_{cJ}} = M_{\chi_{cJ}} - 2 m  $ is the binding energy of $\chi_{cJ}$. 
Equation \eqref{eq:gkrel} is valid up to corrections of $\mathcal{O} (v^4)$ and it is useful for two reasons. 
First, we can solve it for $M_{\chi_{cJ}}$
\begin{equation}
M_{\chi_{cJ}} = 2 m + \frac{1}{m} \frac{\braket{0| \mathcal{P}_1 (^3 P_J)|0}}{\braket{0| \mathcal{O}_1 (^3 P_J)|0}} - \frac{1}{2}  \frac{\braket{0| \mathcal{T}_8 (^3 P_J)|0}}{\braket{0| \mathcal{O}_1 (^3 P_J)|0}},
\label{GKmass}
\end{equation}
and use \Eq\eqref{GKmass} to eliminate the dependence of the matching coefficients on the heavy quarkonium mass in \Eqs\eqref{eq:mcoeff}.
This is a crucial ingredient to check that the matching coefficients that we will list explicitly in \Sec\ref{sec:xsection}
are the same if obtained from the expansion of the QCD amplitudes in the laboratory frame or in the rest frame. 
Second, \Eq\eqref{eq:gkrel} allows us to reduce the number of unknown matrix elements in the final NRQCD-factorized production cross sections 
at the cost of making the numerical predictions depend on the heavy quarkonium binding energy. We will see this in the following \Sec\ref{sec:numerics}.

\section{\label{sec:decay} Electromagnetic decays of \texorpdfstring{$\chi_{cJ}$}{chi cJ} into two photons}
In this section, we show that our matching calculations in the rest frame also contribute to the resolution of the known discrepancy between the results of \cite{Ma:2002eva} and \cite{Brambilla:2006ph}  
on the values of some matching coefficients entering the NRQCD-factorized decay rates for $\chi_{cJ} \to \gamma \gamma$ at $\mathcal{O}(\alpha_{\rm s}^0 v^2)$. 
The formulas for these decay rates read (using a remark made in \Sec\ref{Sec:NRQCDdef} we express eventually the widths in terms of production LDMEs\footnote{With the definitions of the decay and production LDMEs from \cite{Brambilla:2006ph} 
and \Eqs\eqref{eq:cs} respectively, we have $\braket{0| \mathcal{O}_1 (^3 P_J)|0} = \braket{\chi_{cJ}| \mathcal{O}_{1\,\textrm{em}} (^3 P_J)| \chi_{cJ}}$, $\braket{0| \mathcal{P}_1 (^3 P_J)|0} = \braket{\chi_{cJ}| \mathcal{P}_{1\,\textrm{em}} (^3 P_J)| \chi_{cJ}}$ and $\braket{0| \mathcal{T}_8 (^3 P_J)|0} = - \braket{\chi_{cJ}| \mathcal{T}_{8\,\textrm{em}} (^3 P_J)|\chi_{cJ}}$.})
\begin{subequations}
\eq{
& \Gamma (\chi_{c0} \to \gamma \gamma) = \frac{2 \, \textrm{Im} \, f_{\textrm{em}} (^3 P_0)}{3 m^4} \braket{\chi_{c0}| \psi^\dagger \left ( - \frac{i}{2} \overset{\leftrightarrow}{\bfD} \cdot \bfSigma \right ) \chi
| 0} \braket{0| \chi^\dagger  \left ( - \frac{i}{2} \overset{\leftrightarrow}{\bfD} \cdot \bfSigma \right ) \psi
| \chi_{c0}}  \nonumber \\
  &\qquad + \frac{2 \, \textrm{Im} \, g_{\textrm{em}} (^3 P_0)}{6 m^6} \left [ \braket{\chi_{c0}| \psi^\dagger  \left ( - \frac{i}{2} \overset{\leftrightarrow}{\bfD} \cdot \bfSigma \right )  \left ( - \frac{i}{2} \overset{\leftrightarrow}{\bfD} \right  )^2 \chi
| 0}\braket{0| \chi^\dagger \left ( - \frac{i}{2} \overset{\leftrightarrow}{\bfD} \cdot \bfSigma \right )   \psi
| \chi_{c0}}  + \textrm{H.c.} \right ] \nonumber\\
& \qquad + \frac{2 \, \textrm{Im} \, t_{8 \, \textrm{em}} (^3 P_0)}{3 m^5}  \left [i \braket{\chi_{c0}| \psi^\dagger  ( g \bfE \cdot  \bfSigma )  \chi
| 0} \braket{0| \chi^\dagger  \left( - \frac{i}{2} \overset{\leftrightarrow}{\bfD} \cdot \bfSigma \right )   \psi
| \chi_{c0}}  + \textrm{H.c.} \right ] \nonumber \\
& \qquad \equiv \frac{2 \, \textrm{Im} \, f_{\textrm{em}} (^3 P_0)}{m^4} \braket{0| \mathcal{O}_{1} (^3 P_0)|0} 
  + \frac{2 \, \textrm{Im} \, g_{\textrm{em}} (^3 P_0)}{m^6} \braket{0| \mathcal{P}_{1} (^3 P_0)|0} \nonumber \\
& \qquad - \frac{2 \, \textrm{Im} \, t_{8 \, \textrm{em}} (^3 P_0)}{m^5} \braket{0| \mathcal{T}_{8 \, \textrm{em}} (^3 P_0)|0},
\label{eq:decaya}}
\eq{
& \Gamma (\chi_{c2} \to \gamma \gamma) = \frac{2 \, \textrm{Im} \, f_{\textrm{em}} (^3 P_2)}{m^4}  \braket{\chi_{c2}| \psi^\dagger \left  ( - \frac{i}{2} \overleftrightarrow{\bfD}^{(i} \bfSigma^{j)} \right  ) \chi
| 0} \braket{0| \chi^\dagger  \left ( - \frac{i}{2} \overleftrightarrow{\bfD}^{(i} \bfSigma^{j)} \right ) \psi
| \chi_{c2}}  \nonumber \\
&\qquad + \frac{2 \, \textrm{Im} \, g_{\textrm{em}} (^3 P_2)}{2 m^6} \left [ \braket{\chi_{c2}| \psi^\dagger  \left ( - \frac{i}{2} \overleftrightarrow{\bfD}^{(i} \bfSigma^{j)} \right ) \left ( - \frac{i}{2} \overset{\leftrightarrow}{\bfD}   \right )^2 \chi
| 0} \braket{0| \chi^\dagger \left ( - \frac{i}{2} \overleftrightarrow{\bfD}^{(i} \bfSigma^{j)} \right   )   \psi
| \chi_{c2}} + \textrm{H.c.} \right ] \nonumber\\
& \qquad + \frac{2 \, \textrm{Im} \, t_{8 \, \textrm{em}} (^3 P_2)}{m^3}  \left [ i \braket{\chi_{c2}| \psi^\dagger   g \bfE^{(i} \bfSigma^{j)}  \chi
| 0} \braket{0| \chi^\dagger \left ( - \frac{i}{2} \overleftrightarrow{\bfD}^{(i} \bfSigma^{j)} \right )   \psi
| \chi_{c2}} + \textrm{H.c.} \right ] \nonumber \\
& \qquad \equiv \frac{2 \, \textrm{Im} \, f_{\textrm{em} } (^3 P_2)}{m^4} \braket{0| \mathcal{O}_{1} (^3 P_2)|0} 
+ \frac{2 \, \textrm{Im} \, g_{\textrm{em}} (^3 P_2)}{m^6} \braket{0| \mathcal{P}_{1} (^3 P_2)|0} \nonumber \\
& \qquad - \frac{2 \, \textrm{Im} \, t_{8 \, \textrm{em}} (^3 P_2)}{m^5} \braket{0| \mathcal{T}_{8 \, \textrm{em}} (^3 P_2)|0}.
}
\label{eq:decay}
\end{subequations}
While both collaborations obtained the same expressions for
\begin{subequations}
\eq{
\textrm{Im} \, f_{\textrm{em}} (^3 P_0) & = 3 \alpha^2 e_Q^4 \pi, \\
\textrm{Im} \, f_{\textrm{em}} (^3 P_2) & = \frac{4}{5} \alpha^2 e_Q^4 \pi, \\
\textrm{Im} \, g_{\textrm{em}} (^3 P_0) & = - 7 \alpha^2 e_Q^4 \pi,
}
\end{subequations}
they disagree on the values of the other coefficients; \cf Table \ref{tab:decaydiscrepancy}. 
In 2013, it was reported \cite{SangSlides,DongFuture} that an independent investigation of the $\chi_{cJ} \to \gamma \gamma$ decays confirmed the results of~\cite{Brambilla:2006ph}. 
In the course of this study we could repeat the calculation of the $\mathcal{O}(\alpha_{\rm s}^0 v^2)$ corrections to these decay processes. 
Matching at the amplitude level and at the level of the total decay rates we also found agreement with the values of the matching coefficients obtained in \cite{Brambilla:2006ph}. 
These are
\begin{subequations}
\eq{
\textrm{Im} \, g_{\textrm{em}} (^3 P_2) & = -\frac{8}{5} \alpha^2 e_Q^4 \pi, \\
\textrm{Im} \, t_{8\, \textrm{em}} (^3 P_0) & = -\frac{3}{2} \alpha^2 e_Q^4 \pi, \\
\textrm{Im} \, t_{8\, \textrm{em}} (^3 P_2) & = 0.
}
\end{subequations}

\begin{table}[ht]
\begin{tabular}{ c  c  c }
\hline \hline
Matching coefficient & Value from \cite{Ma:2002eva} & Value from \cite{Brambilla:2006ph} \\
    \hline			
$\textrm{Im} \, g_{\textrm{em}} (^3 P_2)$ & $-\frac{86}{105} \alpha^2 e_Q^4 \pi$ & $-\frac{8}{5} \alpha^2 e_Q^4 \pi$ \\
$\textrm{Im} \, t_{8\,\textrm{em}} (^3 P_0)$ & $-2 \alpha^2 e_Q^4 \pi$ & $-\frac{3}{2} \alpha^2 e_Q^4 \pi$ \\
$\textrm{Im} \, t_{8\,\textrm{em}} (^3 P_2)$ & $- \alpha^2 e_Q^4 \pi$ & $0$ \\
  \hline\hline
\end{tabular}
  \caption{Discrepancies in the results of \cite{Ma:2002eva} and \cite{Brambilla:2006ph}.}
\label{tab:decaydiscrepancy}
\end{table}

Note that in \cite{Ma:2002eva} the decay rate of $\chi_{c2}$ contains the additional matrix element
\begin{equation}
\braket{\chi_{c2}| \psi^\dagger  \left (- \frac{i}{2} \right )^2 \overleftrightarrow{\bfD}^{(i} \overleftrightarrow{\bfD}^{j)} \left  ( - \frac{i}{2} \overset{\leftrightarrow}{\bfD} \cdot \bfSigma  \right ) \chi
| 0} \braket{0| \chi^\dagger \left ( - \frac{i}{2} \overleftrightarrow{\bfD}^{(i} \bfSigma^{j)} \right  )   \psi
| \chi_{c2}} + \textrm{H.c.}\,. \label{eq:extraoperator}
\end{equation}
Since in \cite{Brambilla:2006ph} it was shown that the contribution of this operator to the total decay rate is already accounted for
by the inclusion of $\braket{0| \mathcal{P}_{1} (^3 P_2)|0}$, we do not include it in the expression for the decay width.

\section{\label{sec:xsection} Production cross sections at \texorpdfstring{$\mathcal{O}(v^2)$}{second order in velocity}}
Plugging the laboratory frame short-distance coefficients from \Eqs\eqref{eq:sdcoeffs} into \Eqs\eqref{eq:mcoeff} 
we obtain the values of the matching coefficients that enter the factorized cross sections given in \Eqs\eqref{eq:cs}. 
If we define
\eq{
F_1 (^3 P_J) &\equiv 2 \pi M_{\chi_{cJ}} (s - M_{\chi_{cJ}}^2) \int_{-1}^1 d \cos \theta \, \tilde{F}_1 (^3 P_J), \\
G_1 (^3 P_J) &\equiv 2 \pi M_{\chi_{cJ}} (s - M_{\chi_{cJ}}^2) \int_{-1}^1 d \cos \theta \, \tilde{G}_1 (^3 P_J), \\
T_8 (^3 P_J) &\equiv 2 \pi M_{\chi_{cJ}} (s - M_{\chi_{cJ}}^2) \int_{-1}^1 d \cos \theta \, \tilde{T}_8 (^3 P_J), 
}
where $\theta$ denotes the angle between the beam line and the outgoing photon in the laboratory frame, the results read
\begin{subequations}
\eq{
\tilde{F}_1(^3 P_0) &= \frac{\pi \alpha^3 e_Q^4 (1-3r)^2 (1 + \cos^2 \theta)}{3  s^3  (1 - r)^2}, \\
\tilde{G}_1(^3 P_0) &= - \frac{  \pi \alpha^3 e_Q^4 (1-3r) (9 - 24 r + 35 r^2) (1 + \cos^2 \theta)}{15   s^3 (1 - r)^3}, \\
\tilde{T}_8(^3 P_0) &= - \frac{ \pi \alpha^3 e_Q^4 (1-9 r^2) (1 + \cos^2 \theta)}{6   s^3 (1 - r)^2}, \\
\nonumber\\
\tilde{F}_1(^3 P_1) & = \frac{2 \pi \alpha^3 e_Q^4 \left [(1 + \cos^2 \theta)  +  2 r (1 - \cos^2 \theta) \right ] }{s^3 (1 - r)^2}, \\
\tilde{G}_1(^3 P_1) & = -\frac{4 \pi \alpha^3 e_Q^4  \left [(4 - 9r)(1 + \cos^2 \theta)  + 
 (3r - 13r^2)(1 - \cos^2 \theta) \right ]}{5  s^3 (1 - r)^3},\\
\tilde{T}_8(^3 P_1) & = \frac{2 \pi \alpha^3 e_Q^4  \left [(1 + \cos^2 \theta)  + r (1 - \cos^2 \theta) \right ] }{ s^3 (1-r)^2}, \\
\nonumber\\
\tilde{F}_1(^3 P_2) & = \frac{2 \pi \alpha^3 e_Q^4 \left [(1+ 6r^2)(1 + \cos^2 \theta)  +  
6r (1 - \cos^2 \theta) \right ] }{3   s^3 (1-r)^2},  \\
\tilde{G}_1(^3 P_2) & = -\frac{4\pi \alpha^3 e_Q^4 \left [(3 - 6 r - 2 r^2 - 30 r^3)(1 + \cos^2 \theta)  + 15 r (1-3r)(1 - \cos^2 \theta)   \right ]}{15  s^3 (1 - r)^3} , \\
\tilde{T}_8(^3 P_2) &=  \frac{2 \pi \alpha^3 e_Q^4 \left [(1 + \cos^2 \theta)  +  3 r (1 - \cos^2 \theta) \right ]}{3  s^3 (1-r)^2},
}
\end{subequations}
and
\begin{subequations}
\eq{
F_1(^3 P_0) &= \frac{16 \pi^2 \alpha^3 e_Q^4 (1-3r)^2 M_{\chi_{c0}} (s - M_{\chi_{c0}}^2)}{9  s^3  (1 - r)^2}, \\
G_1(^3 P_0) &= - \frac{ 16 \pi^2 \alpha^3 e_Q^4 (1-3r) (9 - 24 r + 35 r^2) M_{\chi_{c0}} (s - M_{\chi_{c0}}^2)}{45   s^3 (1 - r)^3}, \\
T_8(^3 P_0) &= - \frac{8 \pi^2 \alpha^3 e_Q^4 (1-9 r^2) M_{\chi_{c0}}  (s - M_{\chi_{c0}}^2)}{9   s^3 (1 - r)^2}, \\
\nonumber\\
F_1(^3 P_1) & = \frac{32 \pi^2 \alpha^3 e_Q^4 (1+r) M_{\chi_{c1}} (s - M_{\chi_{c1}}^2)}{3  s^3 (1 - r)^2}, \\
G_1(^3 P_1) & = -\frac{32 \pi^2 \alpha^3 e_Q^4 (8 - 15 r - 13 r^2) M_{\chi_{c1}} (s - M_{\chi_{c1}}^2)}{15  s^3 (1 - r)^3},\\
T_8(^3 P_1) & = \frac{16 \pi^2 \alpha^3 e_Q^4 (2+r) M_{\chi_{c1}} (s - M_{\chi_{c1}}^2)}{3 s^3 (1-r)^2}, \\
\nonumber\\
F_1(^3 P_2) & = \frac{32 \pi^2 \alpha^3 e_Q^4 (1 + 3r + 6 r^2) M_{\chi_{c2}} (s - M_{\chi_{c2}}^2)}{9   s^3 (1-r)^2},  \\
G_1(^3 P_2) & = -\frac{32 \pi^2 \alpha^3 e_Q^4 (2+3r) (3 - 3r - 20 r^2) M_{\chi_{c2}} (s - M_{\chi_{c2}}^2)}{45  s^3 (1 - r)^3},  \\
T_8(^3 P_2) &=  \frac{16 \pi^2 \alpha^3 e_Q^4 (2 + 3r) M_{\chi_{c2}} (s - M_{\chi_{c2}}^2)}{9  s^3 (1-r)^2}.
}
\end{subequations}

In order to write the production cross sections, for consistency we combine our findings with the known $\mathcal{O}(\alpha_{\rm s} v^0)$ corrections to $F_1(^3 P_J)$ computed in \cite{Sang:2009jc,Li:2009zu}. 
The explicit values of the coefficients $C^i_j (r)$ can be found in Appendix B of \cite{Sang:2009jc}. 
The $\mathcal{O} ( \alpha_{\rm s} v^2)$ corrections from \cite{Xu:2014zra} are, however, not included. 
The first reason is that the corrections reported in \cite{Xu:2014zra} are incomplete, as they do not include contributions from $\braket{0|\mathcal{T}_8 (^3 P_J)|0}$. 
The second one is that, to be consistent with the NRQCD power counting rules at that order in $\alpha_{\rm s}$ and $v$, 
one would also need to include $\mathcal{O} ( \alpha_{\rm s}^0 v^4)$ and $\mathcal{O} ( \alpha_{\rm s}^2 v^0)$ corrections, which are not available yet. 
The differential and total cross sections read
\begin{subequations}
\eq{
& \frac{d}{d \cos \theta} \sigma(e^+ e^- \to \chi_{c0} + \gamma) =  \frac{ (4\pi \alpha)^3 e_Q^4 (1-3 r)^2 (1 + \cos^2\theta) M_{\chi_{c0}} (s - M_{\chi_{c0}}^2) }{192 \pi^2 m^4 s^3 (1-r)^2} \nonumber \\
  &\qquad  \times \biggl \{   \left ( 1 + \frac{\alpha_{\rm s}}{\pi} C^0_0 (r)  \right ) \braket{0|\mathcal{O}_1 (^3 P_0)|0} - \frac{( 9 - 24 r + 35 r^2)}{5 m^2 (1-3r)(1-r)} \braket{0|\mathcal{P}_1 (^3 P_0)|0} \nonumber \\
& \qquad - \frac{(1 + 3 r)}{ 2 m (1 - 3r)} \braket{0|\mathcal{T}_8 (^3 P_0)|0}  \biggr \}, \\
\nonumber \\
& \frac{d}{d \cos \theta} \sigma(e^+ e^- \to \chi_{c1} + \gamma) =  \frac{(4\pi \alpha)^3 e_Q^4  M_{\chi_{c1}} (s - M_{\chi_{c1}}^2)}{32 \pi^2 m^4 s^3 (1-r)^2} \nonumber \\
 & \qquad \times \biggl \{   \biggl [ (1 + \cos^2 \theta ) \left (1 + \frac{\alpha_{\rm s}}{\pi} C_1^0(r) \right)+ 2r  (1 - \cos^2 \theta  ) \left (1 + \frac{\alpha_{\rm s}}{\pi} C_1^{\pm 1} (r) \right )
\biggr ]  \braket{0|\mathcal{O}_1 (^3 P_1)|0} \nonumber \\
&\qquad - \frac{ 2(4-9r)(1+ \cos^2 \theta) + 2 r (3-13r)(1- \cos^2 \theta)  }{5 m^2 (1-r)} \braket{0|\mathcal{P}_1 (^3 P_1)|0} \nonumber \\
&\qquad + \frac{(1 + \cos^2 \theta) + r (1 - \cos^2 \theta)}{m} \braket{0|\mathcal{T}_8 (^3 P_1)|0}  \biggr \}, \\
\nonumber \\
& \frac{d}{d \cos \theta} \sigma(e^+ e^- \to \chi_{c2} + \gamma) = \frac{(4 \pi \alpha)^3 e_Q^4 M_{\chi_{c2}} (s - M_{\chi_{c2}}^2)}{96 \pi^2  m^4 s^3 (1-r)^2} \nonumber \\
&\qquad  \times \biggl \{  \biggl [
 (1+ \cos^2 \theta) \left (1 + \frac{\alpha_{\rm s}}{\pi} C_2^0(r) \right)  + 6 r (1 - \cos^2 \theta) \left (1 + \frac{\alpha_{\rm s}}{\pi} C_2^{\pm 1}(r) \right) \nonumber \\
&\qquad + 6 r^2 (1 + \cos^2 \theta) \left (1 + \frac{\alpha_{\rm s}}{\pi} C_2^{\pm 2}(r) \right)
\biggr ]  \braket{0|\mathcal{O}_1 (^3 P_2)|0} \nonumber \\
&\qquad - \frac{2(3 - 6r - 2r^2 -30 r^3)(1 +  \cos^2 \theta) + 30 r (1-3r)(1 -  \cos^2 \theta)}{5 m^2 (1-r)} \braket{0|\mathcal{P}_1 (^3 P_2)|0}  \nonumber \\
&\qquad + \frac{(1 + \cos^2 \theta) + 3 r (1 - \cos^2 \theta)}{m} \braket{0|\mathcal{T}_8 (^3 P_2)|0}  \biggr \}
}
\end{subequations}
and 
\begin{subequations}
\eq{
& \sigma(e^+ e^- \to \chi_{c0} + \gamma) =  \frac{(4 \pi \alpha)^3 e_Q^4 (1-3 r)^2 M_{\chi_{c0}} (s - M_{\chi_{c0}}^2)}{36 \pi m^4 s^2 (1-r)^2} \nonumber \\
&\qquad \times \biggl \{  \left ( 1 + \frac{\alpha_{\rm s}}{\pi} C^0_0 (r)  \right ) \braket{0|\mathcal{O}_1 (^3 P_0)|0} - \frac{( 9 - 24 r + 35 r^2 )}{5 m^2 (1 - 3r )(1-r)} \braket{0|\mathcal{P}_1 (^3 P_0)|0} \nonumber \\
&\qquad - \frac{(1 +3 r)}{ 2 m (1 - 3r)} \braket{0|\mathcal{T}_8 (^3 P_0)|0}  \biggr \}, \\
\nonumber\\
& \sigma(e^+ e^- \to \chi_{c1} + \gamma) = \frac{(4 \pi \alpha)^3 e_Q^4 (1 + r) M_{\chi_{c1}} (s - M_{\chi_{c1}}^2)}{6 \pi m^4 s^3 (1-r)^2} \nonumber \\
&\qquad \times \biggl \{ \left ( 1 + \frac{\alpha_{\rm s}}{\pi} \frac{C^0_1 (r) + r C^1_1 (r)}{1+r}  \right ) \braket{0|\mathcal{O}_1 (^3 P_1)|0} 
- \frac{( 8 - 15 r - 13 r^2)}{5 m^2 (1-r^2)} \braket{0|\mathcal{P}_1 (^3 P_1)|0} \nonumber\\ 
&\qquad + \frac{(2 + r)}{2 m (1 + r)} \braket{0|\mathcal{T}_8 (^3 P_1)|0}  \biggr \}, \\
\nonumber\\
& \sigma(e^+ e^- \to \chi_{c2} + \gamma) =  \frac{(4 \pi \alpha)^3 e_Q^4 (1 + 3r + 6 r^2 ) M_{\chi_{c2}} (s - M_{\chi_{c2}}^2)}{18 \pi m^4 s^3 (1-r)^2}  \nonumber \\
&\qquad \times \biggl \{ \left ( 1 + \frac{\alpha_{\rm s}}{\pi} \frac{C^0_2 (r) + 3 r C^1_2 (r) + 6 r^2 C_2^2 (r)}{1+ 3r + 6r^2}  \right )  \braket{0|\mathcal{O}_1 (^3 P_2)|0} \nonumber \\
&\qquad - \frac{(2+3r)(3-3r - 20r^2)}{5 m^2 (1-r) (1 + 3 r + 6 r^2)} \braket{0|\mathcal{P}_1 (^3 P_2)|0}
+ \frac{(2 + 3 r)}{2 m (1 + 3 r + 6 r^2)} \braket{0|\mathcal{T}_8 (^3 P_2)|0}  \biggr \},
} \label{eq:finalCShadronmass}
\end{subequations}
where we recall that $r \equiv 4 m^2/s$.

The dependence on the physical quarkonium mass arises due to the presence of $M_{\chi_{cJ}}$ in \Eq\eqref{eq:xsectionnrqcd} and \Eq\eqref{eq:phasespace}. 
This dependence can easily be eliminated via the Gremm-Kapustin relations. 
In particular, using \Eq\eqref{GKmass} we arrive at cross sections that do not explicitly depend on the mass of the $\chi_{cJ}$. 
These are 
\begin{subequations}
\eq{
& \frac{d}{d \cos \theta} \sigma(e^+ e^- \to \chi_{c0} + \gamma) =  \frac{(4 \pi \alpha)^3 e_Q^4 (1-3 r)^2 (1 + \cos^2\theta)}{48 \pi m^3 s^2 (1-r)} \nonumber \\
&\qquad \times \biggl \{ \left ( 1 + \frac{\alpha_{\rm s}}{\pi} C^0_0 (r)  \right ) \braket{0|\mathcal{O}_1 (^3 P_0)|0} - \frac{( 13 - 18 r + 25 r^2 )}{10 m^2 (1 - r)(1-3r)} \braket{0|\mathcal{P}_1 (^3 P_0)|0} \nonumber \\
&\qquad - \frac{(3 - 2 r + 3 r^2 )}{4 m (1 - r)(1-3r)} \braket{0|\mathcal{T}_8 (^3 P_0)|0}  \biggr \}, \\
\nonumber \\
& \frac{d}{d \cos \theta} \sigma(e^+ e^- \to \chi_{c1} + \gamma) =  \frac{(4 \pi \alpha)^3 e_Q^4}{8 \pi m^3 s^2 (1-r)} \nonumber \\
&\qquad \times \biggl \{ \biggl [ (1 + \cos^2 \theta ) \left (1 + \frac{\alpha_{\rm s}}{\pi} C_1^0(r) \right) + 2r  (1 - \cos^2 \theta  ) \left (1 + \frac{\alpha_{\rm s}}{\pi} C_1^{\pm 1} (r) \right )
\biggr]  \braket{0|\mathcal{O}_1 (^3 P_1)|0} \nonumber \\
&\qquad - \frac{ (11 - 21 r)(1 + \cos^2 \theta) + 2 r (1-11r) (1 - \cos^2 \theta) }{10 m^2 (1-r)} \braket{0|\mathcal{P}_1 (^3 P_1)|0} \nonumber \\
&\qquad + \frac{ (3 - r)(1 + \cos^2 \theta) + 2 r (1+r) (1 - \cos^2 \theta) }{4 m (1 - r)} \braket{0|\mathcal{T}_8 (^3 P_1)|0}  \biggr \}, \\
\nonumber \\
& \frac{d}{d \cos \theta} \sigma(e^+ e^- \to \chi_{c2} + \gamma) = \frac{(4 \pi \alpha)^3 e_Q^4}{24 \pi m^3 s^2 (1-r)} \nonumber \\
&\qquad \times  \biggl \{ \biggl [
(1+ \cos^2 \theta) \left (1 + \frac{\alpha_{\rm s}}{\pi} C_2^0(r) \right) + 6 r (1 - \cos^2 \theta) \left (1 + \frac{\alpha_{\rm s}}{\pi} C_2^{\pm 1}(r) \right) \nonumber \\
&\qquad + 6 r^2 (1 + \cos^2 \theta) \left (1 + \frac{\alpha_{\rm s}}{\pi} C_2^{\pm 2}(r) \right)
\biggr ]  \braket{0|\mathcal{O}_1 (^3 P_2)|0} \nonumber \\
&\qquad - \frac{( 7 - 9r - 38 r^2 - 30 r^3)(1 +  \cos^2 \theta) + 30 (1-3 r) r (1 -  \cos^2 \theta)}{10 m^2 (1-r)} \braket{0|\mathcal{P}_1 (^3 P_2)|0}  \nonumber \\
&\qquad + \frac{(3 - r - 6 r^2 + 18 r^3)(1 + \cos^2 \theta) + 6 r (1+r) (1 - \cos^2 \theta) }{4 m (1 - r)} \braket{0|\mathcal{T}_8 (^3 P_2)|0}  \biggr\},
}
\end{subequations}
and
\begin{subequations}
\eq{
& \sigma(e^+ e^- \to \chi_{c0} + \gamma) =  \frac{(4 \pi \alpha)^3 e_Q^4 (1-3 r)^2}{18 \pi m^3 s^2 (1-r)} \nonumber \\
&\qquad \times \biggl \{ \left ( 1 +  \frac{\alpha_{\rm s}}{\pi} C^0_0 (r)  \right ) \braket{0|\mathcal{O}_1 (^3 P_0)|0} - \frac{( 13 - 18 r + 25 r^2 )}{10 m^2 (1 - 4r + 3r^2)} \braket{0|\mathcal{P}_1 (^3 P_0)|0} \nonumber \\
&\qquad - \frac{(3 - 2 r + 3 r^2)}{4 m (1 - 4r + 3r^2)} \braket{0|\mathcal{T}_8 (^3 P_0)|0}  \biggr \}, \\
\nonumber\\
& \sigma(e^+ e^- \to \chi_{c1} + \gamma) = \frac{(4 \pi \alpha)^3 e_Q^4 (1 + r)}{3 \pi m^3 s^2 (1-r)} \nonumber \\
&\qquad \times \biggl \{ \left ( 1 + \frac{\alpha_{\rm s}}{\pi}  \frac{C^0_1 (r) + r C^1_1 (r)}{1+r} \right ) \braket{0|\mathcal{O}_1 (^3 P_1)|0} 
- \frac{( 11 - 20 r - 11 r^2)}{10 m^2 (1-r^2)} \braket{0|\mathcal{P}_1 (^3 P_1)|0} \nonumber\\
&\qquad + \frac{(3+r^2)}{4 m (1 - r^2)} \braket{0|\mathcal{T}_8 (^3 P_1)|0}  \biggr \}, \\
\nonumber\\
& \sigma(e^+ e^- \to \chi_{c2} + \gamma) =  \frac{(4 \pi \alpha)^3 e_Q^4 (1 + 3r + 6 r^2 )}{9 \pi m^3 s^2 (1-r)}  \nonumber \\
&\qquad \times  \biggl \{ \left ( 1 + \frac{\alpha_{\rm s}}{\pi} \frac{C^0_2 (r) + 3 r C^1_2 (r) + 6 r^2 C_2^2 (r)}{1+ 3r + 6r^2}  \right )  \braket{0|\mathcal{O}_1 (^3 P_2)|0} \nonumber \\
&\qquad - \frac{( 7 + 6r - 83 r^2 - 30 r^3)}{10 m^2 (1-r) (1 + 3 r + 6 r^2)} \braket{0|\mathcal{P}_1 (^3 P_2)|0}
+ \frac{(3 + 2 r - 3 r^2 + 18 r^3)}{4 m (1-r) (1 + 3 r + 6 r^2)} \braket{0|\mathcal{T}_8 (^3 P_2)|0}  \biggr \}. \label{eq:finalCS3}
} \label{eq:finalCS}
\end{subequations}

Comparing these results with the literature, we observe that the matching coefficients $F_1(^3 P_J)$ agree with those reported in \cite{Chung:2008km} and in all subsequent studies. 
Furthermore, our values for $G_1(^3 P_J)$ are consistent with the $K$-factors given in \cite{Li:2013nna}. The values of the matching coefficients $\tilde{T}_8(^3 P_J)$ and $T_8(^3 P_J)$ are original of this work.

\section{\label{sec:numerics} Numerical results}
The main limitation in the determination of the $e^+ e^- \to \gamma^\ast \to \chi_{cJ} + \gamma$ production cross sections comes from the uncertainties in the LDMEs.
Nevertheless we will see that we can express the LDMEs in such a way to reduce those uncertainties in the final results.

For $\braket{0| \mathcal{O}_1 (^3 P_J)|0}$ we can use the value quoted in \cite{Braaten:2002fi}, which follows from a Buchm\"uller-Tye potential model calculation \cite{Eichten:1995ch}
\begin{equation}
\braket{0| \mathcal{O}_1 (^3 P_J)|0} = ( 0.107 \pm 0.032 )\,\textrm{GeV}^5. \label{eq:BT}
\end{equation}
The error corresponds to an $\mathcal{O}(v^2)$ uncertainty that we estimate to be $30\%$ of the central value.\footnote{
The authors of \cite{Eichten:1995ch} do not provide an uncertainty for their result.}  
This uncertainty accounts also for the breaking of the heavy-quark spin symmetry.
Further, we notice that within the error bounds the choice in \Eq\eqref{eq:BT} is consistent with the most accurate values of $\braket{0| \mathcal{O}_1 (^3P_{0,2})|0}$ determined in Table I of the recent study \cite{Sang:2015uxg} devoted to $\mathcal{O}(\alpha_{\textrm{s}}^2 v^0)$ corrections of the $\chi_{cJ} \to 2 \gamma$ decay.

The number of the remaining independent LDMEs can be reduced by applying the heavy-quark spin symmetry \cite{Bodwin:1994jh}, 
which relates matrix elements with the same orbital momenta but different spin values and is valid up to corrections of $\mathcal{O}(v^2)$. 
Since $\braket{0| \mathcal{P}_1 (^3 P_J)|0}$ and $\braket{0| \mathcal{T}_8 (^3 P_J)|0}$ are already $v^2$ suppressed as compared to $\braket{0| \mathcal{O}_1 (^3 P_J)|0}$, 
the corrections from the breaking of the heavy-quark spin symmetry to the total cross section are of $\mathcal{O}(v^4)$ and therefore irrelevant for the accuracy that we are aiming at:
\eq{
\braket{0| \mathcal{P}_1 (^3 P_J)|0} &= \braket{0| \mathcal{P}_1 (^3 P_0)|0} \left  (1 + \mathcal{O}(v^2) \right ), \\
\braket{0| \mathcal{T}_8 (^3 P_J)|0} &= \braket{0| \mathcal{T}_8 (^3 P_0)|0} \left  (1 + \mathcal{O}(v^2) \right ).
\label{eq:ldmevals}
}
This is clearly different from the case of $\braket{0| \mathcal{O}_1 (^3 P_J)|0}$ where spin symmetry breaking effects, 
being relevant at $\mathcal{O}(v^2)$, cannot be neglected at our accuracy and have been included in the uncertainty of \Eq\eqref{eq:BT}.

The values of $\braket{0| \mathcal{P}_1 (^3 P_0)|0}$ and $\braket{0| \mathcal{T}_8 (^3 P_0)|0}$ can be determined
from the electromagnetic decays $\chi_{c0} \to \gamma \gamma$ and $\chi_{c2} \to \gamma \gamma$ that have recently been measured by the BES-III experiment~\cite{Ablikim:2012xi}, 
\eq{
\Gamma (\chi_{c0} \to \gamma \gamma) &= (2.33 \pm 0.20 \pm 0.22) \times 10^{-3} \, \textrm{MeV}, \\
\Gamma (\chi_{c2} \to \gamma \gamma) &= (0.63 \pm 0.04 \pm 0.06) \times 10^{-3} \, \textrm{MeV},
}
where the first error is statistical, and the second one consists of systematic errors and the error in the branching fraction combined in quadrature. 
Combining the NRQCD results for the electromagnetic decay rates in \Eqs\eqref{eq:decay} with the Gremm-Kapustin relation in \Eq\eqref{eq:gkrel} we obtain
\begin{subequations}
\eq{
 \Gamma (\chi_{c0} \to \gamma \gamma) &= \frac{6 \alpha^2 e_Q^4 \pi}{m^4} \left ( 1 + \frac{3 \pi^2 - 28}{9} \frac{\alpha_{\rm s}}{\pi} \right )
 \braket{0| \mathcal{O}_1 (^3 P_0)|0}- \frac{14 \alpha^2 e_Q^4 \pi}{m^6} \braket{0| \mathcal{P}_1 (^3 P_0)|0}
 \nonumber \\
 &  +\frac{3\alpha^2 e_Q^4 \pi}{m^5} \braket{0| \mathcal{T}_8 (^3 P_0)|0},
 \label{eq:decayWithCorrs1} \\
\braket{0| \mathcal{P}_1 (^3 P_0)|0}  &= m E_{\chi_{c0}} \braket{0| \mathcal{O}_1 (^3 P_0)|0} + \frac{m}{2} \braket{0| \mathcal{T}_8 (^3 P_0)|0},
} \label{eq:linSystem1}
\end{subequations}
and
\begin{subequations}
\eq{
 \Gamma (\chi_{c2} \to \gamma \gamma) &= \frac{8 \alpha^2 e_Q^4 \pi}{5 m^4} \left ( 1 - \frac{16}{3} \frac{\alpha_{\rm s}}{\pi} \right )
 \braket{0| \mathcal{O}_1 (^3 P_2)|0} - \frac{16 \alpha^2 e_Q^4 \pi}{5 m^6} \braket{0| \mathcal{P}_1 (^3 P_2)|0},
 \label{eq:decayWithCorrs2} \\
\braket{0| \mathcal{P}_1 (^3 P_2)|0}  &= m E_{\chi_{c2}} \braket{0| \mathcal{O}_1 (^3 P_2)|0} + \frac{m}{2} \braket{0| \mathcal{T}_8 (^3 P_2)|0}.
} \label{eq:linSystem2}
\end{subequations}
The decay rate formulas in \Eqs\eqref{eq:decayWithCorrs1} and \eqref{eq:decayWithCorrs2} contain not only corrections of $\mathcal{O}(\alpha_{\rm s}^0 v^0)$ and $\mathcal{O}(\alpha_{\rm s}^0 v^2)$, 
but also the $\mathcal{O}(\alpha_{\rm s})$ correction to $\textrm{Im} \, f_{\textrm{em}} (^3 P_{0,2})$ \cite{Barbieri:1975am,Barbieri:1980yp}. 
This is required for consistency with the NRQCD power counting, where corrections of $\mathcal{O}(\alpha_{\rm s} v^0)$ and $\mathcal{O}(\alpha_{\rm s}^0 v^2)$ are assumed to be equally important.

Equations \eqref{eq:linSystem1} and \eqref{eq:linSystem2} each describe a system of two linear equations with three unknowns:
$\braket{0| \mathcal{P}_1 (^3 P_0)|0}$, $\braket{0| \mathcal{T}_8 (^3 P_0)|0}$, $E_{\chi_{c0}}$
and  $\braket{0| \mathcal{P}_1 (^3 P_2)|0}$, $\braket{0| \mathcal{T}_8 (^3 P_2)|0}$, $E_{\chi_{c2}}$ respectively.
Since we employ the heavy-quark spin symmetry, both systems can be used to determine the unknown LDMEs,
which we choose to be $\braket{0| \mathcal{P}_1 (^3 P_0)|0}$, $\braket{0| \mathcal{T}_8 (^3 P_0)|0}$, as a function of the binding energy.
We will therefore use both \Eqs\eqref{eq:linSystem1} and \Eqs\eqref{eq:linSystem2} and present the combined results.

The mass $m$ appearing in the above equations is the charm pole mass. The pole mass is a poorly known quantity that, however, 
can be traded for the physical masses of the $\chi_{cJ}$, which are experimentally known, and the binding energy, which is one of our unknowns, according to
\begin{equation}
m = \frac{M_{\chi_{cJ}} -E_{\chi_{cJ}}}{2}.
\end{equation}
In order to be consistent with the NRQCD power counting, we replace $m$ with $(M_{\chi_{cJ}} - E_{\chi_{cJ}})/2$ in contributions that are of order $v^0$ 
and use $m  = M_{\chi_{cJ}}/2$ in contributions that are of order $v^2$ (with $\alpha_{\rm s} \sim v^2$).
Solving the system for the unknown LDMEs and expanding in $E_{\chi_{cJ}} \sim v^2$ up to relative order $v^2$ yield
\begin{subequations}
\eq{
\braket{0| \mathcal{P}_1 (^3 P_0)|0} &= \frac{3}{16} M_{\chi_{c0}} \braket{0| \mathcal{O}_1 (^3 P_0)|0} 
\left(  M_{\chi_{c0}} + \frac{(3 \pi^2 - 28)}{9 \pi} M_{\chi_{c0}} \frac{\alpha_{\rm s}}{\pi} + 2 E_{\chi_{c0}} \right )  \nonumber \\
& - \frac{M_{\chi_{c0}}^6}{512 \pi \alpha^2 e_Q^4} \Gamma (\chi_{c0} \to \gamma \gamma),  \label{eq:ldmes01} \\
\braket{0| \mathcal{T}_8 (^3 P_0)|0} &= \frac{3}{4} \braket{0| \mathcal{O}_1 (^3 P_0)|0} 
\left(  M_{\chi_{c0}} + \frac{(3 \pi^2 - 28)}{9 \pi} M_{\chi_{c0}} \frac{\alpha_{\rm s}}{\pi} - \frac{2}{3}E_{\chi_{c0}} \right )  \nonumber \\
&- \frac{M_{\chi_{c0}}^5}{128 \pi \alpha^2 e_Q^4} \Gamma (\chi_{c0} \to \gamma \gamma), \label{eq:ldmes11}
} \label{eq:ldmes1}
\end{subequations}
for the  determination using \Eqs\eqref{eq:linSystem1} and 
\begin{subequations}
\eq{
\braket{0| \mathcal{P}_1 (^3 P_0)|0} &= \frac{1}{8} M_{\chi_{c2}} \braket{0| \mathcal{O}_1 (^3 P_0)|0} 
\left(  M_{\chi_{c2}} - \frac{16}{3 \pi} M_{\chi_{c2}} \frac{\alpha_{\rm s}}{\pi} + 4 E_{\chi_{c2}} \right )  \nonumber \\
& - \frac{5 M_{\chi_{c2}}^6}{1024 \pi \alpha^2 e_Q^4} \Gamma (\chi_{c2} \to \gamma \gamma), \label{eq:ldmes02} \\
\braket{0| \mathcal{T}_8 (^3 P_0)|0} &= \frac{1}{2} M_{\chi_{c2}} \braket{0| \mathcal{O}_1 (^3 P_0)|0} 
\left(  1 - \frac{16}{3 \pi} \frac{\alpha_{\rm s}}{\pi} \right ) - \frac{5 M_{\chi_{c2}}^5}{256 \pi \alpha^2 e_Q^4} \Gamma (\chi_{c2} \to \gamma \gamma),\label{eq:ldmes12}
} \label{eq:ldmes2}
\end{subequations}
for the  determination using \Eqs\eqref{eq:linSystem2}. Furthermore, we choose
\eq{
M_{\chi_{c0}}                 & = \left (3414.75 \pm 0.31\right) \textrm{MeV}, \\
M_{\chi_{c2}}                 & = \left (3556.20 \pm 0.09\right) \textrm{MeV}, \\
\alpha( M_{\chi_{c0}} /2 ) = \alpha( M_{\chi_{c2}} /2 )    & = 1/133, \\
\alpha_{\rm s} ( M_{\chi_{c0}} /2 ) & = 0.285, \\
\alpha_{\rm s} ( M_{\chi_{c2}} /2 ) & = 0.280,
}
where for $M_{\chi_{cJ}}$ we use the most recent PDG values \cite{Olive:2016xmw}, 
while the values of the strong coupling constant (at 1-loop accuracy) were obtained with \textsc{RunDec} \cite{Chetyrkin:2000yt,Schmidt:2012az,Herren:2017osy}. 
The errors of $\alpha_{\rm s}$ from varying $M_{\chi_{cJ}}$ are of relative order $10^{-5}$ and hence can safely be neglected in comparison to other error sources in the present analysis.

\begin{figure}[ht]
\centering
\begin{subfigure}[b]{0.45\textwidth}
\includegraphics[width=\textwidth,clip]{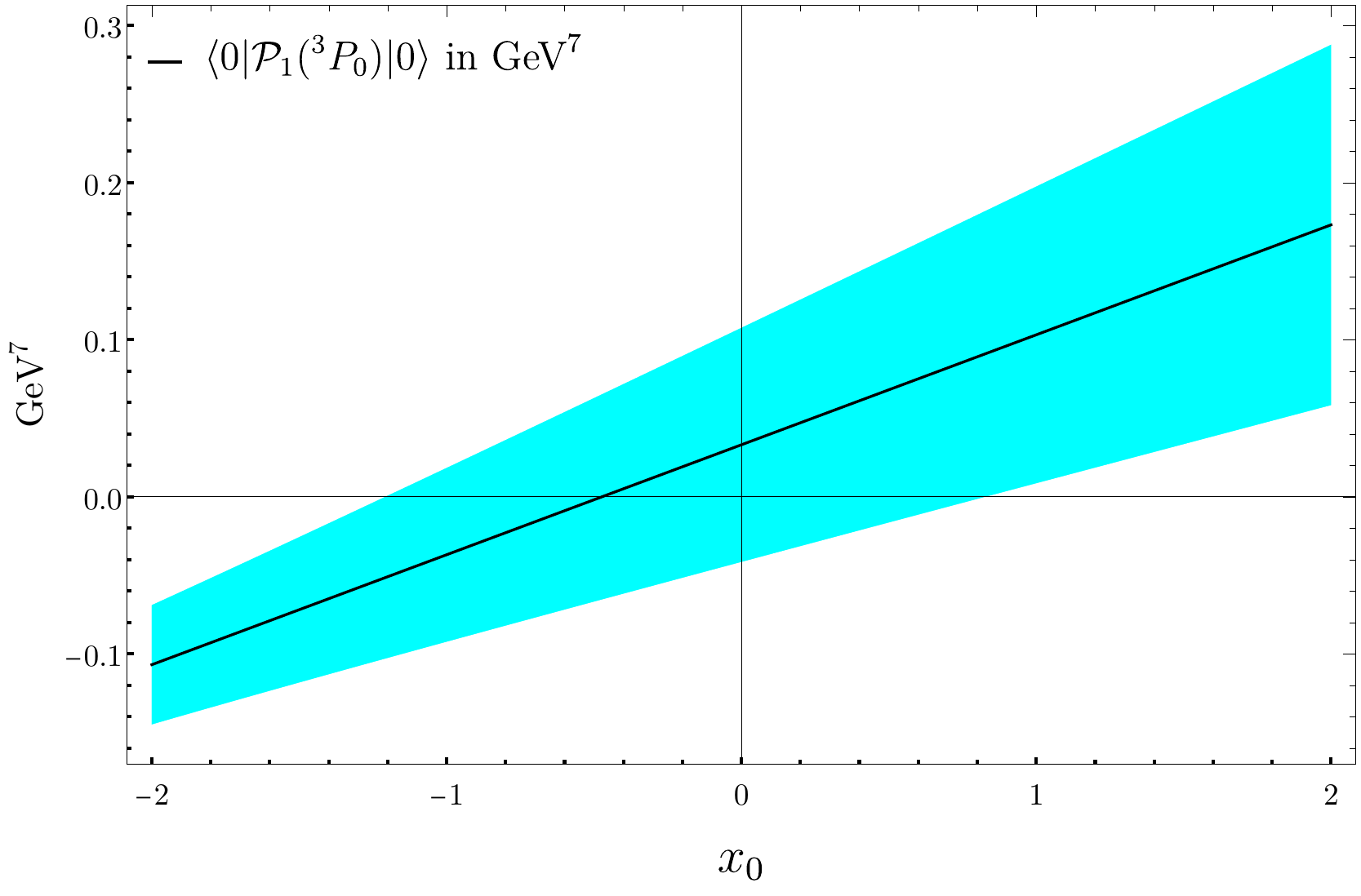}
    \end{subfigure}
\begin{subfigure}[b]{0.45\textwidth}
\includegraphics[width=\textwidth,clip]{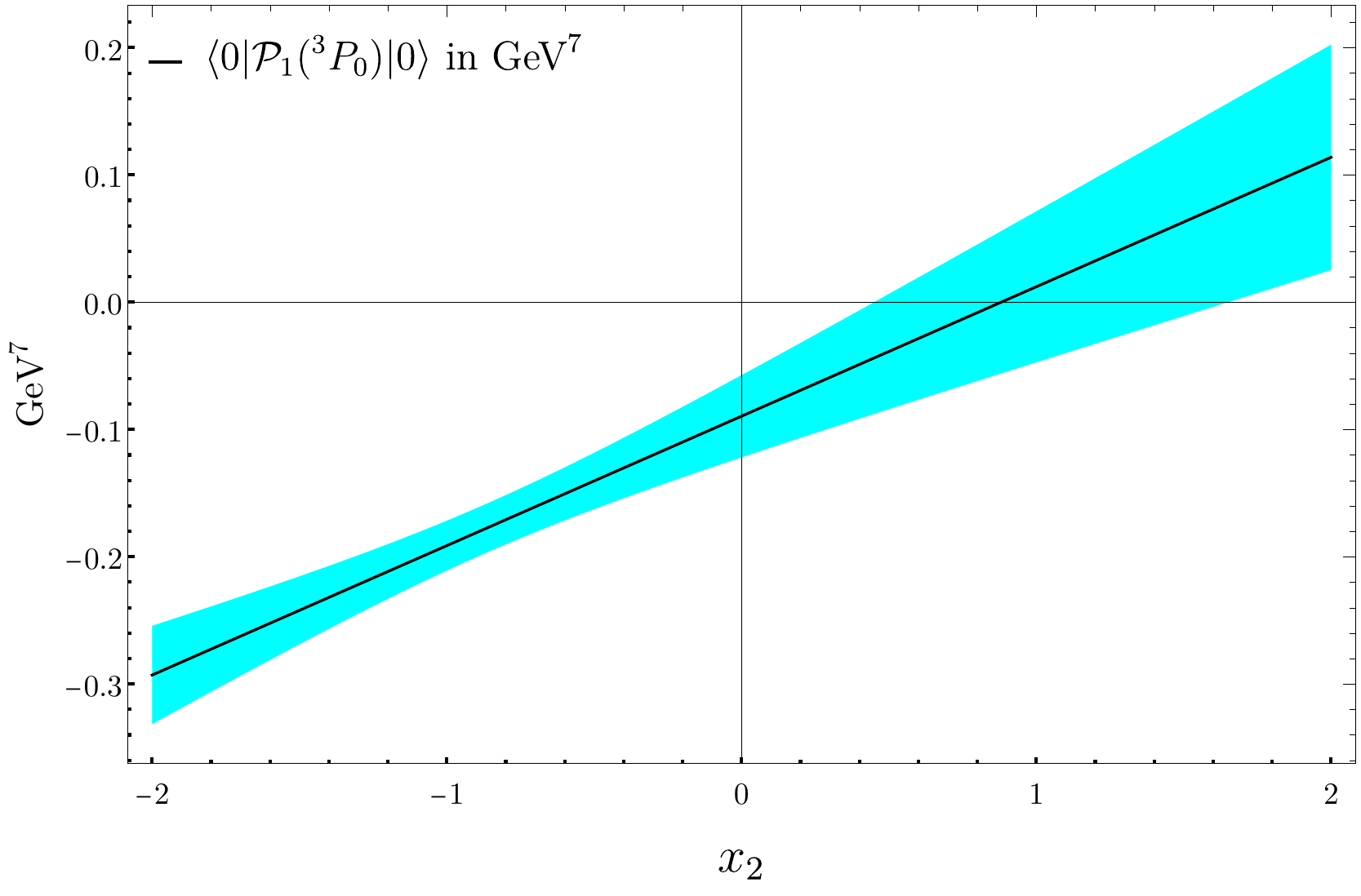}
    \end{subfigure}
\begin{subfigure}[b]{0.45\textwidth}
\includegraphics[width=\textwidth,clip]{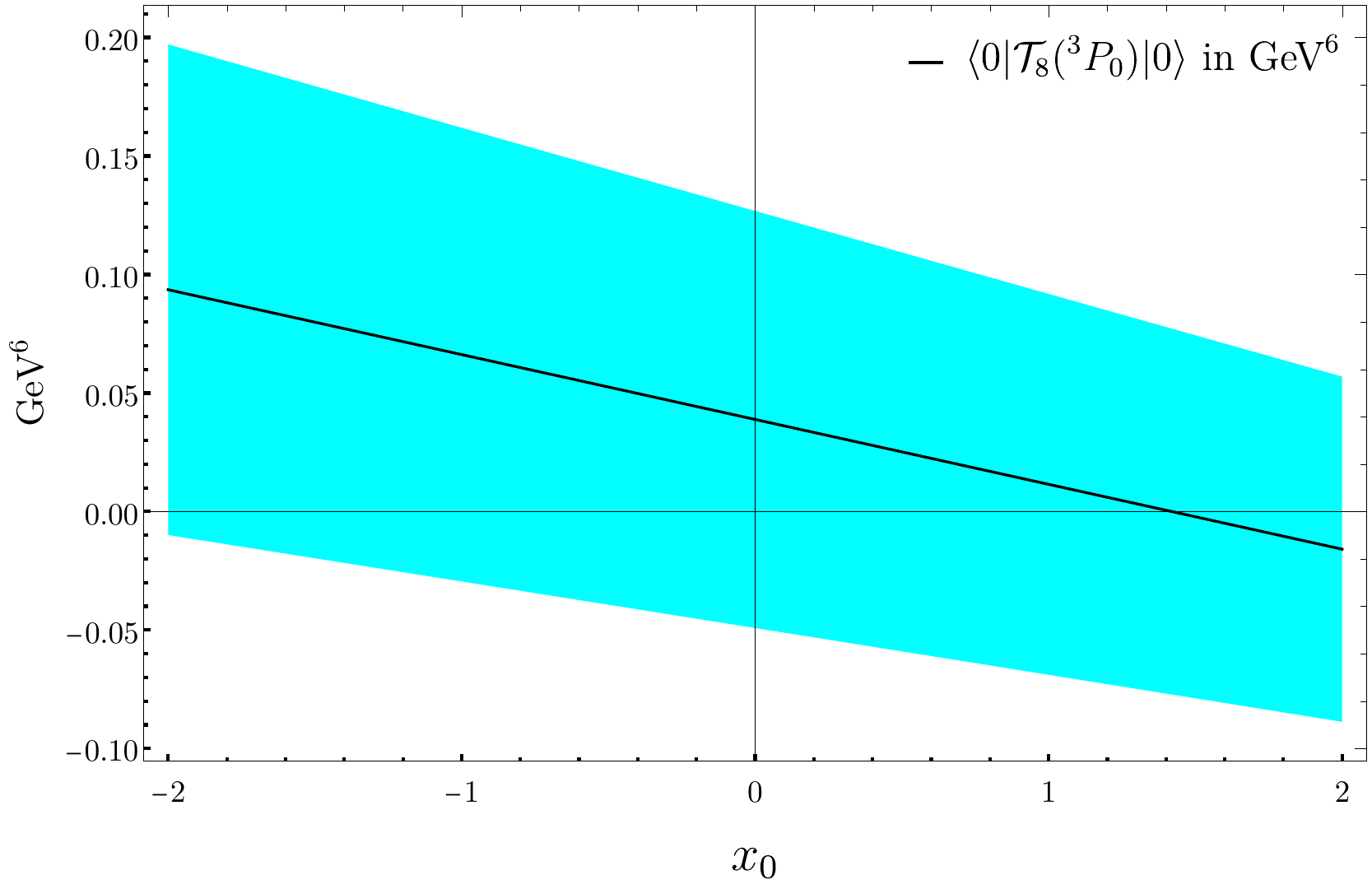}
    \end{subfigure}
\begin{subfigure}[b]{0.45\textwidth}
\includegraphics[width=\textwidth,clip]{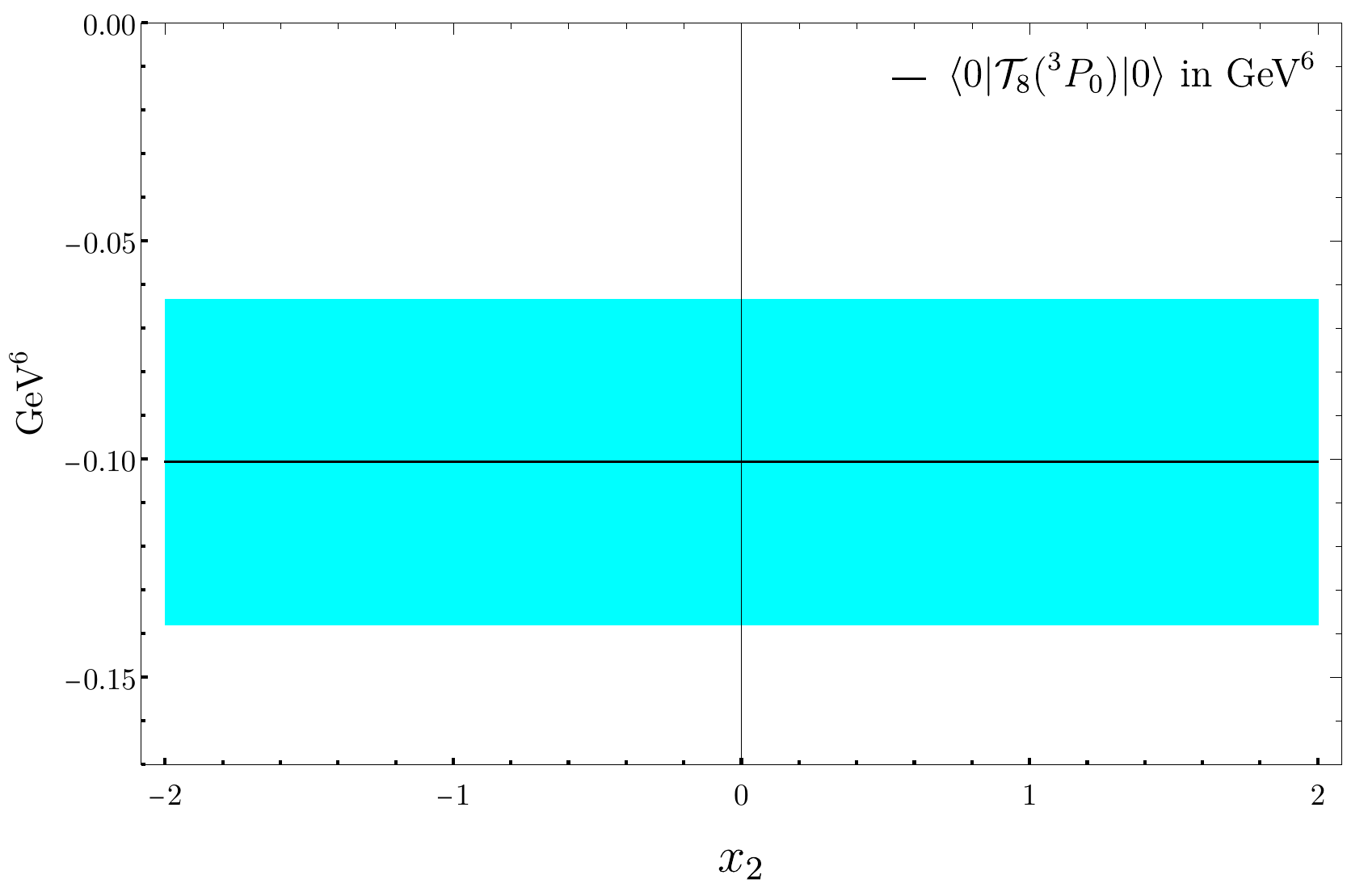}
    \end{subfigure}    
\caption{The matrix elements $\braket{0| \mathcal{P}_1 (^3 P_0)|0}$  and $\braket{0| \mathcal{T}_8 (^3 P_0)|0}$ as a function of $x_0$
  [determined from \Eqs\eqref{eq:ldmes1}] or $x_2$ [determined from \Eqs\eqref{eq:ldmes2}]. 
The cyan bands indicate the total uncertainty obtained from the quadrature of the uncertainty in $\braket{0| \mathcal{O}_1 (^3 P_0)|0}$ 
and the combined (in quadrature) uncertainty in $M_{\chi_{cJ}}$ and $\Gamma (\chi_{cJ} \to \gamma \gamma)$. 
The small uncertainty of $\braket{0| \mathcal{P}_1 (^3 P_0)|0}$ in the region around $x_2=-1.0$ results from the fact that, for this value of $x_2$, 
the terms proportional to $\braket{0| \mathcal{O}_1 (^3 P_0)|0}$ in \Eq\eqref{eq:ldmes02} become very small.
Furthermore, when $\braket{0| \mathcal{T}_8 (^3 P_0)|0}$ is determined from \Eqs\eqref{eq:ldmes12}, it does not depend on the binding energy and hence on $x_2$.}
\label{fig:ldme-x-dep}
\end{figure}

The matrix elements $\braket{0| \mathcal{P}_1 (^3 P_0)|0}$ and $\braket{0| \mathcal{T}_8 (^3 P_0)|0}$ depend on the binding energy, $E_{\chi_{cJ}}$.
Since of the binding energy we only know the nonrelativistic scaling, $E_{\chi_{cJ}} \sim m v^2$, where $v^2 \approx 0.3$ for charmonia \cite{Bodwin:1994jh}, 
it may be useful to parametrize the binding energy in terms of a dimensionless parameter $x_J$ of $\mathcal{O}(1)$ and write it as 
\begin{equation}
E_{\chi_{cJ}} = 0.3 \, x_J \frac{M_{\chi_{cJ}}}{2}.
\label{eq:ebind}
\end{equation}
The LDMEs $\braket{0| \mathcal{P}_1 (^3 P_0)|0}$ and $\braket{0| \mathcal{T}_8 (^3 P_0)|0}$ as a function of $x_J$ are shown in \Fig\ref{fig:ldme-x-dep}.
The uncertainties coming from $M_{\chi_{cJ}}$, $\Gamma (\chi_{cJ} \to \gamma \gamma)$ and $\braket{0| \mathcal{O}_1 (^3 P_0)|0}$
have been estimated using Gaussian error propagation and are shown by the band.
Because the Buchm\"uller-Tye potential or the Cornell potential or a purely confining potential would all give a positive 
binding energy,\footnote{
A purely Coulombic potential, which would give a negative binding energy, seems inappropriate for the loosely bound $\chi_{cJ}$.}
we will restrict $x_J$ to positive values centered around the natural value $x_J=1.0$.
Allowing for a $100\%$ uncertainty, we take 
\begin{equation}
x_J = 1.0 \pm 1.0\,.
\label{uncertaintyxJ}
\end{equation}
With this choice we obtain
\begin{subequations}
\eq{
\braket{0| \mathcal{P}_1 (^3 P_0)|0}  &= (0.103 \pm 0.092 \pm 0.070 \pm 0.026)\, \textrm{GeV}^7, \label{eq:ldmes-11} \\
\braket{0| \mathcal{T}_8 (^3 P_0)|0}  &= (0.0115 \pm 0.0750 \pm 0.0274 \pm 0.0305) \, \textrm{GeV}^6,
}
\label{eq:ldmes-1}
\end{subequations}
from the determination using \Eqs\eqref{eq:linSystem1} and
\begin{subequations}
\eq{
\braket{0| \mathcal{P}_1 (^3 P_0)|0}  &= (0.012 \pm 0.057 \pm 0.102 \pm 0.020)\, \textrm{GeV}^7, \label{eq:ldmes-21} \\
\braket{0| \mathcal{T}_8 (^3 P_0)|0}  &= (-0.101 \pm 0.030 \pm 0.000 \pm 0.023) \, \textrm{GeV}^6,
}
\label{eq:ldmes-2}
\end{subequations}
from the determination using \Eqs\eqref{eq:linSystem2}, where the uncertainties were estimated using the Gaussian error propagation formula.
The first error originates from the uncertainty in $\braket{0| \mathcal{O}_1 (^3 P_0)|0}$, 
the second one originates from the uncertainty in $x_J$, and the third one is the combined (in quadrature) uncertainty in  $M_{\chi_{cJ}}$ and $\Gamma (\chi_{cJ} \to \gamma \gamma)$.

The values of the LDMEs given in \Eqs\eqref{eq:ldmes-1} and \Eqs\eqref{eq:ldmes-2} are consistent with each other within the error bounds, 
as expected from the heavy-quark spin symmetry. 
Under the assumption that the heavy-quark spin symmetry holds at leading order in the velocity expansion, 
we can average the two determinations, which yields our best determination for these LDMEs
\begin{subequations}
\eq{
\braket{0| \mathcal{P}_1 (^3 P_0)|0}  &= (0.058 \pm 0.074 \pm 0.086 \pm 0.033)\, \textrm{GeV}^7, \\
\braket{0| \mathcal{T}_8 (^3 P_0)|0}  &= (-0.045 \pm 0.052 \pm 0.014 \pm 0.039) \, \textrm{GeV}^6,
}
\label{eq:ldmes-added}
\end{subequations}
where the uncertainties in $\braket{0| \mathcal{O}_1 (^3 P_0)|0}$ and $x_J$ were added linearly [first and second uncertainties in \Eqs\eqref{eq:ldmes-added} respectively],
while those  in  $M_{\chi_{cJ}}$ and $\Gamma (\chi_{cJ} \to \gamma \gamma)$ (third uncertainty) are regarded as independent and were therefore added quadratically. The values given in \Eqs\eqref{eq:ldmes-added} represent our best determination of  $\braket{0| \mathcal{P}_1 (^3 P_0)|0}$ and $\braket{0| \mathcal{T}_8 (^3 P_0)|0}$ using the heavy-quark spin symmetry.
\begin{figure}[ht]
\centering
\begin{subfigure}[b]{0.45\textwidth}
\includegraphics[width=\textwidth,clip]{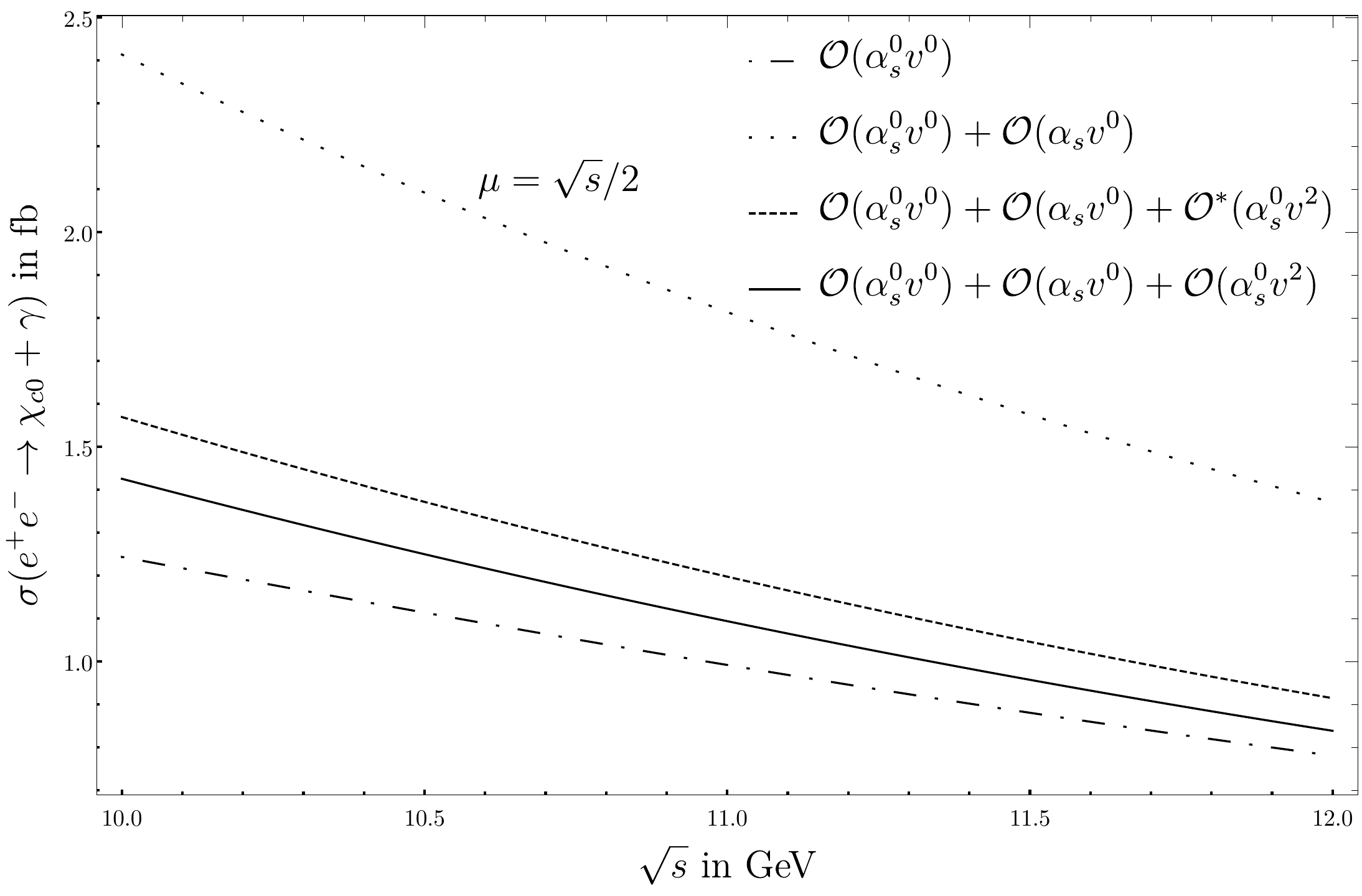}
    \end{subfigure}
\begin{subfigure}[b]{0.45\textwidth}
\includegraphics[width=\textwidth,clip]{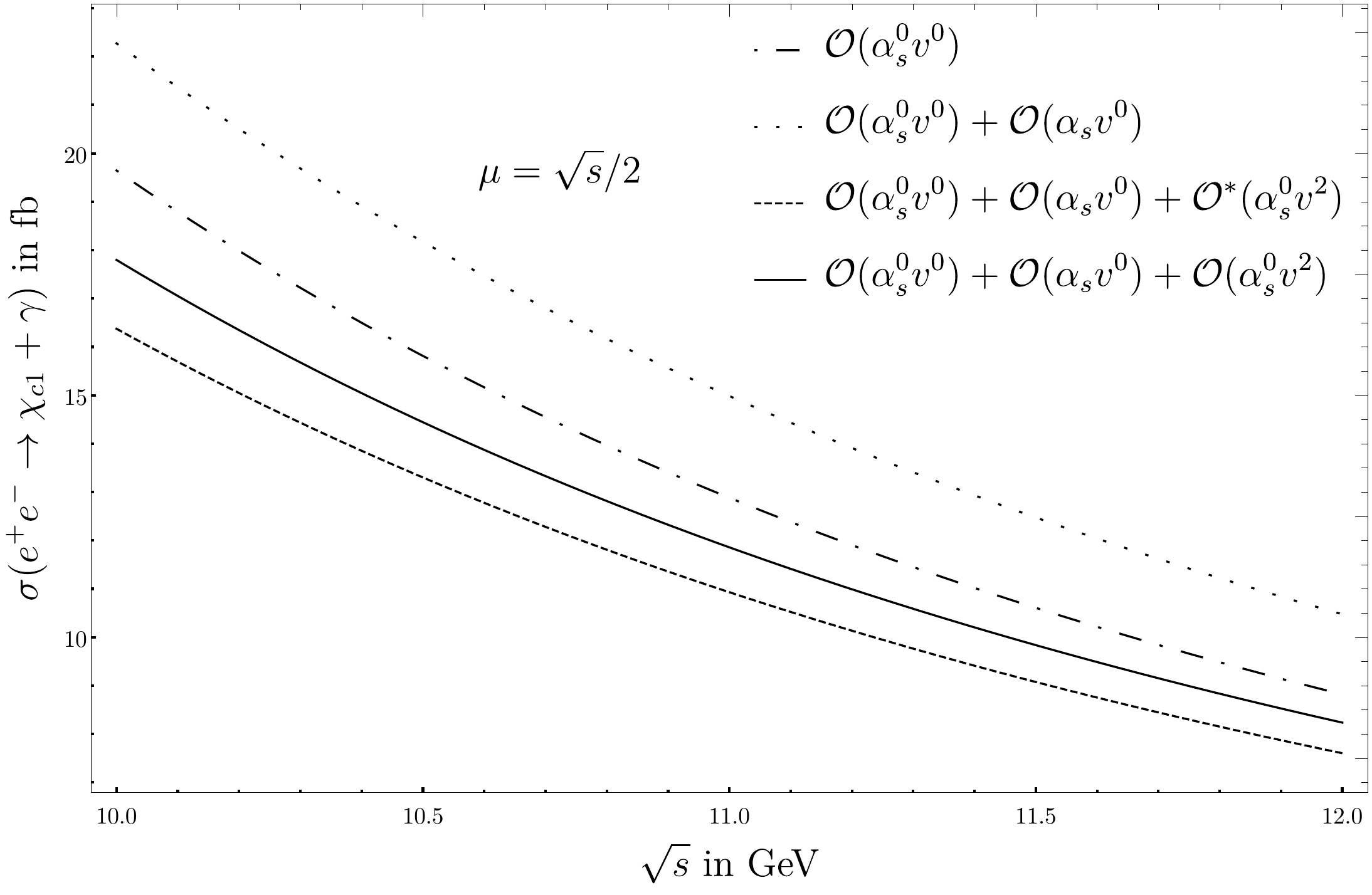}
    \end{subfigure}
\begin{subfigure}[b]{0.45\textwidth}
\includegraphics[width=\textwidth,clip]{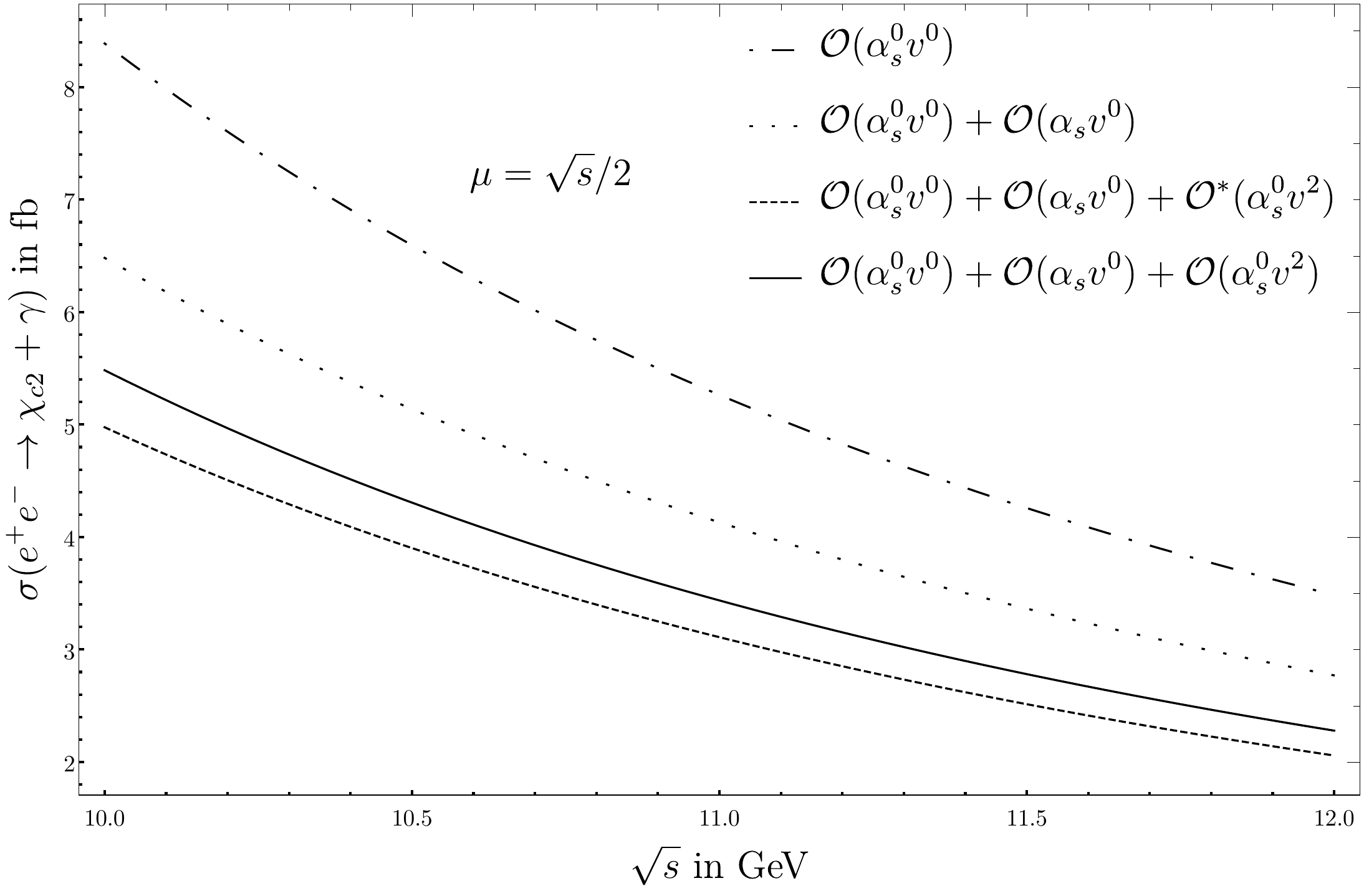}
    \end{subfigure}
\caption{Cross sections for the exclusive production of a $\chi_{cJ}$ and a hard photon in the energy region of Belle II. The matrix elements $\braket{0| \mathcal{P}_1 (^3 P_0)|0}$ and $\braket{0| \mathcal{T}_8 (^3 P_0)|0}$ were determined from \Eqs\eqref{eq:linSystem1}. The dash-dotted curve shows only the contribution from the $\braket{0| \mathcal{O}_1 (^3 P_0)|0}$ matrix element multiplied with the tree level matching coefficient. 
The dotted curve also includes the loop correction to the matching coefficient of $\braket{0| \mathcal{O}_1 (^3 P_0)|0}$. 
The dashed curve incorporates the contribution from  $\braket{0| \mathcal{P}_1 (^3 P_0)|0}$ but not $\braket{0| \mathcal{T}_8 (^3 P_0)|0}$. 
Finally, the solid curve displays our final result that contains all the relevant contributions.}
\label{fig:cs}
\end{figure}

\begin{figure}[ht]
\centering
\begin{subfigure}[b]{0.45\textwidth}
\includegraphics[width=\textwidth,clip]{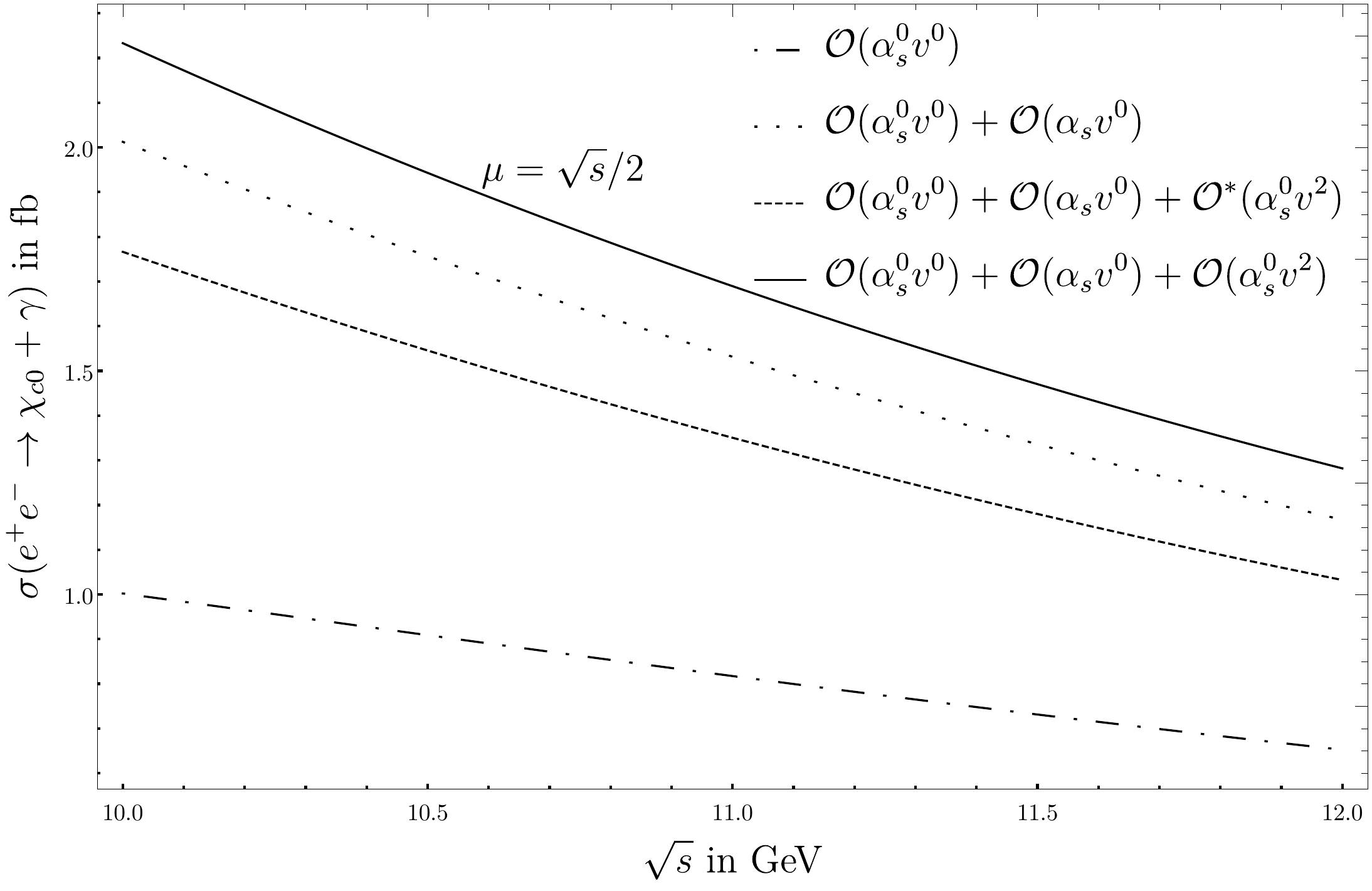}
    \end{subfigure}
\begin{subfigure}[b]{0.45\textwidth}
\includegraphics[width=\textwidth,clip]{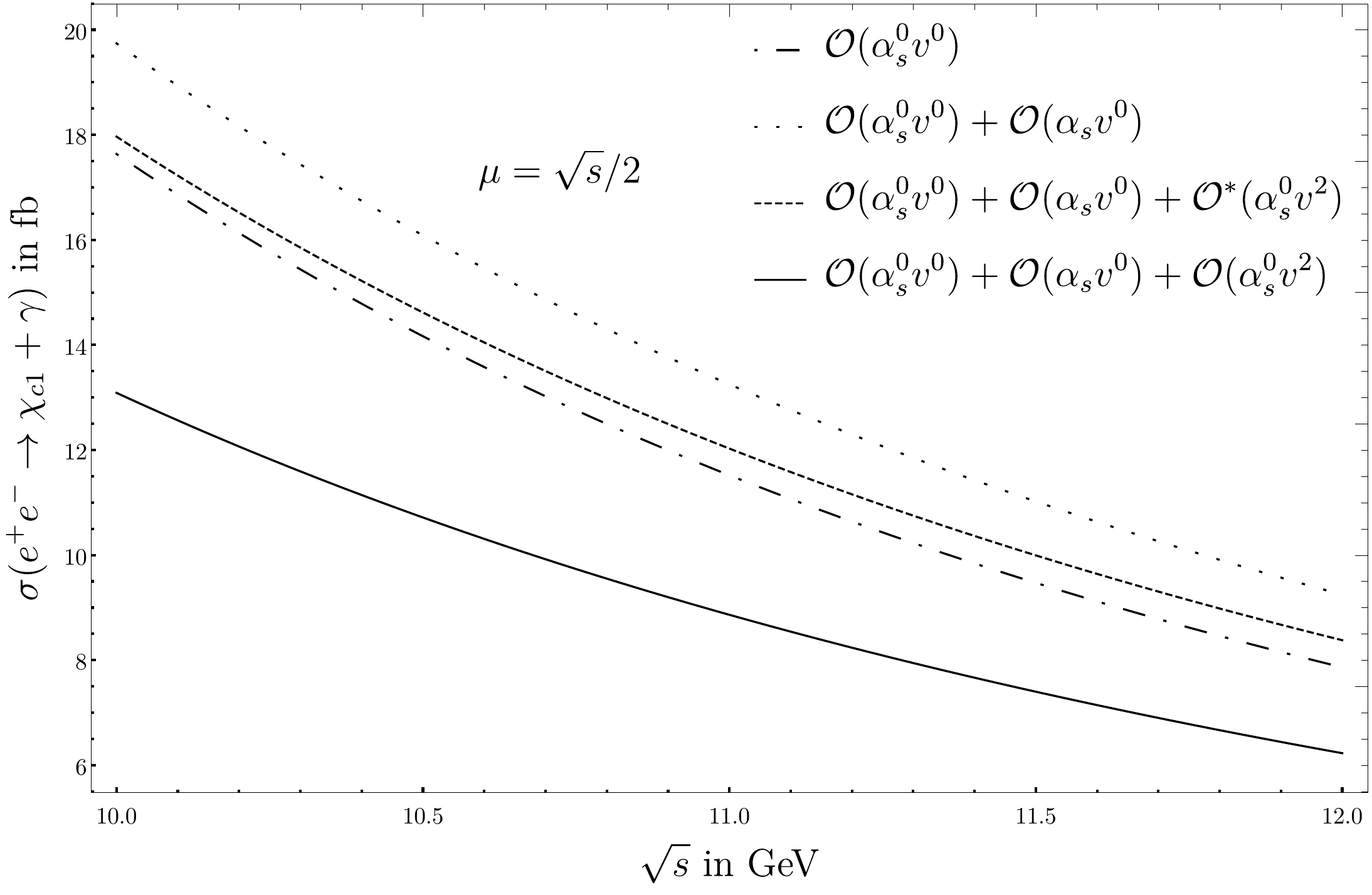}
    \end{subfigure}
\begin{subfigure}[b]{0.45\textwidth}
\includegraphics[width=\textwidth,clip]{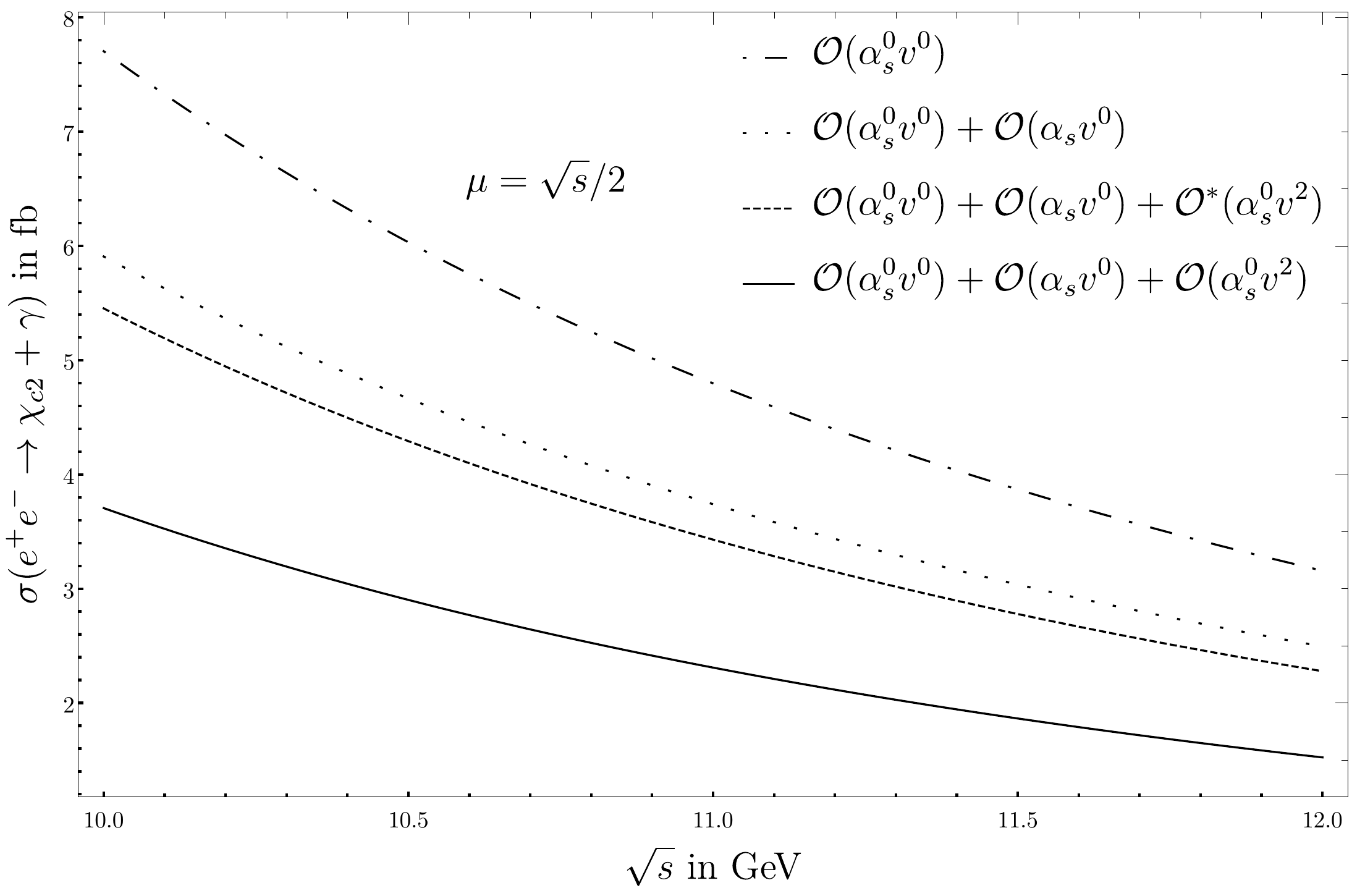}
    \end{subfigure}
\caption{Cross sections for the exclusive production of a $\chi_{cJ}$ and a hard photon in the energy region of Belle II. The matrix elements $\braket{0| \mathcal{P}_1 (^3 P_0)|0}$ and $\braket{0| \mathcal{T}_8 (^3 P_0)|0}$ were determined from \Eqs\eqref{eq:linSystem2}. The dash-dotted curve shows only the contribution from the $\braket{0| \mathcal{O}_1 (^3 P_0)|0}$ matrix element multiplied with the tree level matching coefficient. 
The dotted curve also includes the loop correction to the matching coefficient of $\braket{0| \mathcal{O}_1 (^3 P_0)|0}$. 
The dashed curve incorporates the contribution from  $\braket{0| \mathcal{P}_1 (^3 P_0)|0}$ but not $\braket{0| \mathcal{T}_8 (^3 P_0)|0}$. 
Finally, the solid curve displays our final result that contains all the relevant contributions.}
\label{fig:csc}
\end{figure}

In order to reduce the uncertainties in the total cross sections $\sigma (e^+ e^- \to \chi_{ cJ} +  \gamma)$, it is useful to compute them 
by replacing the matrix elements $\braket{0| \mathcal{P}_1 (^3 P_0)|0}$ and $\braket{0| \mathcal{T}_8 (^3 P_0)|0}$ with their expressions from \Eqs\eqref{eq:ldmes1} or \Eqs\eqref{eq:ldmes2}.
In this way each cross section will have five independent sources of uncertainty, which are $M_{\chi_{cJ}}$, $\Gamma (\chi_{cJ} \to \gamma \gamma)$, 
$\braket{0| \mathcal{O}_1 (^3 P_0)|0}$), $x_J$ and the renormalization scale $\mu$ that enters through $\alpha_{\rm s}$. 

The process $e^+ e^- \to \chi_{cJ} + \gamma$ can be measured at B-factories with a sufficiently high CM energy to produce a $\chi_{cJ}$ and a hard photon, e.\,g., at Belle II.
Hence, we select our input parameters as
\eq{
\sqrt{s} &= 10.6\,\textrm{GeV}, \\
\alpha (10.6\,\textrm{GeV})&= 1/131.
}
In principle, one power of $\alpha$ that originates from the photon emission of the heavy quark (\cf \Fig\ref{fig:qcd}) could run as low as  $\alpha(M_{\chi_{cJ}}/2)$.  
However, for simplicity we will ignore this potential uncertainty and evaluate all $\alpha$ at the same value.
To estimate the uncertainty in the choice of the renormalization scale of the strong coupling, we use four possible values of $\mu$
\eq{
\alpha_{\rm s} (\sqrt{s}) &= 0.171, \\
\alpha_{\rm s} (\sqrt{s}/2) &= 0.200,   \\
\alpha_{\rm s} (M_{\chi_{c0}}/2) &= 0.226, \\
\alpha_{\rm s} (M_{\chi_{c2}}/2) &= 0.223,
}
all obtained with \textsc{RunDec};
$\alpha_{\rm s} (\sqrt{s}/2)$ is taken as the central value, while $\alpha_{\rm s} (\sqrt{s})$ and $\alpha_{\rm s} (M_{\chi_{c0,2}}/2)$ are used to estimate the uncertainty in $\mu$. 

\begin{table}[ht]
\renewcommand{\arraystretch}{4.0}
\begin{tabular}{ c  c  c  c  c  c}
\hline \hline  \rule{0pt}{1.9\normalbaselineskip}
 & \pbox{20cm}{ $\mathcal{O}(\alpha_{\rm s}^0 v^0)$} 
 & \pbox{20cm}{ $\mathcal{O}(\alpha_{\rm s}^0 v^0)$ \\ and $\mathcal{O}(\alpha_{\rm s} v^0)$}   &
\pbox{20cm}{
$\mathcal{O}(\alpha_{\rm s}^0 v^0)$, $\mathcal{O}(\alpha_{\rm s} v^0)$ \\ and $\mathcal{O}^{\ast}(\alpha_{\rm s}^0 v^2)$} &
\pbox{20cm}{$\mathcal{O}(\alpha_{\rm s}^0 v^0)$, $\mathcal{O}(\alpha_{\rm s} v^0)$ \\ and $\mathcal{O}(\alpha_{\rm s}^0 v^2)$} & \pbox{20cm}{$\frac{\sigma_8}{\sigma_1}$} \\
\hline			
$\quad \sigma(\chi_{ c_0}) \quad$
& \pbox{20cm}{$1.09 \pm 0.33$ \\ \phantom{tmp} \mbox{}} & \pbox{20cm}{$2.03 \pm 0.61$ \\ $\pm 0.75 \pm 0.04$ \\ $\pm 0.00$} & 
\pbox{20cm}{$1.34 \pm 0.02$ \\ $\pm 0.30 \pm 0.04$ \\ $\pm 0.17$} & \pbox{20cm}{$1.22 \pm 0.49$ \\ $\pm 0.48 \pm 0.04$ \\ $\pm 0.37$} & 
\pbox{20cm}{$0.91 \pm 0.38$ \\ $\pm 0.16 \pm 0.00$ \\ $\pm 0.16$} \\
$\sigma(\chi_{ c_1})$
& \pbox{20cm}{$15.16 \pm 4.53$ \\ \phantom{tmp} \mbox{}} & \pbox{20cm}{$17.47 \pm 5.22$ \\ $\pm 5.87 \pm 0.69$ \\ $\pm 0.00$} & 
\pbox{20cm}{$12.78 \pm 1.25$ \\ $\pm 2.83 \pm 0.71$ \\ $\pm 1.10$} & \pbox{20cm}{$13.87 \pm 5.97$ \\ $\pm 1.10 \pm 0.68$ \\ $\pm 0.77$} & 
\pbox{20cm}{$1.09 \pm 0.36$ \\ $\pm 0.15 \pm 0.01$ \\ $\pm 0.15$} \\
$\sigma(\chi_{ c_2})$
& \pbox{20cm}{$6.30 \pm 1.88$ \\ \phantom{tmp} \mbox{}} & \pbox{20cm}{$4.91 \pm 1.47$ \\ $\pm 2.01 \pm 0.66$ \\ $\pm 0.00$} & 
\pbox{20cm}{$3.72 \pm 0.46$ \\ $\pm 1.24 \pm 0.66$ \\ $\pm 0.28$} & \pbox{20cm}{$4.11 \pm 2.13$ \\ $\pm 0.63 \pm 0.65$ \\ $\pm 0.38$} & 
\pbox{20cm}{$1.10 \pm 0.44$ \\ $\pm 0.20 \pm 0.02$ \\ $\pm 0.19$} \\
  \hline \hline
\end{tabular}
\caption{Estimates for $\sigma (e^+ e^- \to \chi_{ cJ} + \gamma)$ (in fb) at $\sqrt{s} = 10.6\,\textrm{GeV}$ with $\mu = \sqrt{s}/2$,
  where $\braket{0| \mathcal{P}_1 (^3 P_0)|0}$ and $\braket{0| \mathcal{T}_8 (^3 P_0)|0}$ are determined from \Eqs\eqref{eq:ldmes1}.
  The first column shows results obtained by including only $\braket{0| \mathcal{O}_1 (^3 P_0)|0}$ multiplied by its tree level matching coefficient.
  Here the only relevant error source is the uncertainty in $\braket{0| \mathcal{O}_1 (^3 P_0)|0}$. The loop corrections are added in the second column.
  In the third column we include contributions from $\braket{0| \mathcal{P}_1 (^3 P_0)|0}$ but not from $\braket{0| \mathcal{T}_8 (^3 P_0)|0}$.
  In the fourth column all three LDMEs are included.
  The last column displays the ratio of the two final cross sections, when $\braket{0| \mathcal{T}_8 (^3 P_0)|0}$ is included ($\sigma_8$) or omitted ($\sigma_1$).}
\label{tab:finalPredictionsJ0}
\end{table}

\begin{table}[ht]
\renewcommand{\arraystretch}{4.0}
\begin{tabular}{ c  c  c  c  c  c}
\hline \hline  \rule{0pt}{1.9\normalbaselineskip}
 & \pbox{20cm}{ $\mathcal{O}(\alpha_{\rm s}^0 v^0)$} 
 & \pbox{20cm}{ $\mathcal{O}(\alpha_{\rm s}^0 v^0)$ \\ and $\mathcal{O}(\alpha_{\rm s} v^0)$}   &
\pbox{20cm}{
$\mathcal{O}(\alpha_{\rm s}^0 v^0)$, $\mathcal{O}(\alpha_{\rm s} v^0)$ \\ and $\mathcal{O}^{\ast}(\alpha_{\rm s}^0 v^2)$} &
\pbox{20cm}{$\mathcal{O}(\alpha_{\rm s}^0 v^0)$, $\mathcal{O}(\alpha_{\rm s} v^0)$ \\ and $\mathcal{O}(\alpha_{\rm s}^0 v^2)$} & \pbox{20cm}{$\frac{\sigma_8}{\sigma_1}$} \\
\hline			
$\quad \sigma(\chi_{ c_0}) \quad$
& \pbox{20cm}{$0.89 \pm 0.27$  \\ \phantom{tmp} \mbox{}} & \pbox{20cm}{$1.71 \pm 0.51$ \\ $\pm 0.64 \pm 0.03$ \\ $\pm 0.00$} & 
\pbox{20cm}{$1.50 \pm 0.19$ \\ $\pm 0.13 \pm 0.09$ \\ $\pm 0.10$} & \pbox{20cm}{$1.89 \pm 0.02$ \\ $\pm 0.13 \pm 0.15$ \\ $\pm 0.22$} & 
\pbox{20cm}{$1.26 \pm 0.17$ \\ $\pm 0.02 \pm 0.03$ \\ $\pm 0.07$} \\
$\sigma(\chi_{ c_1})$
& \pbox{20cm}{$13.58 \pm 4.06$  \\ \phantom{tmp} \mbox{}} & \pbox{20cm}{$15.46 \pm 4.62$ \\ $\pm 5.18 \pm 0.61$ \\ $\pm 0.00$} & 
\pbox{20cm}{$14.05 \pm 2.39$ \\ $\pm 1.63 \pm 0.24$ \\ $\pm 0.69$} & \pbox{20cm}{$10.31 \pm 4.44$ \\ $\pm 1.63 \pm 0.89$ \\ $\pm 0.52$} & 
\pbox{20cm}{$0.73 \pm 0.19$ \\ $\pm 0.03 \pm 0.05$ \\ $\pm 0.07$} \\
$\sigma(\chi_{ c_2})$
& \pbox{20cm}{$5.75 \pm 1.72$  \\ \phantom{tmp} \mbox{}} & \pbox{20cm}{$4.46 \pm 1.33$ \\ $\pm 1.77 \pm 0.57$ \\ $\pm 0.00$} & 
\pbox{20cm}{$4.10 \pm 0.77$ \\ $\pm 0.87 \pm 0.47$ \\ $\pm 0.18$} & \pbox{20cm}{$2.77 \pm 1.49$ \\ $\pm 0.87 \pm 0.70$ \\ $\pm 0.25$} & 
\pbox{20cm}{$0.68 \pm 0.24$ \\ $\pm 0.07 \pm 0.09$ \\ $\pm 0.09$} \\
  \hline \hline
\end{tabular}
\caption{Estimates for $\sigma (e^+ e^- \to \chi_{ cJ} + \gamma)$ (in fb) at $\sqrt{s} = 10.6\,\textrm{GeV}$ with $\mu = \sqrt{s}/2$,
  where $\braket{0| \mathcal{P}_1 (^3 P_0)|0}$ and $\braket{0| \mathcal{T}_8 (^3 P_0)|0}$ are determined from \Eqs\eqref{eq:ldmes2}.
  The first column shows results obtained by including only $\braket{0| \mathcal{O}_1 (^3 P_0)|0}$ multiplied by its tree level matching coefficient.
  Here the only relevant error source is the uncertainty in $\braket{0| \mathcal{O}_1 (^3 P_0)|0}$. The loop corrections are added in the second column.
  In the third column we include contributions from $\braket{0| \mathcal{P}_1 (^3 P_0)|0}$ but not from $\braket{0| \mathcal{T}_8 (^3 P_0)|0}$.
  In the fourth column all three LDMEs are included.
  The last column displays the ratio of the two final cross sections, when $\braket{0| \mathcal{T}_8 (^3 P_0)|0}$ is included ($\sigma_8$) or omitted ($\sigma_1$).}
\label{tab:finalPredictionsJ2}
\end{table}

\begin{figure}[ht]
\centering
\begin{subfigure}[b]{0.45\textwidth}
\includegraphics[width=\textwidth,clip]{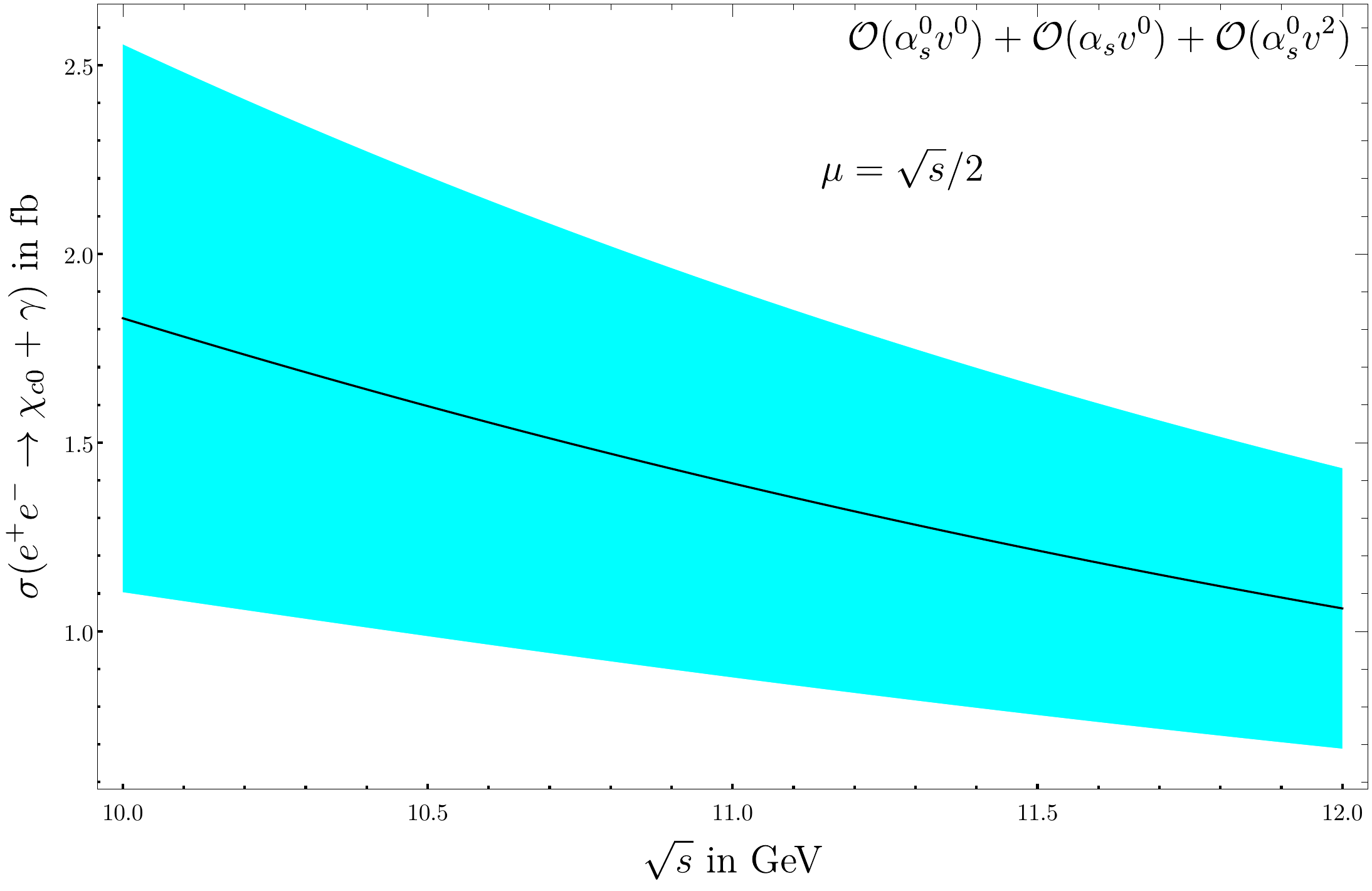}
    \end{subfigure}
\begin{subfigure}[b]{0.45\textwidth}
\includegraphics[width=\textwidth,clip]{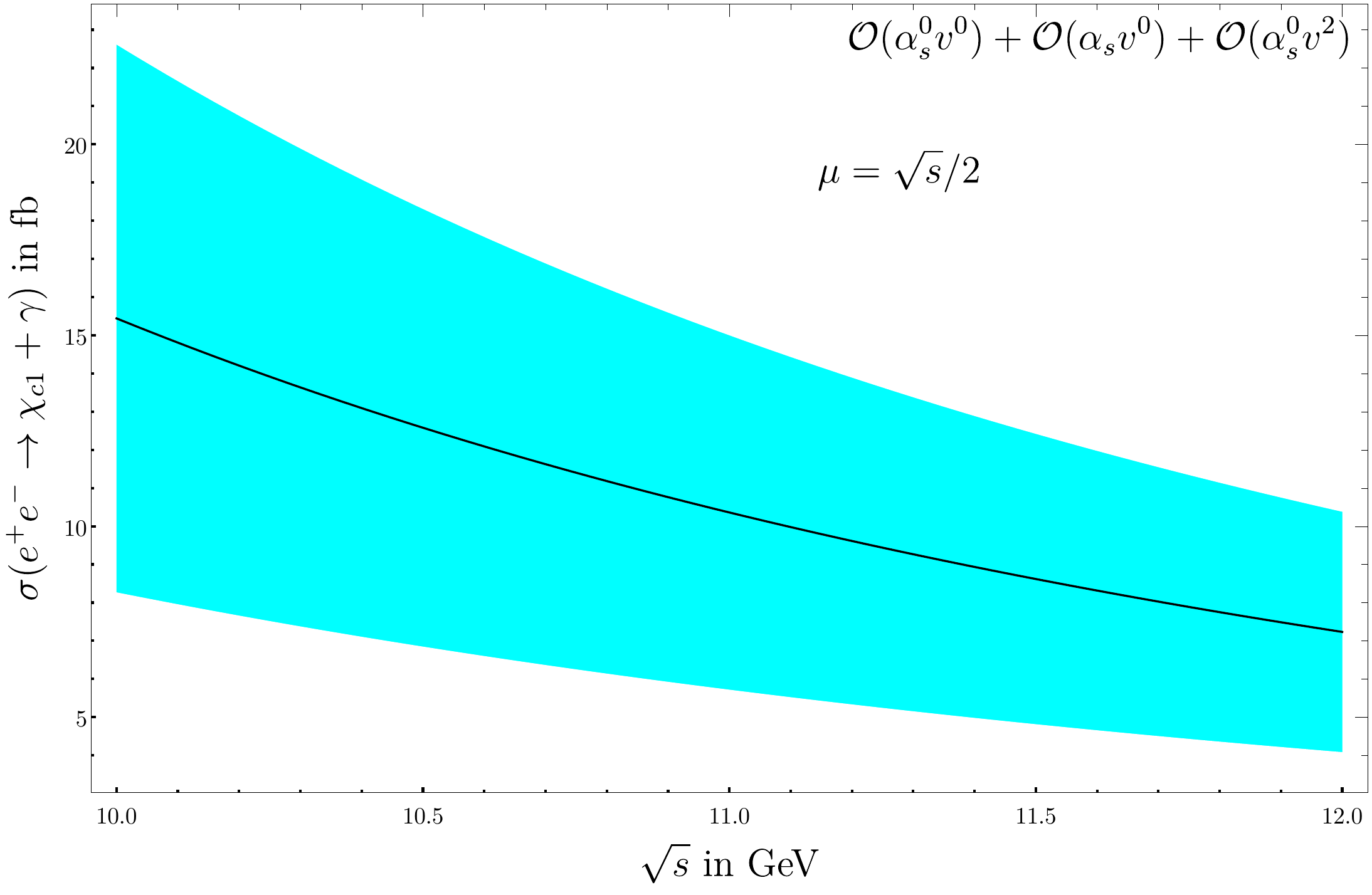}
    \end{subfigure}
\begin{subfigure}[b]{0.45\textwidth}
\includegraphics[width=\textwidth,clip]{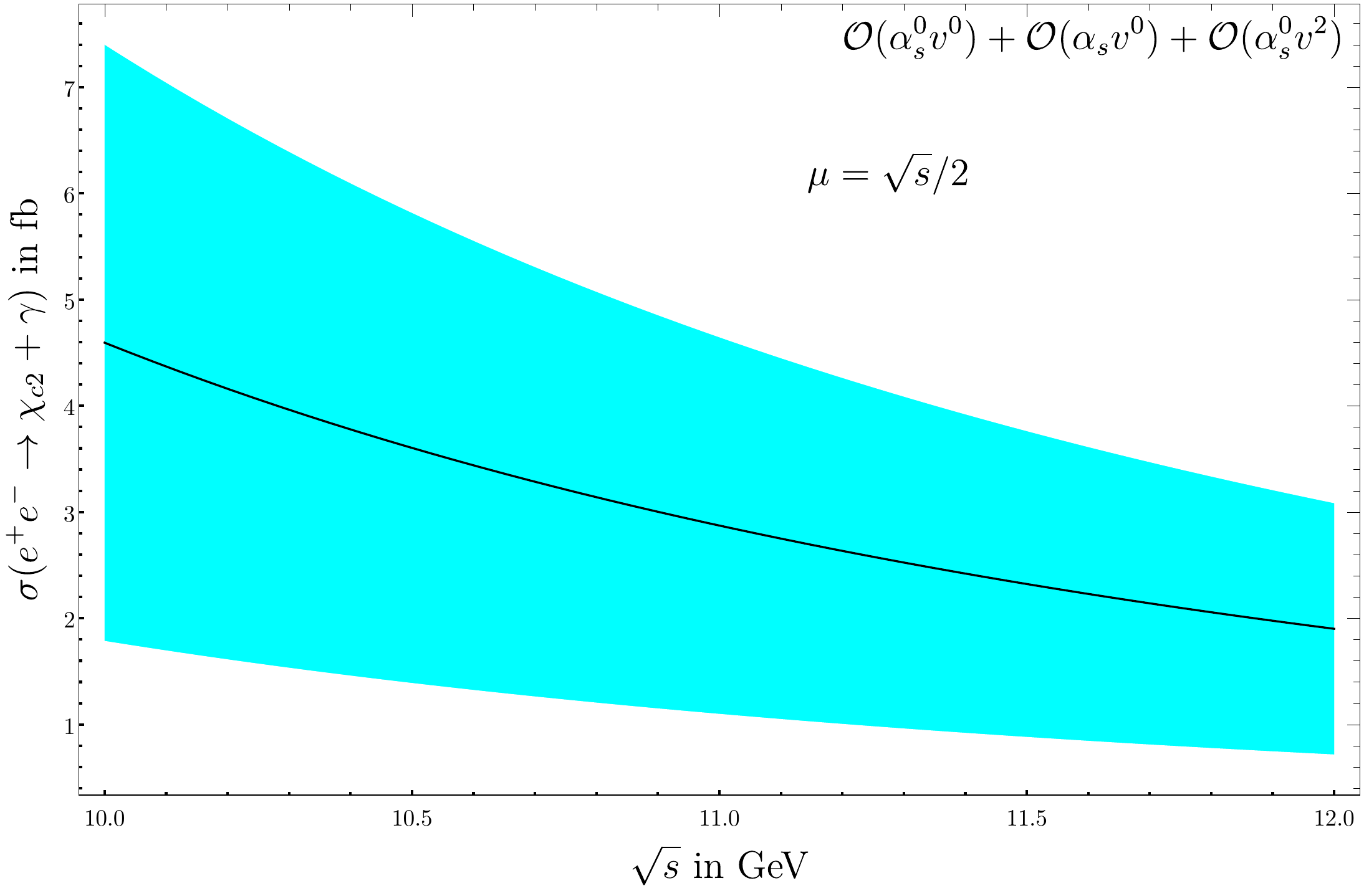}
    \end{subfigure}
\caption{Final cross sections for the exclusive production of $\chi_{cJ}$ and a hard photon in the energy region of Belle II.
  The results are obtained from averaging the cross sections with the subleading LDMEs obtained from \Eqs\eqref{eq:ldmes1} and \Eqs\eqref{eq:ldmes2}.
  The error bands are obtained from the quadrature of all four sources of uncertainties listed in Tables \ref{tab:finalPredictionsJ0} and \ref{tab:finalPredictionsJ2}.}
\label{fig:cs2}
\end{figure}

\begin{figure}[ht]
\centering
\begin{subfigure}[b]{0.45\textwidth}
\includegraphics[width=\textwidth,clip]{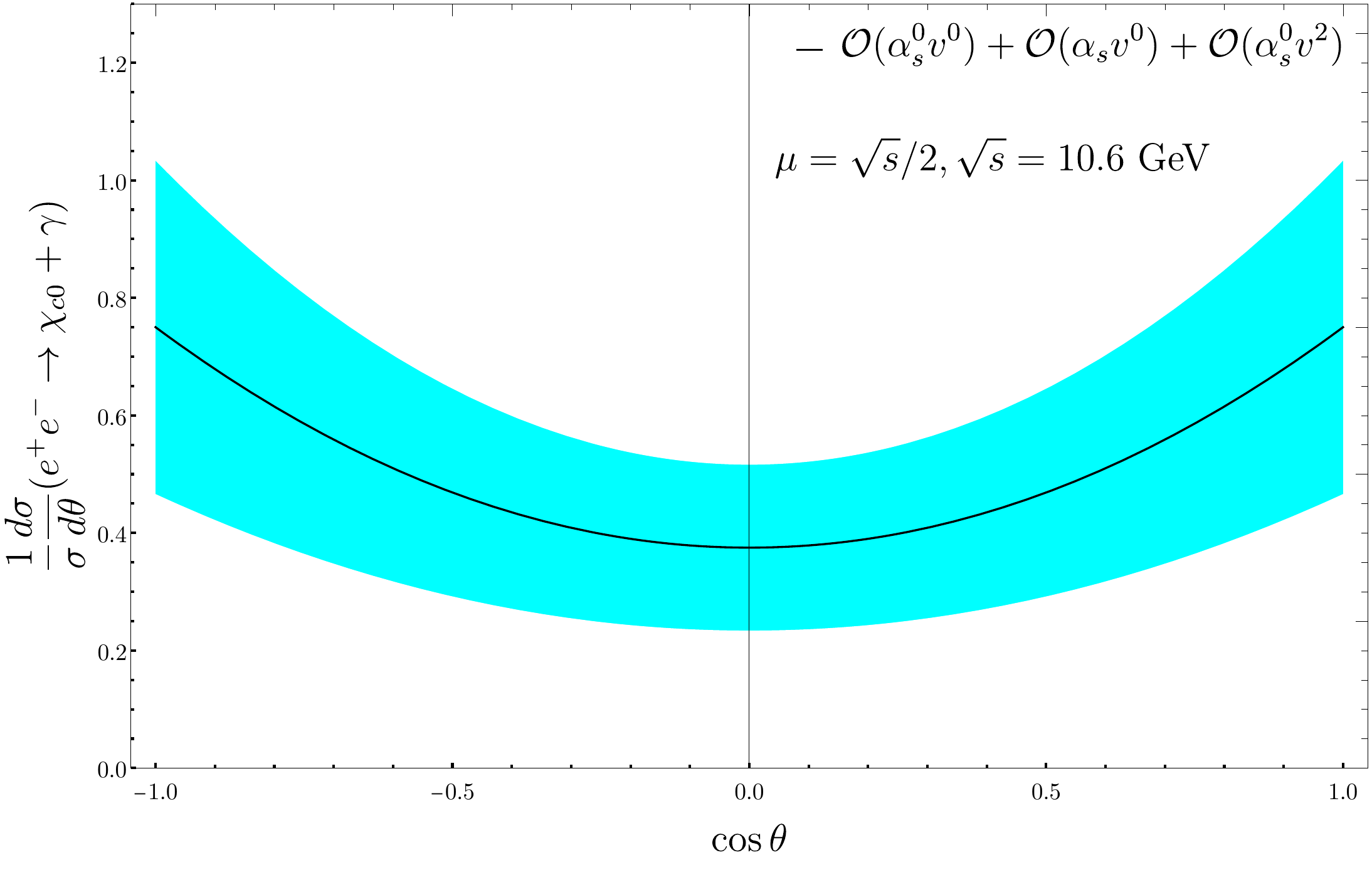}
    \end{subfigure}
\begin{subfigure}[b]{0.45\textwidth}
\includegraphics[width=\textwidth,clip]{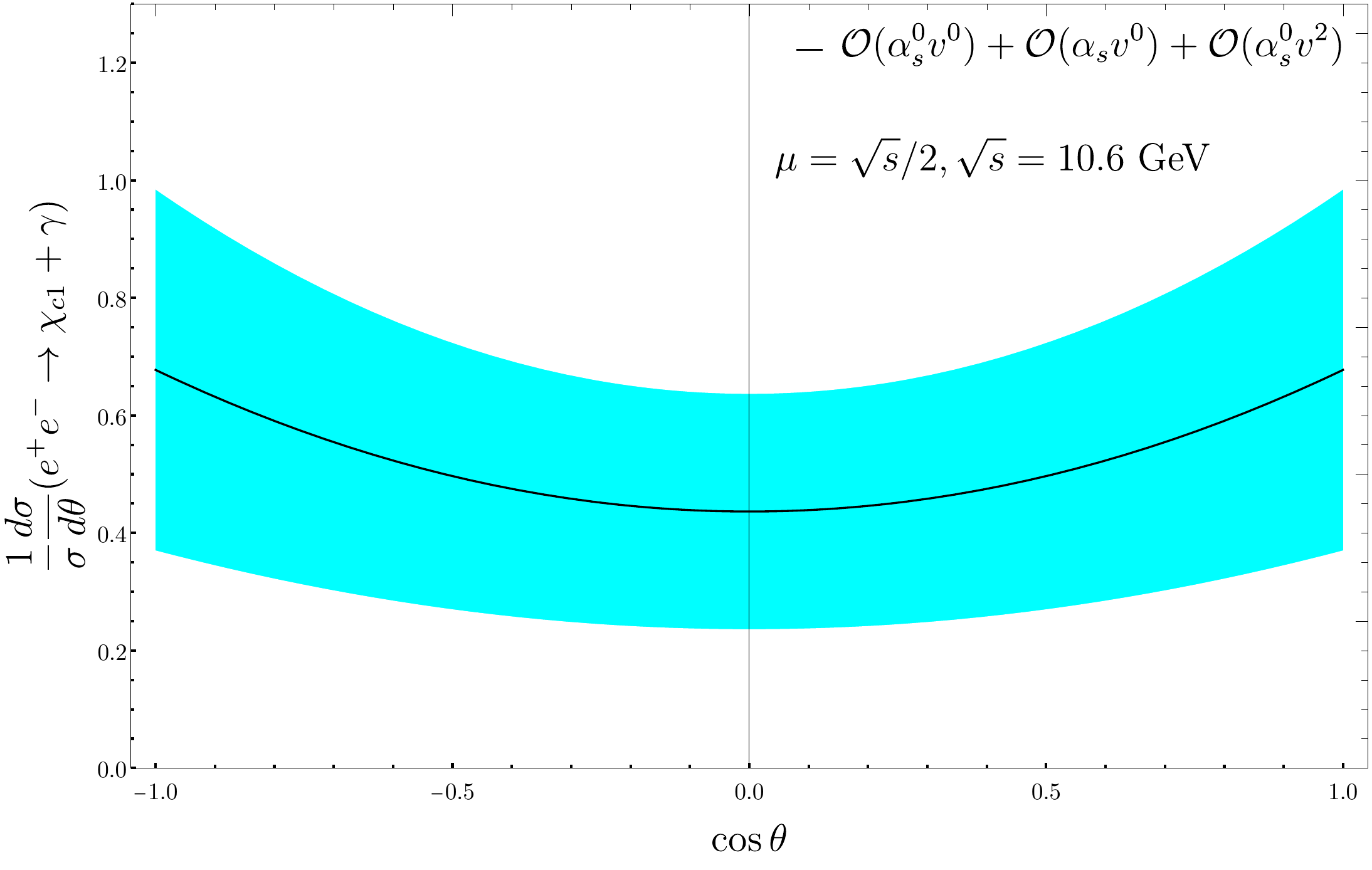}
    \end{subfigure}
\begin{subfigure}[b]{0.45\textwidth}
\includegraphics[width=\textwidth,clip]{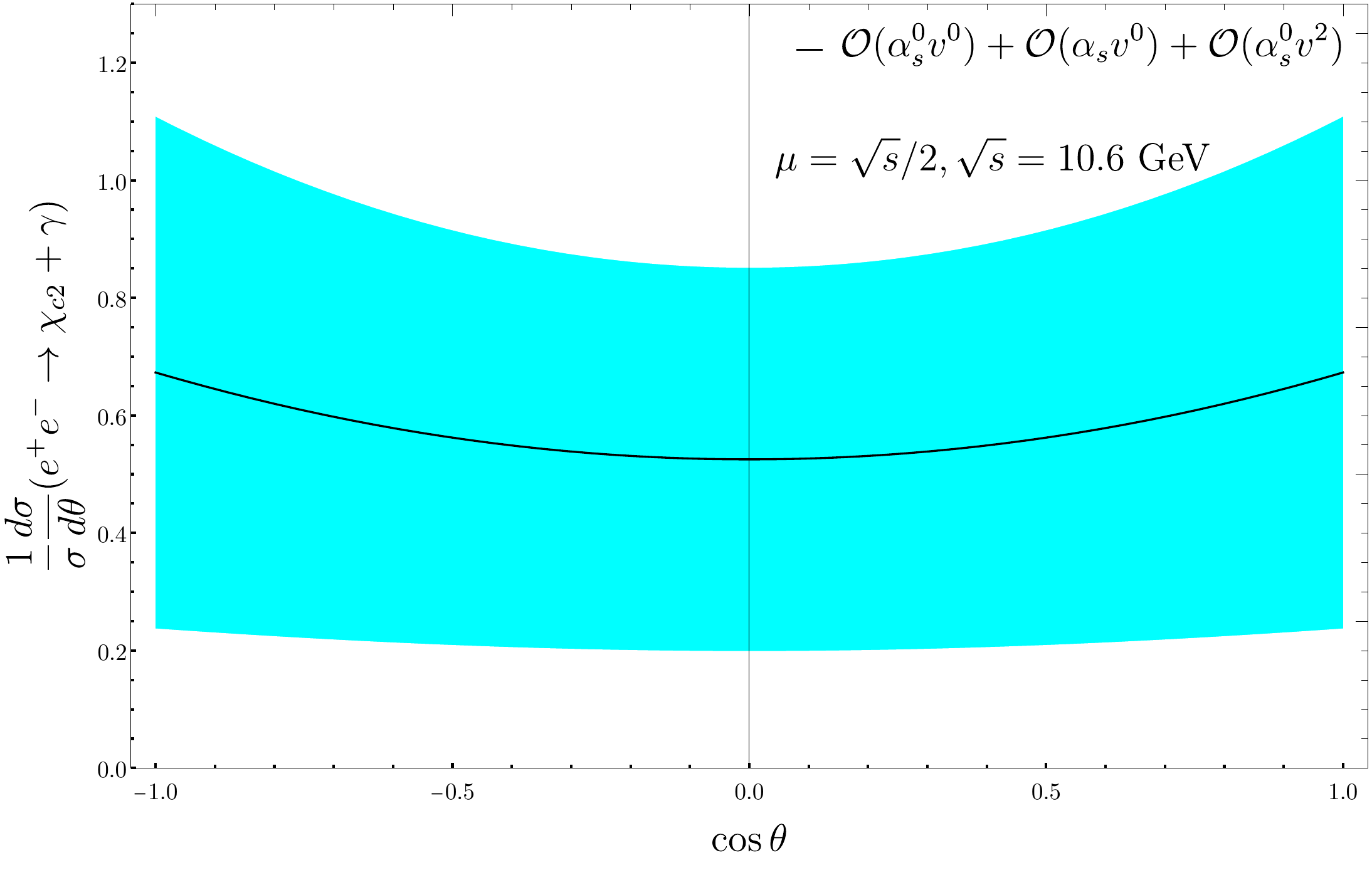}
    \end{subfigure}
\caption{Differential cross sections for the exclusive production of $\chi_{cJ}$ and a hard photon at $\sqrt{s} = 10.6 \textrm{ GeV}$.
  The results are obtained from averaging the cross sections with the subleading LDMEs obtained from \Eqs\eqref{eq:ldmes1} and \Eqs\eqref{eq:ldmes2}.
  The error bands are obtained from the quadrature of all four sources of uncertainties listed in Tables \ref{tab:finalPredictionsJ0}
  and \ref{tab:finalPredictionsJ2}.
  The central values and the errors are normalized with respect to the average
  of the central values of the total cross sections at $\sqrt{s} = 10.6 \textrm{ GeV}$.}
\label{fig:csdiff}
\end{figure}

\begin{figure}[ht]
\centering
\begin{subfigure}[b]{0.45\textwidth}
\includegraphics[width=\textwidth,clip]{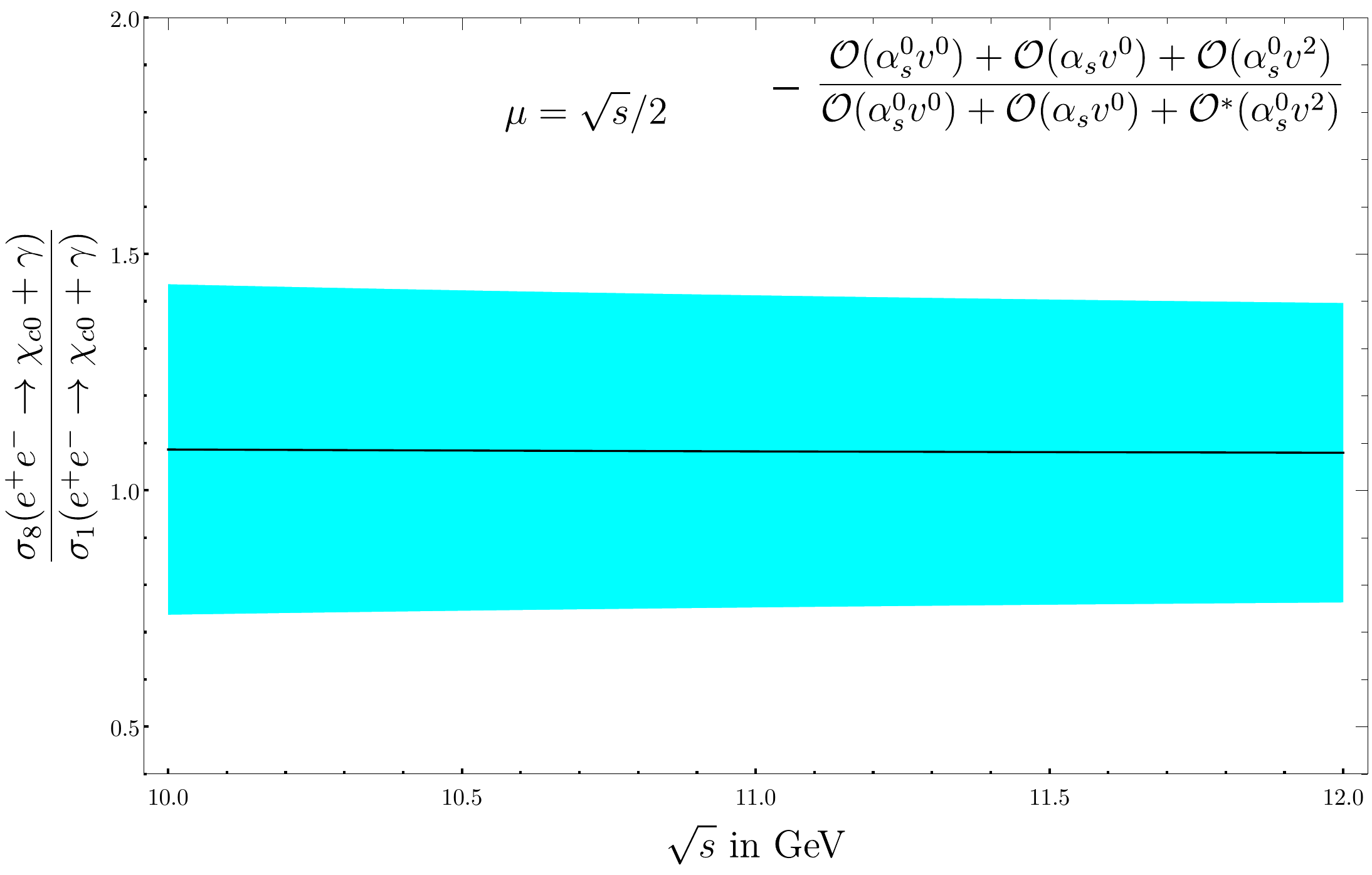}
    \end{subfigure}
\begin{subfigure}[b]{0.45\textwidth}
\includegraphics[width=\textwidth,clip]{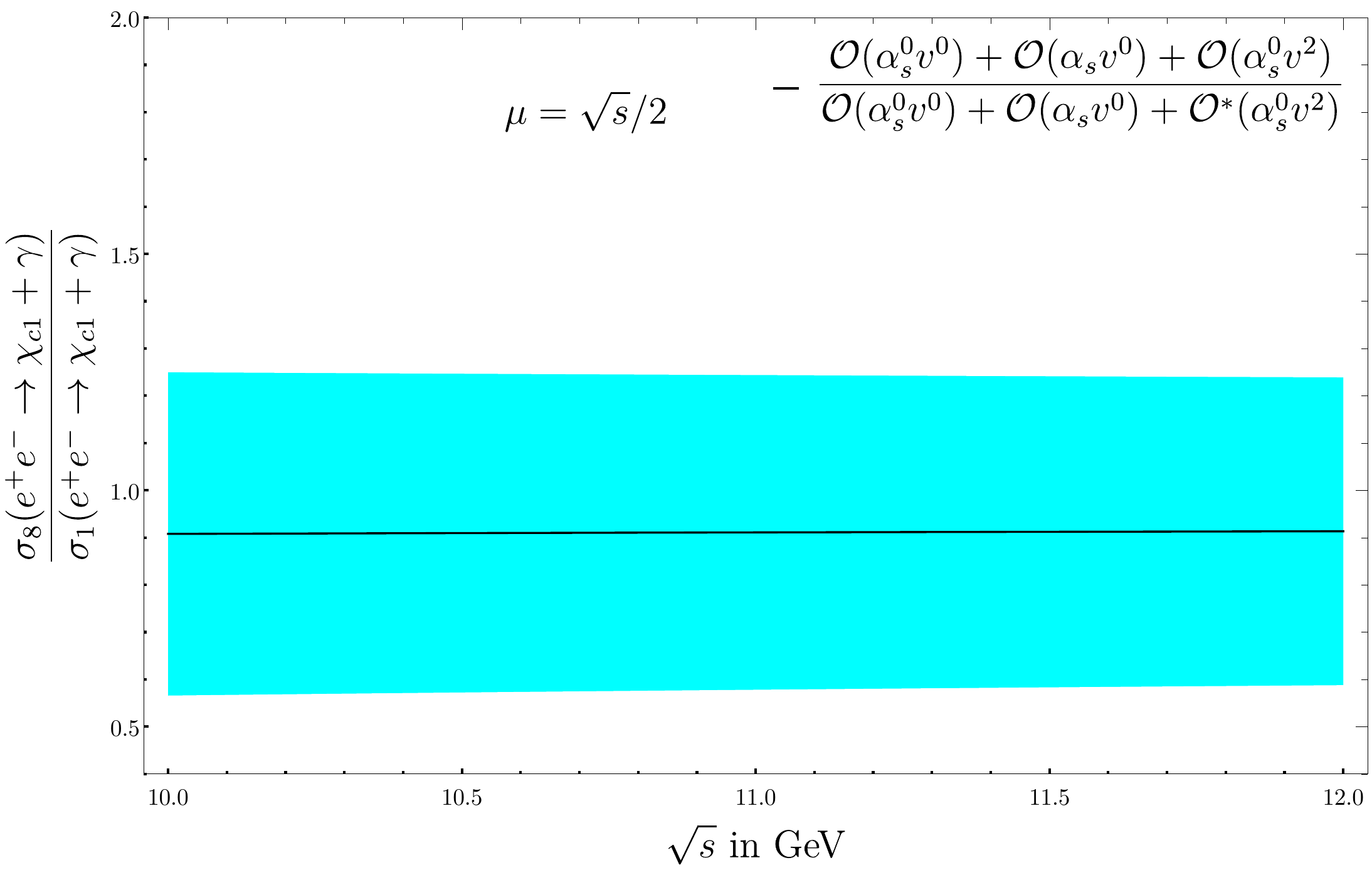}
    \end{subfigure}
\begin{subfigure}[b]{0.45\textwidth}
\includegraphics[width=\textwidth,clip]{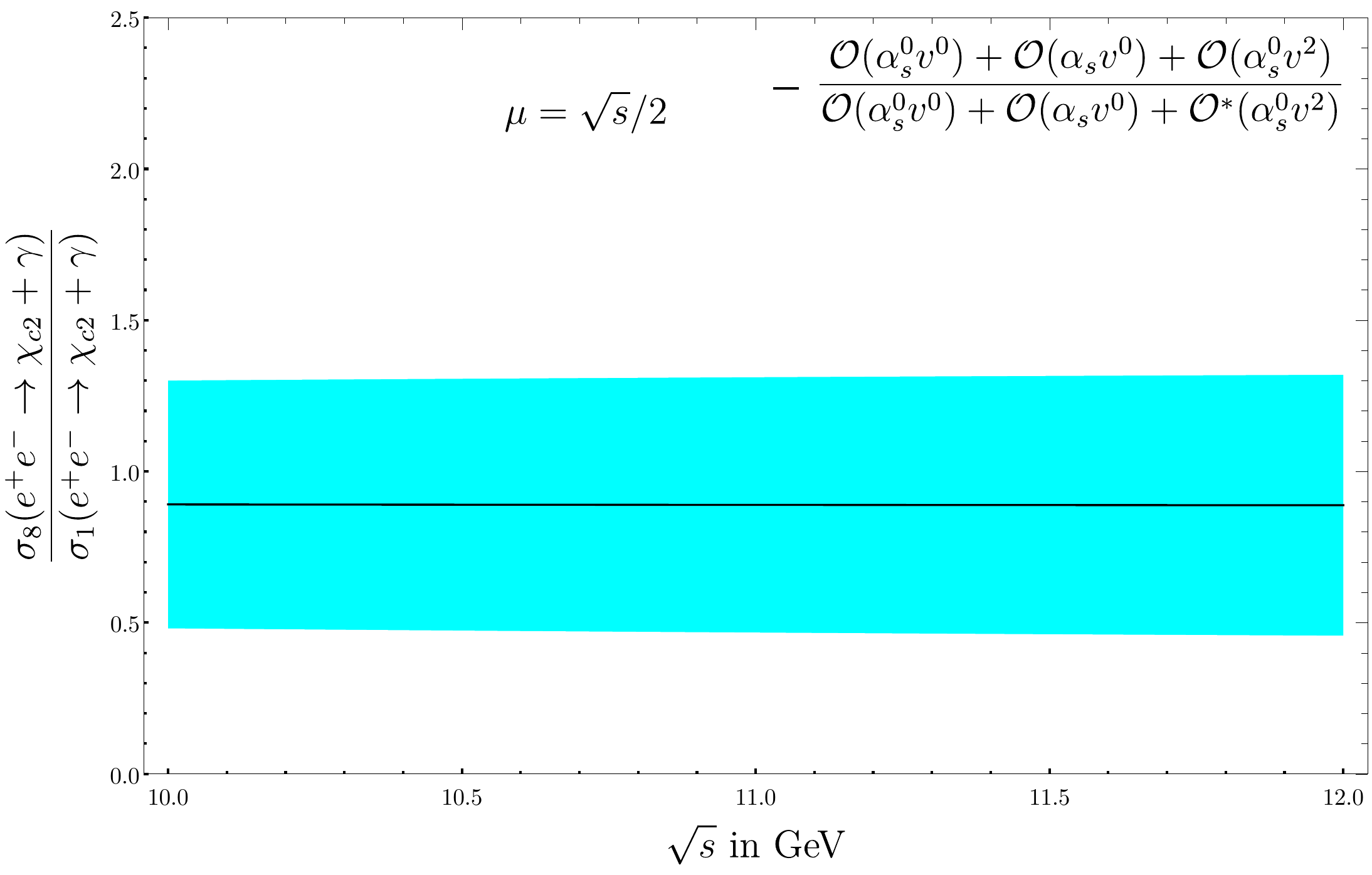}
    \end{subfigure}
\caption{Ratios of the total cross sections with $\braket{0| \mathcal{T}_8 (^3 P_0)|0}$ included ($\sigma_8$) or omitted ($\sigma_1$) in the energy region of Belle II.
  The results are obtained from averaging the ratios with the subleading LDMEs obtained from \Eqs\eqref{eq:ldmes1} and \Eqs\eqref{eq:ldmes2}.
  The error bands are obtained from the quadrature of all four sources of uncertainties listed in Tables \ref{tab:finalPredictionsJ0} and \ref{tab:finalPredictionsJ2}.}
\label{fig:cs-ratio}
\end{figure}

The contributions of different LDMEs to the total cross section in the energy region of Belle II are shown in \Figs \ref{fig:cs} and \ref{fig:csc}, 
and summarized in Tables \ref{tab:finalPredictionsJ0} and \ref{tab:finalPredictionsJ2} for $\sqrt{s} = 10.6\,\textrm{GeV}$.
The full cross sections with the corresponding error bands are presented in \Fig\ref{fig:cs2} (total cross sections as functions of $\sqrt{s}$) 
and \Fig\ref{fig:csdiff} (differential cross sections for $\sqrt{s} = 10.6\,\textrm{GeV}$ as functions of the angle 
between the beam line and the outgoing photon, $\theta$, in the laboratory frame).
In \Fig\ref{fig:cs-ratio} we also show the ratios of the total cross sections,
where the LDME $\braket{0| \mathcal{T}_8 (^3 P_0)|0}$ is included (denoted as $\sigma_8$) or omitted (denoted as $\sigma_1$) as a function of~$\sqrt{s}$. 
The only difference between $\sigma_8$ and $\sigma_1$ is that in the latter the value of $\braket{0| \mathcal{T}_8 (^3 P_0)|0}$ is set to zero, 
while the values of $\braket{0| \mathcal{O}_1 (^3 P_0)|0}$ and $\braket{0| \mathcal{P}_1 (^3 P_0)|0}$ entering $\sigma_8$ and $\sigma_1$ are identical and given by \Eqs\eqref{eq:BT}, \eqref{eq:ldmes01}, and \eqref{eq:ldmes02}.\footnote{Our $\sigma_1$ estimate does not require a different determination of $\braket{0| \mathcal{P}_1 (^3 P_0)|0}$ to be consistent. Owing to \Eq\eqref{eq:decayWithCorrs2}, the value of $\braket{0| \mathcal{P}_1 (^3 P_0)|0}$ quoted in \Eq\eqref{eq:ldmes-21} does not depend on the size of $\braket{0| \mathcal{T}_8 (^3 P_0)|0}$ and hence does not change for $\braket{0| \mathcal{T}_8 (^3 P_0)|0} = 0$. Furthermore, determining $\braket{0| \mathcal{P}_1 (^3 P_0)|0}$ from \Eq\eqref{eq:decayWithCorrs1} with $\braket{0| \mathcal{T}_8 (^3 P_0)|0}$ set to zero, leads to 
$\braket{0| \mathcal{P}_1 (^3 P_0)|0}  = (0.099 \pm 0.064 \pm 0.080 \pm 0.015)\, \textrm{GeV}^7$ which is almost identical to the value quoted in \Eq\eqref{eq:ldmes-11}.}
The numerical values of the ratios at $\sqrt{s} = 10.6\,\textrm{GeV}$ can be read off the last columns of Tables \ref{tab:finalPredictionsJ0} and \ref{tab:finalPredictionsJ2}.

All the uncertainties are estimated assuming Gaussian error propagation. 
The first uncertainty originates from $\braket{0| \mathcal{O}_1 (^3 P_0)|0}$. 
It is clearly a large contribution to the total uncertainty of the production cross section for $\chi_{cJ}$. The second uncertainty stems from the uncertainty in $x_J$ and therefore in the binding energy. It is equally a large source of uncertainty for all produced states. The third uncertainty is obtained from varying the renormalization scale $\mu$ and is, with the exception of the $\chi_{c0}$ production in Table \ref{tab:finalPredictionsJ0}, comparable with the uncertainty in the binding energy. Finally, the fourth source of uncertainty comes from the uncertainties in the experimental values of $M_{\chi_{cJ}}$ and $\Gamma (\chi_{cJ} \to \gamma \gamma)$. It leads to a large error in the case of $\chi_{c0}$ but is much less important for $\chi_{c1}$ and $\chi_{c2}$. The surprisingly small errors for the production cross section of $\chi_{c0}$ in the determination from \Eqs\eqref{eq:ldmes2} stem from a large numerical cancellation between the contributions from $\braket{0| \mathcal{O}_1 (^3 P_0)|0}$ and $\braket{0| \mathcal{T}_8 (^3 P_0)|0}$. This cancellation is accidental as it does not appear in the determination from \Eqs\eqref{eq:ldmes1}. Note that having attributed to the quarkonium binding energy a 100\% uncertainty [see \Eqs\eqref{eq:ebind} and \eqref{uncertaintyxJ}] 
makes that uncertainty parametrically of the same size as the uncertainty of $\braket{0| \mathcal{O}_1 (^3 P_0)|0}$.

Averaging the results stemming from \Eqs\eqref{eq:ldmes1} and \Eqs\eqref{eq:ldmes2}, we obtain
\eq{
\sigma (e^+ e^- \to \chi_{ c0} + \gamma) &= (1.55 \pm 0.26 \pm 0.30 \pm 0.09 \pm 0.43) \textrm{ fb}  , \\
\sigma (e^+ e^- \to \chi_{ c1} + \gamma) &= (12.09 \pm 5.20 \pm 1.37 \pm 0.78 \pm 0.93) \textrm{ fb}, \\
\sigma (e^+ e^- \to \chi_{ c2} + \gamma) &= (3.44 \pm 1.81 \pm 0.75 \pm 0.68 \pm 0.46) \textrm{ fb},
}
where the errors in $\braket{0| \mathcal{O}_1 (^3 P_0)|0}$ (first uncertainty), in $x_J$ (second uncertainty), and in $\mu$ (third uncertainty) were added linearly,
while those  in  $M_{\chi_{cJ}}$ and $\Gamma (\chi_{cJ} \to \gamma \gamma)$ (fourth uncertainty), regarded as independent, were added quadratically.

Adding the uncertainties in quadrature, our final predictions at $\sqrt{s} = 10.6\textrm{ GeV}$ read
\eq{
\sigma (e^+ e^- \to \chi_{ c0} + \gamma) &= (1.6 \pm 0.6) \textrm{ fb}  , \\
\sigma (e^+ e^- \to \chi_{ c1} + \gamma) &= (12.1 \pm 5.5) \textrm{ fb}, \\
\sigma (e^+ e^- \to \chi_{ c2} + \gamma) &= (3.4 \pm 2.1) \textrm{ fb}.
}
As for the influence of the additional LDME $\braket{0| \mathcal{T}_8 (^3 P_J)|0}$ on the total cross section,
we refer to the last columns of Table \ref{tab:finalPredictionsJ0} and \ref{tab:finalPredictionsJ2} and to \Fig\ref{fig:cs-ratio}.
We see that the contribution of $\braket{0| \mathcal{T}_8 (^3 P_J)|0}$ is small, but, at its central value, not negligible and thus potentially significant for all $\chi_{cJ}$.

While our analytic results for the matching coefficients are unambiguous,
the numerical results strongly depend on the estimated size of the quarkonium binding energy and the assumed value of $\braket{0| \mathcal{O}_1 (^3 P_J)|0}$.
Moreover, the experimental errors in the knowledge of $\Gamma (\chi_{cJ} \to \gamma \gamma)$ are also not negligible.
A reliable determination of  $\braket{0| \mathcal{O}_1 (^3 P_J)|0}$, $\braket{0| \mathcal{P}_1 (^3 P_0)|0}$ and $\braket{0| \mathcal{T}_8 (^3 P_0)|0}$
from experimental data or lattice calculations is therefore highly desirable and would provide a major improvement of the present cross section determinations.

\section{\label{sec:summary} Summary}
NRQCD factorization provides a systematic prescription to compute production cross sections of heavy quarkonia order by order in $\alpha_{\rm s}$ and $v$. 
Higher order corrections in $\alpha_{\rm s}$ arise from loop effects, while higher order corrections in $v$ are generated by the inclusion of higher-dimensional operators. 
A consistent computation of higher order corrections requires both loop corrections to matching coefficients of lower dimensional operators  
and relativistic corrections from higher dimensional operators.

In this work we have computed the $\mathcal{O}(\alpha_{\rm s}^0 v^2)$ contribution induced by the higher Fock state \qqbarg to the $e^+ e^- \to \gamma^\ast \to \chi_{ cJ} + \gamma$ cross section,
in this way completing the order $v^2$ corrections to this $P$-wave quarkonium production process.
The new pieces of information contained in this work are the leading order expressions of the matching coefficients $T_8(^3 P_J)$ multiplying LDMEs that depend on the chromoelectric field.

We match QCD to NRQCD perturbatively at the amplitude level, where we treat the final state gluon as a part of the quarkonium system, together with the heavy quark pair. 
In the course of the calculation we encountered infrared singularities induced by the emission of the gluon from a heavy quark line. 
We explicitly verified that such terms cancel exactly in the matching. 
As a further consistency check we performed the matching calculation in two different reference frames:
the CM frame of the colliding leptons and the rest frame of the heavy quarkonium. 
Although these two calculations are technically very different, they lead to the same matching coefficients.

The process $e^+ e^- \to \gamma^\ast \to \chi_{cJ} + \gamma$ should be observable at B-factories with sufficiently high CM energy, \eg at Belle II. 
As no published experimental results are available so far, the phenomenological relevance of contributions from higher Fock states remains to be tested.

\begin{acknowledgments}
N.~B., V.~S. and A.~V. thank Geoffrey Bodwin for correspondence and useful discussions. 
Their work was supported by the DFG and the NSFC through funds provided to the Sino-German CRC 110 ``Symmetries and the Emergence of Structure in QCD'' (NSFC Grant No. 11621131001) and by the DFG grant BR 4058/2-1.
They also acknowledge support from the DFG cluster of excellence ``Origin and structure of the universe'' (\url{www.universe-cluster.de}). 
The work of W.~C. and Y.~J. is supported in part by the National Natural Science Foundation of China under Grants No. 11475188 and No. 11621131001 (CRC110 by DFG and NSFC), 
by the IHEP Innovation Grant under Contract No. Y4545170Y2, and by the State Key Lab for Electronics and Particle Detectors.
The Feynman diagrams in this paper were created with \textsc{JaxoDraw} \cite{Binosi:2003yf}. 
In the preparation of the plots we also employed the \textit{Mathematica} package \textsc{MaTeX} (\url{github.com/szhorvat/MaTeX}).
\end{acknowledgments}

\FloatBarrier

\appendix

\section{Additional short-distance coefficients} \label{sec:appendix0}
As discussed in \Sec\ref{sec:nrqcdside}, not all short-distance coefficients that are obtained in the matching of the amplitude are relevant for our final NRQCD-factorized production cross section. 
Therefore, as a by-product of our calculations we could also determine short-distance coefficients that appear up to $\mathcal{O}(v^4)$
and contribute through the dominant Fock state \qqbar to the amplitude $\mathcal{A}_{\textrm{pert QCD}}$:
\eq{
  \mathcal{A}_{\textrm{NRQCD}}^{J=0} &+  \frac{c_4^{J=0}}{m^5} \braket{H |\psi^\dagger \left ( - \frac{i}{2} \overleftrightarrow{\bfD}  \right )^4 \chi|0},}
for $J=0$ states, and 
\eq{
\mathcal{A}_{\textrm{NRQCD}}^{J=2} & +   \frac{(c_4^{J=2})^{ij}}{M^5} \braket{H |\psi^\dagger  \left ( - \frac{i}{2} \right)^2 \overleftrightarrow{\bfD}^{(i}  
\overleftrightarrow{\bfD}^{j)} \left ( - \frac{i}{2} \overleftrightarrow{\bfD}  \right )^2 \chi|0} \nonumber \\
& +   \frac{(c_5^{J=2})^{ij}}{M^4} \left ( \braket{H |\psi^\dagger  \left ( - \frac{i}{2} \right)^2 \overleftrightarrow{\bfD}^{(i}  
\overleftrightarrow{\bfD}^{j)} \left ( - \frac{i}{2} \overleftrightarrow{\bfD} \cdot \bfSigma \right )  \chi|0} \right. \nonumber \\
 & \left. - \frac{2}{5} \braket{H |\psi^\dagger  \left ( - \frac{i}{2} \right)^2 \overleftrightarrow{\bfD}^{(i}  
\overleftrightarrow{\bfSigma}^{j)} \left ( - \frac{i}{2} \overleftrightarrow{\bfD}  \right )^2 \chi|0} \right ),
}
for $J=2$ states.
We provide their values in the laboratory frame
\eq{
c_4^{J=0} & =- \frac{8}{15}  \lambda (\boldsymbol{L} \cdot (\hat{\bfk} \times \bfeps^{\ast}_{\gamma})), \\
(c_4^{J=2})^{ij} & = \frac{8}{7} \lambda (\boldsymbol{L} \cdot (\hat{\bfk} \times \bfeps^{\ast}_{\gamma})) \hat{\bfk}^i \hat{\bfk}^j, \\
(c_5^{J=2})^{ij} & = \frac{1}{42} \lambda r_{-} \left ( 5 \sqrt{r} (1 - 5\sqrt{r})  (\boldsymbol{L} \cdot \hat{\bfk}) (\hat{\bfk}^i \bfeps^{\ast j}_{\gamma} + i \leftrightarrow j ) \right. \nonumber \\
& \left. + 25 r ( \boldsymbol{L}^i  \bfeps^{\ast j}_{\gamma} + i \leftrightarrow j ) -
 (3 + 17r ) (\boldsymbol{L} \cdot \bfeps^{\ast}_{\gamma}) \hat{\bfk}^i \hat{\bfk}^j
 \right ),
}
and in the rest frame
\eq{
c_{4,R}^{J=0} & = - \frac{8}{15}\lambda (\boldsymbol{L}_R\cdot({\hat{\bfk}_{R}}\times \bfeps_{\gamma,R}^\ast)), \\
(c_{4,R}^{J=2})^{ij} & = \frac{8}{7}\lambda (\boldsymbol{L}_R\cdot({\hat{\bfk}_{R}}\times \bfeps_{\gamma,R}^\ast)) \hat{\bfk}_{R}^i\hat{\bfk}_{R}^j, \\
(c_{5,R}^{J=2})^{ij} &=\frac{i}{42} \lambda r_{-} \left ( 25r (\boldsymbol{L}_R^i  \bfeps_{\gamma,R}^{\ast j} + i \leftrightarrow j) 
-5r r_{+}(5r+3)(\boldsymbol{L}_R\cdot\hat{\bfk}_{R}) (\hat{\bfk}_{R}^i \bfeps_{\gamma,R}^{\ast j} +i \leftrightarrow j)  \right. \nonumber \\
 & \left.  - (3+ 17r) (\boldsymbol{L}_R\cdot\bfeps_{\gamma,R}^\ast)\hat{\bfk}_{R}^i \hat{\bfk}_{R}^j \right),
}
as a reference for future studies. For the notation we refer to \Sec\ref{sec:qcdside}.

\section{Threshold expansion method} \label{sec:appendix1}
\subsection{2-body system}
We consider, first, the case of a quarkonium state described by a $Q \overline{Q}$ pair whose quark and antiquark carry momenta $p_1$ and $p_2$.  
We follow \cite{Braaten:1996jt}.
These momenta can be written in terms of the Jacobi momenta of a 2-body system as 
\begin{equation}
p_1 = \frac{1}{2} P + \mathcal{Q}, \quad \quad p_2 = \frac{1}{2} P - \mathcal{Q},
\end{equation}
where $P= p_1+p_2$ denotes the total momentum of the quarkonium,
while $\mathcal{Q}= (p_1-p_2)/2$ stands for the relative momentum between the heavy quark and the heavy antiquark.

In the laboratory frame, where the heavy quarkonium is moving (\ie $\bfp_{1} + \bfp_{2} \neq 0$), we have
\begin{equation}
P = (\sqrt{P^2+ \bfP^2}, \bfP), \quad \quad \mathcal{Q} = (\sqrt{\mathcal{Q}^2+ \bm{\mathcal{Q}}^2}, \bm{\mathcal{Q}}).
\end{equation}
In the rest frame of the heavy quarkonium, where $\bfp_{1,R} + \bfp_{2,R} = 0$, $P_R$ and $\mathcal{Q}_R$ have a particularly simple form
\begin{equation}
P_R = (2\sqrt{\bfq^2 + m^2},0) \equiv (2 E_q,0) ,
  \quad \quad
\mathcal{Q}_{R}  \equiv q = (0,\bfq),
\end{equation}
with
\begin{equation}
\bfq \equiv \bfp_{1,R} = - \bfp_{2,R}.
\end{equation}
Here $\bfq$ is indeed soft and can be used as an expansion parameter. 
The two frames are related by a Lorentz transformation, which means that we can express $\mathcal{Q}$ and $P$ in terms of~$\bfq$. 
According to \cite{Braaten:1996jt} the Lorentz transformation reads 
\begin{equation}
\mathcal{Q}^\mu = \Lambda^\mu_{\phantom{\mu} i} \bfq^i,
\end{equation}
\begin{equation}
\Lambda^0_{\phantom{0} i} = \frac{\bfP^i}{2 E_q}, \quad \quad
\Lambda^i_{\phantom{i} j} = \delta^{ij} +  \left ( \frac{P^0}{2 E_q} - 1 \right ) \hat{\bfP}^i \hat{\bfP}^j,
\end{equation}
where
\eq{
P^0 = \sqrt{4 E_q^2 + \bfP^2}.
}

The Dirac spinors in the rest frame are given by
\begin{equation}
u_R(p_{1,R}) = N \begin{pmatrix}
  \xi  \\  \frac{\bfq \cdot \bfSigma }{E_q + m}\xi
\end{pmatrix}, \quad  \quad
v_R(p_{2,R}) = N \begin{pmatrix}
  \frac{ -\bfq \cdot \bfSigma }{E_q + m} \eta  \\  \eta
\end{pmatrix},
\end{equation}
where $N$ is the normalization factor.
In case of relativistic normalization, it is given by $\sqrt{E_q+m}$, while in the nonrelativistic case one has $N = \sqrt{(E_q+m)/(2 E_q)}$. 
When boosted to the laboratory frame they become
\eq{
u(p_1) &= \frac{1}{\sqrt{4 E_q (P^0 + 2 E_q) }} (2 E_q + \slashed{P} \gamma^0) N \begin{pmatrix}
  \xi  \\  \frac{\bfq \cdot \bfSigma }{E_q + m}\xi
\end{pmatrix}, \\
v(p_2) &= \frac{1}{\sqrt{4 E_q (P^0 + 2 E_q) }} (2 E_q + \slashed{P} \gamma^0) N \begin{pmatrix}
  \frac{ -\bfq \cdot \bfSigma }{E_q + m} \eta  \\  \eta
\end{pmatrix}.
}
For practical purposes, it is more convenient to work with boosted Dirac bilinears.
Only vector and axial vector bilinears appear in our process of interest; they can be written as
\eq{
\bar{u}(p_1) \gamma^\mu v(p_2) &= N^2 \Lambda^\mu_{\phantom{0} i} \left (\frac{2 E_q}{E_q + m} \xi^\dagger \bfSigma^i \eta - \frac{2}{(E_q+m)^2} \bfq^i \xi^\dagger \bfq \cdot \bfSigma \eta \right), \\
\bar{u}(p_1) \gamma^\mu \gamma_5 v(p_2) &= N^2 \left (\frac{m}{E_q} \frac{1}{E_q+m} P^\mu \xi^\dagger \eta -
2 i \frac{1}{E_q + m} \Lambda^\mu_{\phantom{0} i}  \xi^\dagger (\bfq \times \bfSigma)^i \eta\right).
}

\subsection{3-body system} \label{expansion-co}
We now turn to the case of a quarkonium state composed by a heavy quark, a heavy antiquark and a gluon.
The momenta of the three constituents can be written in terms of the Jacobi momenta of a 3-body system as
\begin{equation}
p_1 = \frac{1}{3} P + \mathcal{Q}_1 - \mathcal{Q}_2, \quad \quad
p_2 = \frac{1}{3} P - \mathcal{Q}_1 - \mathcal{Q}_2, \quad \quad
p_g = \frac{1}{3} P + 2 \mathcal{Q}_2.
\end{equation}
In the rest frame, where $\bfp_{1,R} + \bfp_{2,R} + \bfp_{g,R} = \bfP_R  = 0$, we have
\begin{equation}
p_{1,R} =  \frac{1}{3} P_R + q_1 - q_2, \quad \quad
p_{2,R} =  \frac{1}{3} P_R- q_1 - q_2, \quad \quad
p_{g,R} =  \frac{1}{3} P_R  + 2 q_2,
\end{equation}
or
\begin{equation}
q_1 = \frac{1}{2} \left ( p_{1,R} -p_{2,R} \right ), \quad \quad
q_2 = \frac{1}{6} \left ( 2 p_{g,R} -p_{1,R} - p_{2,R}\right ),
\end{equation}
and, in particular,
\begin{equation}
\bfp_{1,R} = \bfq_1 -  \bfq_2, \quad \quad
\bfp_{2,R} = -\bfq_1 - \bfq_2, \quad \quad
\bfp_{g,R} = 2 \bfq_2,
\end{equation}
with $\bfq_1$ and $\bfq_2$ not being hard momenta.
Furthermore,
\eq{
q_1^0  &= \frac{1}{2} \left ( \sqrt{ (\bfq_1 - \bfq_2)^2 + m^2} - \sqrt{(\bfq_1 +  \bfq_2)^2 + m^2} \right ), \\
q_2^0 &=  \frac{1}{6} \left (4 |\bfq_2| - \sqrt{ (\bfq_1 - \bfq_2)^2 + m^2} - \sqrt{(\bfq_1 + \bfq_2)^2 + m^2} \right ),
}
and
\begin{equation}
P_R^0 =  \left ( 2 |\bfq_2| +\sqrt{ (\bfq_1 -  \bfq_2)^2 + m^2} + \sqrt{(\bfq_1 +  \bfq_2)^2 + m^2}   \right ).
\end{equation}
Since scalar products of 4-vectors are Lorentz invariant, we have
\begin{equation}
\mathcal{Q}_1^2  =  (q_1^0)^2 -\bfq_1^2, \quad 
\mathcal{Q}_2^2  =   (q_2^0)^2 -\bfq_2^2, \quad 
P^2 = (P^0_R)^2 \textrm{ that implies } 
P^0  = \sqrt{(P_R^0)^2 + \bfP^2}.
\end{equation}
To rewrite the laboratory frame 4-momenta, $\mathcal{Q}_1$ and $\mathcal{Q}_2$, in terms of the soft expansion parameters from the rest frame, $\bfq_1$ and $\bfq_2$, we use that
\begin{equation}
  \mathcal{Q}_{1}^\mu = \Lambda^\mu_{\phantom{\mu} \nu} q_{1}^\nu,  \qquad 
  \mathcal{Q}_{2}^\mu = \Lambda^\mu_{\phantom{\mu} \nu} q_{2}^\nu,
\end{equation}
where the boost matrix now reads
\begin{equation}
\Lambda^0_{\phantom{0} 0} = \sqrt{1 + \frac{\bfP^2}{(P_R^0)^2}}, \quad \quad
\Lambda^0_{\phantom{0} i} = \Lambda^i_{\phantom{0} 0} =  \frac{\bfP^i}{P^0_R}, \quad \quad
\Lambda^i_{\phantom{i} j} = \delta^{ij} + \left ( \sqrt{1 + \frac{\bfP^2}{(P_R^0)^2}} - 1 \right ) \hat{\bfP}^i \hat{\bfP}^j.
\end{equation}

As far as the Dirac spinors are concerned, in the rest frame they read
\eq{
u_R(p_{1,R}) = N_1 \begin{pmatrix}
  \xi  \\  \frac{\bfp_{1,R} \cdot \bfSigma }{E_{1,R} + m}\xi
\end{pmatrix}, \quad  \quad
v_R(p_{2,R}) = N_2 \begin{pmatrix}
  \frac{ \bfp_{2,R} \cdot \bfSigma }{E_{2,R} + m} \eta  \\  \eta
\end{pmatrix},
}
with $E_{i,R} = \sqrt{\bfp_{i,R}^2 + m^2}$ and $N_i = \sqrt{E_{i,R} + m}$ for relativistic and $N_i = \sqrt{(E_{i,R} + m)/(2 E_{i,R})}$ for nonrelativistic normalization respectively. 
The boost matrix for these spinors is given by
\begin{equation}
S (\Lambda) = \sqrt{ \frac{P^0 + P_R^0}{2 P_R^0}} \begin{pmatrix}
  1  & \frac{\bfSigma \cdot \bfP}{P^0 + P_R^0}  \\  \frac{\bfSigma \cdot \bfP}{P^0 + P_R^0}  &  1
\end{pmatrix},
\end{equation}
so that we arrive at
\eq{
u(p_1) = \frac{N_1}{\sqrt{2 P_R^0 (P^0 + P_R^0)}} (P_R^0 + \slashed{P} \gamma^0) \begin{pmatrix}
  \xi  \\  \frac{\bfp_{1,R} \cdot \bfSigma }{E_{1,R} + m}\xi
\end{pmatrix}, \\
v(p_2) = \frac{N_2}{\sqrt{2 P_R^0 (P^0 + P_R^0)}} (P_R^0 + \slashed{P} \gamma^0) \begin{pmatrix}
  \frac{ \bfp_{2,R} \cdot \bfSigma }{E_{2,R} + m} \eta  \\  \eta
\end{pmatrix}.
}
Again, for practical purposes it is more useful to have explicit expressions for the boosted Dirac bilinears, which read
\eq{
\bar{u}(p_1) \gamma^\mu v(p_2) = \Lambda^\mu_{\phantom{\mu} \nu} \bar{u}_R(p_{1,R}) \gamma^\nu v_R (p_{2,R}), \\
\bar{u}(p_1) \gamma^\mu \gamma_5 v(p_2) = \Lambda^\mu_{\phantom{\mu} \nu} \bar{u}_R(p_{1,R}) \gamma^\nu \gamma_5 v_R (p_{2,R}),
}
where the rest frame bilinears are given by
\eq{
\bar{u}_R(p_{1,R}) \gamma^0 v_R (p_{2,R}) &= N_1 N_2 \xi^\dagger \left  (  \frac{\bfp_{1,R} \cdot \bfSigma}{E_{1,R} + m} + \frac{\bfp_{2,R} \cdot \bfSigma}{E_{2,R} + m} \right ) \eta, \\
\bar{u}_R(p_{1,R}) \gamma^i v_R (p_{2,R}) &= N_1 N_2 \xi^\dagger \biggl  [ \bfSigma^i + \frac{1}{E_{1,R}+m} \frac{1}{E_{2,R}+m} \biggl ( \bfp_{1,R}^i (\bfp_{2,R} \cdot \bfSigma)
+ \bfp_{2,R}^i (\bfp_{1,R} \cdot \bfSigma) \nonumber \\
&  - (\bfp_{1,R} \cdot \bfp_{2,R}) \bfSigma^i - i (\bfp_{1,R}\times \bfp_{2,R})^i \biggr )\biggr ] \eta, \\
\bar{u}_R(p_{1,R}) \gamma^0 \gamma_5 v_R (p_{2,R}) &= N_1 N_2 \xi^\dagger \biggl  [ 1 +  \frac{1}{E_{1,R}+m} \frac{1}{E_{2,R}+m} \left( \bfp_{1,R} \cdot \bfp_{2,R}
  + i \bfSigma \cdot (\bfp_{1,R} \times \bfp_{2,R}) \right) \biggr ] \eta, \\
\bar{u}_R(p_{1,R}) \gamma^0 \gamma_5 v_R (p_{2,R}) &= N_1 N_2 \xi^\dagger \biggl [  \frac{\bfp_{1,R}^i - i (\bfp_{1,R} \times \bfSigma)^i}{E_{1,R}+m}  +
\frac{\bfp_{2,R}^i + i (\bfp_{2,R} \times \bfSigma)^i}{E_{2,R}+m} \biggr ] \eta.
}

\section{Polarization sums for a hadron with arbitrary spin} \label{sec:appendix3}
The helicity of a hadron state $\left|H(\lambda)\right\rangle$ with spin $J$ and spin third component $\lambda$ can be represented by a symmetric traceless rank-$J$ tensor $\varepsilon_{\lambda}^{i_1i_2\cdots i_J}$, 
constructed from polarization vectors as
\begin{equation}
\varepsilon_J^{i_1i_2\cdots i_J}\equiv\varepsilon_+^{(i_1}\varepsilon_+^{i_2}\cdots\varepsilon_+^{i_J)}\equiv E^{(J)i_1i_2\cdots i_J}_{~~~~i_1^{\prime}i_2^{\prime}\cdots i_J^{\prime}}\varepsilon_+^{i_1^{\prime}}\varepsilon_+^{i_2^{\prime}}\cdots\varepsilon_+^{i_J^{\prime}},
\end{equation}
where $E^{(J)i_1i_2\cdots i_J}_{~~~~i_1^{\prime}i_2^{\prime}\cdots i_J^{\prime}}$ is the natural projection 
that can be computed following \cite{Coope1970} and 
\begin{equation}
  \varepsilon_{\pm}^i= \frac{1}{\sqrt{2}}\begin{pmatrix}\mp 1\\-i\\0\end{pmatrix},\qquad
  \varepsilon_0^i=\begin{pmatrix}0\\0\\1\end{pmatrix}.
\end{equation}
$\varepsilon_{\lambda}^{i_1i_2\cdots i_J}$ with $\lambda<J$ is easily constructed from $\varepsilon_J^{i_1i_2\cdots i_J}$ by the trick of raising and lowering operators.
Polarization tensors are normalized such that
\begin{equation}
\varepsilon_{\lambda}^{i_1i_2\cdots i_J}\varepsilon_{\lambda^{\prime}}^{i_1i_2\cdots i_J}=\delta^{\lambda\lambda^{\prime}}.
\end{equation}
The polarization tensor transforms under a rotation $R$ according to
\begin{equation}
	R^{i_1i_2\cdots i_J}_{i_1^{\prime}i_2^{\prime}\cdots i_J^{\prime}}\varepsilon_{\lambda}^{i_1^{\prime}i_2^{\prime}\cdots i_J^{\prime}}=D^J_{\lambda^{\prime}\lambda}\varepsilon_{\lambda^{\prime}}^{i_1i_2\cdots i_J},
\end{equation}
where $D^J_{\lambda^{\prime}\lambda}$ is the rotational matrix in a $2J+1$ dimensional representation.
Because $D^J_{\lambda^{\prime}\lambda}$ is a unitary matrix,
\begin{equation}
	 \Pi_{i_1^{\prime}i_2^{\prime}\cdots i_J^{\prime}}^{i_1i_2\cdots i_J}\equiv\sum_{\lambda=-J}^{J}\varepsilon_{\lambda}^{i_1i_2\cdots i_J}\varepsilon_{\lambda}^{i_1^{\prime}i_2^{\prime}\cdots i_J^{\prime}}
\label{Pieps}
\end{equation}
is invariant under rotation.
Moreover, $\Pi_{i_1^{\prime}i_2^{\prime}\cdots i_J^{\prime}}^{i_1i_2\cdots i_J}$ is symmetric and traceless about both upper and lower indices.
Then due to theorem 4 in Ref.\,\cite{Coope1970},
\begin{equation}
\Pi_{i_1^{\prime}i_2^{\prime}\cdots i_J^{\prime}}^{i_1i_2\cdots i_J}\sim E^{(J)i_1i_2\cdots i_J}_{~~~~i_1^{\prime}i_2^{\prime}\cdots i_J^{\prime}}.
\end{equation}
The proportionality constant can easily be determined by observing that
\begin{equation}
\Pi_{i_1^{\prime\prime}i_2^{\prime\prime}\cdots i_J^{\prime\prime}}^{i_1i_2\cdots i_J}\Pi_{i_1^{\prime}i_2^{\prime}\cdots i_J^{\prime}}^{i_1^{\prime\prime}i_2^{\prime\prime}\cdots i_J^{\prime\prime}}=\Pi_{i_1^{\prime}i_2^{\prime}\cdots i_J^{\prime}}^{i_1i_2\cdots i_J},
\end{equation}
which is exactly the one satisfied by the natural projection operator.
Thus, we have
\begin{equation}\label{ea1.3}
\Pi_{i_1^{\prime}i_2^{\prime}\cdots i_J^{\prime}}^{i_1i_2\cdots i_J}= E^{(J)i_1i_2\cdots i_J}_{~~~~i_1^{\prime}i_2^{\prime}\cdots i_J^{\prime}}.
\end{equation}

Let us consider an NRQCD operator $\mathcal{O}^{i_1i_2\cdots i_J}$ that is irreducible under space rotation.
According to the Wigner-Eckart theorem,
\begin{equation}\label{ea1.4}
  \left\langle0\right|\mathcal{O}^{i_1i_2\cdots i_J}\left|H(\lambda)\right\rangle=\mathcal{N}(\mathcal{O})\varepsilon_{\lambda}^{i_1i_2\cdots i_J},
\end{equation}
where $\mathcal{N}(\mathcal{O})$ is a constant that can readily be determined.
By virtue of \Eqs \eqref{Pieps} and \eqref{ea1.3}, we have
\begin{equation}
\sum_{\lambda}\left\langle0\right|\mathcal{O}^{\prime i_1^{\prime}i_2^{\prime}\cdots i_J^{\prime}\dagger}\left|H(\lambda)\right\rangle
\left\langle H(\lambda)\right|\mathcal{O}^{i_1i_2\cdots i_J}\left|0\right\rangle=\mathcal{N}(\mathcal{O}^{\prime})^*\mathcal{N}(\mathcal{O})E_{~~~~i_1i_2\cdots i_J}^{(J)i_1^{\prime}i_2^{\prime}\cdots i_J^{\prime}}.
\end{equation}
Taking the trace of both sides with respect to the upper and lower indices, we get
\begin{equation}
\mathcal{N}(\mathcal{O}^{\prime})^*\mathcal{N}(\mathcal{O})=\frac{1}{2J+1}\sum_{\lambda}\left\langle0\right|\mathcal{O}^{\prime i_1i_2\cdots i_J\dagger}
\left|H(\lambda)\right\rangle\left\langle H(\lambda)\right|\mathcal{O}^{i_1i_2\cdots i_J}\left|0\right\rangle.
\end{equation}
Then we have
\begin{equation}
\begin{split}
&\sum_{\lambda}\left\langle0\right|\mathcal{O}^{\prime i_1i_2\cdots i_J\dagger}\left|H(\lambda)\right\rangle\left\langle H(\lambda)\right|\mathcal{O}^{i_1^{\prime}i_2^{\prime}\cdots i_J^{\prime}}\left|0\right\rangle\\
&=\frac{1}{2J+1}\sum_{\lambda}\left\langle0\right|\mathcal{O}^{\prime i_1^{\prime\prime}i_2^{\prime\prime}\cdots i_J^{\prime\prime}\dagger}
\left|H(\lambda)\right\rangle\left\langle H(\lambda)\right|\mathcal{O}^{i_1^{\prime\prime}i_2^{\prime\prime}\cdots i_J^{\prime\prime}}\left|0\right\rangle E^{(J)i_1i_2\cdots i_J}_{~~~~i_1^{\prime}i_2^{\prime}\cdots i_J^{\prime}},
\end{split}
\end{equation}
which for $\chi_c$ states becomes 
\begin{eqnarray}
\sum_{\lambda}\left\langle0\right|\mathcal{O}^{\prime i\dagger}\left|\chi_{c1}(\lambda)\right\rangle\left\langle \chi_{c1}(\lambda)\right|\mathcal{O}^{i^{\prime}}\left|0\right\rangle&=&
\frac{1}{3}\sum_{\lambda}\left\langle0\right|\mathcal{O}^{\prime j\dagger}\left|\chi_{c1}(\lambda)\right\rangle\left\langle \chi_{c1}(\lambda)\right|\mathcal{O}^{j}\left|0\right\rangle\delta^{ii^{\prime}},\\
\sum_{\lambda}\left\langle0\right|\mathcal{O}^{\prime ij\dagger}\left|\chi_{c2}(\lambda)\right\rangle\left\langle \chi_{c2}(\lambda)\right|\mathcal{O}^{i^{\prime}j^{\prime}}\left|0\right\rangle&=&
\frac{1}{5}\sum_{\lambda}\left\langle0\right|\mathcal{O}^{\prime kl\dagger}\left|\chi_{c1}(\lambda)\right\rangle\left\langle \chi_{c2}(\lambda)\right|\mathcal{O}^{kl}\left|0\right\rangle\nonumber\\
&&\quad \times\left(\frac{1}{2} \delta ^{i i^{\prime}} \delta ^{j j^{\prime}}+\frac{1}{2} \delta ^{i j^{\prime}} \delta ^{j i^{\prime}}-\frac{1}{3} \delta ^{i j} \delta ^{i^{\prime}j^{\prime}}\right).
\end{eqnarray}

\section{Generalized Gremm-Kapustin relations for \texorpdfstring{$^3 P_J$}{3PJ} quarkonia}\label{a3}
The Gremm-Kapustin relations \cite{Gremm:1997dq} follow from
\begin{equation}
  \left\langle H\right|\left[\mathcal{O},H_{\rm eff}\right]\left|0\right\rangle=-\left\langle H\right|H_{\rm eff}\mathcal{O}\left|0\right\rangle =
  (2m-M_H)\left\langle H\right|\mathcal{O}\left|0\right\rangle,
\end{equation}
where $\mathcal{O}$ is an operator with a nonvanishing matrix element $\left\langle H\right|\mathcal{O}\left|0\right\rangle$. 
For instance, taking $H=\chi_{c0}$, and $\mathcal{O}=-i\psi^{\dagger}\mbf{D}\cdot\bm{\sigma}\chi$,
we get up to order $1/m$ (notice that $-\mbf{E}$ is the canonical momentum conjugate to $\mbf{A}$ in the Hamiltonian formalism)
\begin{equation}
(M_{\chi_{c0}}-2m)\left\langle\chi_{c0}\right|i\psi^{\dagger}\mbf{D}\cdot\bm{\sigma}\chi\left|0\right\rangle = 
-\frac{1}{m}\left\langle\chi_{c0}\right|i\psi^{\dagger}\mbf{D}\cdot\bm{\sigma}\mbf{D}^2\chi\left|0\right\rangle
+i\left\langle\chi_{c0}\right|\psi^{\dagger}(g\mbf{E}\cdot\bm{\sigma})\chi\left|0\right\rangle.
\end{equation}
Similarly, for $\eta_c$, $\chi_{c1}$, and $\chi_{c2}$, we have up to order $1/m$
\eq{
(M_{\eta_c}-2m)\left\langle\eta_c\right|\psi^{\dagger}\chi\left|0\right\rangle & =
-\frac{1}{m}\left\langle\eta_c\right|\psi^{\dagger}\mbf{D}^2\chi\left|0\right\rangle-\frac{1}{m}\left\langle\eta_c\right|\psi^{\dagger}(g\mbf{B}\cdot\bm{\sigma})\chi\left|0\right\rangle ,\\
(M_{\chi_{c1}}-2m)\left\langle\chi_{c1}\right|\psi^{\dagger}i(\mbf{D}\times\bm{\sigma})^i\chi\left|0\right\rangle & =
-\frac{1}{m}\left\langle\chi_{c1}\right|\psi^{\dagger}i(\mbf{D}\times\bm{\sigma})^i\mbf{D}^2\chi\left|0\right\rangle \nonumber \\
& \quad + i\left\langle\chi_{c1}\right|\psi^{\dagger}g(\mbf{E}\times\bm{\sigma})^i\chi\left|0\right\rangle ,\\
(M_{\chi_{c2}}-2m)\left\langle\chi_{c2}\right|\psi^{\dagger}i\mbf{D}^{(i}\bm{\sigma}^{j)}\chi\left|0\right\rangle & = 
-\frac{1}{m}\left\langle\chi_{c2}\right|\psi^{\dagger}i\mbf{D}^{(i}\bm{\sigma}^{j)}\mbf{D}^2\chi\left|0\right\rangle
+i\left\langle\chi_{c2}\right|\psi^{\dagger}(g\mbf{E}^{(i}\bm{\sigma}^{j)})\chi\left|0\right\rangle.
}

\bibliography{biblio}

\end{document}